%% file: PIP_coma.tex
\documentclass[a4paper,traditabstract,longauth]{aa}  
\usepackage{graphicx}
\usepackage{lscape}
\usepackage{txfonts}
\usepackage[breaklinks, colorlinks, citecolor=blue]{hyperref}
\usepackage{url}
\usepackage{natbib}
\usepackage{color}
\usepackage{amssymb}
\usepackage{booktabs}
\usepackage{subfig}

\def\xmm{XMM-{\it Newton} }

\def\xmm{{\it XMM-Newton}}

\def\planck{{\it Planck}}
\def\Planck{{\it \Planck }}
\def\rosat{{\it ROSAT}}

\newfont{\gwpfont}{cmssq8 scaled 1000}

\def\lesssim{\mathrel{\hbox{\rlap{\hbox{\lower4pt\hbox{$\sim$}}}\hbox{$<$}}}}
\def\gtrsim{\mathrel{\hbox{\rlap{\hbox{\lower4pt\hbox{$\sim$}}}\hbox{$>$}}}}

\newcommand{\propsim}{\lower 3pt \hbox{$\, \buildrel {\textstyle
       \propto}\over {\textstyle \sim}\,$}}

\begin{document}
%

\input{PIP_10_authors_and_institutes.tex}

\title{\textit{Planck} Intermediate Results. X. Physics of the hot gas in the Coma cluster}
 \date{Received ; accepted}
  \abstract
  {We present an analysis of  \planck\ satellite data on the Coma Cluster observed via the Sunyaev-Zeldovich effect.
Thanks to its great sensitivity, \planck\ is able, for the first time, to detect SZ emission up to $r\approx 3 \times R_{500}$.
 We test previously proposed spherically symmetric models for the pressure distribution in clusters against the azimuthally averaged data.
In particular,  we find that the Arnaud et al. (2010) ``universal'' pressure profile
does not fit Coma, and that their pressure profile for merging systems provides a reasonable fit to the data only at $r<R_{500}$;
by $r=2\times R_{500}$ it  underestimates the observed $y$ profile by a factor of $\simeq 2$.  
This may  indicate that at these larger radii either: i) the cluster SZ emission is contaminated by unresolved SZ sources along the line of sight; or
ii)  the pressure profile of Coma is higher at $r>R_{500}$ than the mean pressure profile predicted by  the simulations used to constrain the  models.  The  \planck\ image shows
 significant local steepening of
the $y$ profile in two regions about half a degree to the west and to
the south-east of the cluster centre.  These features are consistent with
the presence of shock fronts at these radii, and indeed the western
feature was previously  noticed in the \rosat~PSPC mosaic as well as in the radio. 
Using \planck\ $y$ profiles extracted from  corresponding sectors we find 
 pressure jumps of  $4.9^{+0.4}_{-0.2}$ and  
$5.0^{+1.3}_{-0.1}$ in the west and south-east, respectively. 
 Assuming Rankine-Hugoniot pressure jump conditions, we
deduce that  the shock waves should propagate with Mach number $M_{\rm w}=2.03^{+0.09}_{-0.04}$ and  $M_{\rm se}=2.05^{+0.25}_{-0.02}$ in  the west and south-east, respectively. Finally, we find that the $y$ and radio-synchrotron signals are quasi-linearly correlated on Mpc scales, with 
 small intrinsic scatter. This implies either that the energy density of cosmic-ray electrons is relatively constant throughout the cluster, or that the magnetic fields fall off much more slowly with radius than previously thought. 
}
  

\authorrunning{Planck Collaboration}
\titlerunning{The physics of the Coma cluster}
  \maketitle

\input{Planck.tex}

\color{black}
\section{Introduction}\label{sec:intro}

 The Coma cluster is the most spectacular Sunyaev-Zeldovich (SZ)  source in the \planck\ sky.  It is a low-redshift, massive, and  hot cluster, and is sufficiently extended that ~\planck\ can resolve it well spatially. 
Its intracluster medium (ICM)  was observed in SZ for the first time  with the  $5.5{\rm m}$ {\it OVRO} 
telescope \citep{1992AAS...181.8901H, 1995ApJ...449L...5H}.
Later, it was also observed with {\it MSAM1} \citep{1997ApJ...485...22S}, {\it MITO} \citep{2002ApJ...574L.119D}, {\it VSA} \citep{2005MNRAS.359...16L} and 
{\it WMAP} \citep{2011ApJS..192...18K} which detected the cluster with  signal-to-noise ratio of $S/N=3.6$.
As reported in the   all-sky early Sunyaev-Zeldovich cluster paper, \planck\ 
 detected  the Coma cluster with a $S/N>22$ \citep{planck2011-5.1a}.

Coma  has also been  extensively observed in the X-rays from the \rosat~all-sky survey and pointed observations \citep{1992A&A...259L..31B, 1993MNRAS.261L...8W}, as well as via a huge mosaic by  \xmm\  \citep[e.g.][]{2001A&A...365L..74N, 2003A&A...400..811N, 2004A&A...426..387S}. The X-ray emission reveals many spatial features indicating infalling sub-clusters such as NGC4839 \citep{1995ApJ...439..113D,1997ApJ...474L...7V, 2001A&A...365L..74N, 2003A&A...400..811N} , turbulence \citep[e.g.][]{2004A&A...426..387S, 2012MNRAS.tmp.2290C}  and further signs of accretion and strong dynamical activity. 

Moreover, the  Coma cluster hosts a remarkable giant radio halo extending over $1\, {\rm Mpc}$,  which traces the non-thermal emission from relativistic electrons and magnetic fields \citep[e.g.][]{1993ApJ...406..399G, 2011MNRAS.412....2B}.  The radio halo's spectrum and extent require an ongoing, distributed mechanism for acceleration of the relativistic electrons, since their radiative lifetimes against synchrotron and inverse Compton losses are short, even compared to their diffusion time across the cluster \citep[e.g.][]{1999ApJ...520..529S, 2001MNRAS.320..365B}.
 The radio halo also appears to exhibit a shock front in the west, also seen in the X-ray image, and is connected at larger scales with a huge  radio relic in the south-west \citep[][]{1998A&A...332..395E, 2011MNRAS.412....2B}.

In this paper we present a detailed radial  and sector analysis of the  Coma cluster as observed by \planck . 
These  results are compared with 
 X-ray and radio observations obtained with \xmm\  and the Westerbork Synthesis Radio Telescope.

\begin{figure}[t]
\begin{centering}
\includegraphics[width=\linewidth,angle=0,keepaspectratio]{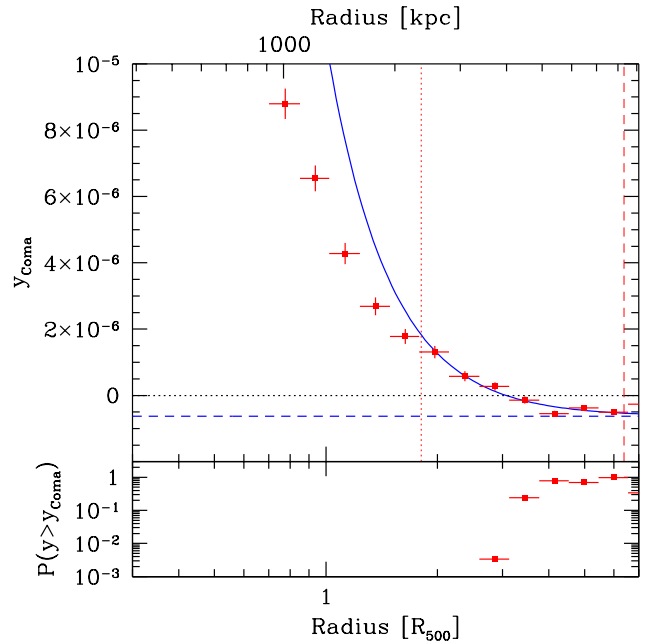}
\end{centering}
\caption{{\it Upper panel:} Radial profile of $y$ in a set of circular
annuli centred on Coma.  The blue curve is the best fitting simple model to the
profile over the radial range from 85~arcmin to 300~arcmin. The model consists
of a power law plus a constant $y_{\rm off}$. The best fitting value of
$y_{\rm off}$ is shown with the dashed horizontal line. Two vertical lines
indicate the range of radii used for fitting.
{\it Lower panel:}  The probability of finding an observed value of
$y>y_{\rm Coma}$ in a given annulus. The probability was estimated by
measuring $y$ in a set of annuli with random centres in any part of
the image outside $5 \times R_{500}$, where $R_{500}=47\, {\rm arcmin}$.  
}
\label{fig:zeroth_lev}
\end{figure}

We use $H_{0} =70\rm{km\, s^{-1}Mpc^{-1}}$, $\Omega_{\rm m}=0.3$ and $\Omega_\Lambda=0.7$, which imply a linear scale of 
$27.7\, \rm{kpc}\, {\rm arcmin}^{-1}$ at the distance of the Coma cluster ($z=0.023$). All the maps are in Equatorial J2000 coordinates.

\section {The \textit {Planck}  frequency maps}\label{s:data}
 
\planck\footnote{\Planck\ (http://www.esa.int/planck) is a project of the European Space Agency (ESA) with instruments provided by two scientific consortia funded by ESA member states (in particular the lead countries: France and Italy) with contributions from NASA (USA), and telescope reflectors provided in a collaboration between ESA and a scientific consortium led and funded by Denmark.} \citep{tauber2010a, planck2011-1.1} is the third-generation space mission to measure the anisotropy of the cosmic microwave background (CMB).  It observes the sky in nine frequency bands covering 30--857\,GHz with high sensitivity and angular resolution from 31\arcm\ to 5\arcm.  The Low Frequency Instrument (LFI; \citealt{Mandolesi2010, Bersanelli2010, planck2011-1.4}) covers the 30, 44, and 70\,GHz bands with amplifiers cooled to 20\,\hbox{K}.  The High Frequency Instrument (HFI; \citealt{Lamarre2010, planck2011-1.5}) covers the 100, 143, 217, 353, 545, and 857\,GHz bands with bolometers cooled to 0.1\,\hbox{K}.  Polarisation is measured in all but the highest two bands \citep{Leahy2010, Rosset2010}.  A combination of radiative cooling and three mechanical coolers produces the temperatures needed for the detectors and optics \citep{Planck2011-1.3}.  Two data processing centres (DPCs) check and calibrate the data and make maps of the sky \citep{planck2011-1.7, planck2011-1.6}.  \planck's sensitivity, angular resolution, and frequency coverage make it a powerful instrument for Galactic and extragalactic astrophysics as well as for cosmology. Early astrophysics results are given in Planck Collaboration VIII-XXVI 2011, based on data taken between 13 August 2009 and 7 June 2010.

This paper is based on the \planck\  nominal survey of 14 months, i.e. taken between 13 August 2009 and 27 November 2010. The whole sky has been covered two times. We refer to \citet{planck2011-1.7} and \citet{planck2011-1.6} for the generic scheme of time ordered information (TOI)
 processing and map making, as well as for the technical characteristics of the maps used. We adopt a circular Gaussian beam pattern for each frequency as described in these papers.
We use the full-sky maps in the nine \planck\  frequency bands provided in HEALPix \citep{gorski2005} $N_{\rm side}=2048$ resolution. An error map is associated with each frequency band and is obtained from the difference of the first half and second half of the \planck\  rings for a given position of the satellite, but are basically free from astrophysical emission. However, they are a good representation of the statistical instrumental noise and systematic errors.
Uncertainties in flux measurements due to beam corrections, map calibrations and uncertainties in bandpasses are expected to be small, as discussed extensively in \citet{planck2011-5.1a,Planck2011-5.1b,Planck2011-5.2a}.

\section{Reconstruction and analysis of the $y$ map}\label{sec:y_map}
The Comptonisation parameter $y$  maps used in this work have been obtained
using the MILCA (Modified Internal Linear Combination Algorithm) method~\citep{2010arXiv1007.1149H}  on the \planck\ frequency maps from 100~GHz to 857~GHz in a region centred on the Coma cluster. MILCA  is a component separation approach aimed at extracting a chosen component (in our case the thermal Sunyaev Zeldovich, tSZ, signal) from a multi-channel set of input maps. It is based mainly  on the well known ILC approach (see for example \citealt{2004ApJ...612..633E}), which searches for the linear combination of  input maps that minimises the variance of the final reconstructed map while imposing spectral constraints. 
For this work, we apply MILCA using two constraints, the first  to preserve the $y$ signal and the second  to remove CMB contamination in the final $y$ map.  Furthermore, we correct for the bias induced by the instrumental noise, and
we simultaneously use the extra degrees of freedom (dof) to minimise residuals from other components (2 dof) and from the instrumental noise (2 dof). These would otherwise increase the variance of the final reconstructed $y$ map.
The noise covariance matrix is estimated from jack-knife maps. To improve the efficiency of the algorithm we perform our separation independently on several bins in the spatial-frequency plane.
The final $y$ map has an effective point spread function (PSF) with a resolution of $10\arcm$ FWHM.
Finally, to characterise the noise properties, such as correlation and inhomogeneities, we use jack-knife and redundancy maps for each frequency and apply the same linear transformation as  used to compute the MILCA $y$ map.
The MILCA procedure  provides us with a data map $y$
together with random realisations of an additive noise 
model $dy$, which is Gaussian, correlated, and may present some non-stationary behaviour across the field of view.
These maps are used to derive radial profiles  and to perform the image analysis, as described below.

We verified that  the reconstruction methods GMCA \citep{2008StMet...5..307B} and NILC \citep{2011MNRAS.410.2481R} give results that are consistent within the errors with the MILCA method  \cite[see][]{planck2012-V}. \\

\subsection {Analysis of radial profiles}\label{sec:y-analysis}
 In this paper, we present various radial profiles $y(r)$ of the 2D distribution of the Comptonisation parameter $y$.
These  allow us to study the underlying pressure distribution of the intracluster medium of Coma.
The  $y$ parameter is proportional to the gas pressure $P=n_{\rm e}{\rm k}T$ integrated along the line of sight: 

\begin{equation}
y=\frac{\sigma_{\rm T}}{m_{\rm e} c^2}\int{{P}(l)dl},
\label{e:yp}
\end{equation}

\noindent where $n_{\rm e}$ and $T$ are the gas electron density and temperature, $\sigma_{\rm T}$ is the Thomson cross-section, $k$ the Boltzmann constant, $m_{\rm e}$ the mass of the electron and $c$ the speed of light.  All the radial profiles $y(r)$ are extracted from the  $y$ map after masking out  bright radio sources. 
In this work we model the observed $y(r)$ projected profiles using the forward  approach described in detail by e.g.  \citet{2008A&A...479..307B}. 
We assume that the three-dimensional pressure profiles
can be adequately  represented by some analytic functions that have the freedom to describe a wide range of  possible profiles. 
The 3D model is  projected along the line of sight, assuming spherical symmetry and  convolved with the \planck\  PSF to produce a projected model function $f(r)$. Finally we fit $f(r)$   to the data
using a $\chi^2$ minimisation of its distance from the radial profiles $y(r) + dy(r)$ derived from the MILCA map ($y(r)$) 
and 1000 realization of its additive noise model ($dy(r)$). 
The $\chi^2$ is calculated in the principal component basis of these noise realisations. This procedure uses an orthogonal transformation to diagonalise the 
noise covariance matrix which, thus, decorrelates  the additive noise fluctuation. 

It is important to say that, as the parameters of the fitting functions are highly degenerate, we adopt two techniques to quantify 
the uncertainties, i) for each individaul parameter, and ii) for the overall model (that is, the global model envelope).

More specifically, the confidence intervals on each parameter are calculated using the 
 percentile method; i.e.,
we rank the 
fitted values and select the value corresponding to the chosen percentile. 
Suppose that our 1000 realizations for a specific  parameter $\zeta$ are already ranked from bottom to top, 
the 
percentile confidence  interval at 68.4\% corresponds to $[\zeta_{158^{th}},\zeta_{842^{th}}]$.
Notice that in this work the confidence intervals are reported with respect to the best-fit value obtained by fitting 
the model to the initial data set.

The envelope of the profiles shown in Figs.~\ref{fig:conbined_y},~\ref{fig:shock},~\ref{fig:p_shock}, \ref{fig:slopes}, \ref{fig:comparison_p_prof},~\ref{fig:total_pressure}, and \ref{fig:sector_pressure} delimit, instead, the  first 684 out of the 1000 model profiles with the lowest  $\chi^2$.
Note that, by design, the forward approach tests the capability of a specific functional to globally reproduce the observed data. For this reason, the error estimates 
represent the uncertainties on the parameters of the fitting function rather than the local uncertainties of the deprojected quantity. This technique has been fully tested on hydrodynamic simulations \citep[e.g.][]{2007ApJ...655...98N, 2010A&A...514A..93M}.

\begin{figure*}[th]
\begin{centering}
\begin{minipage}[t]{\textwidth}
\resizebox{\hsize}{!} {
\includegraphics[height=3cm,angle=0,keepaspectratio]{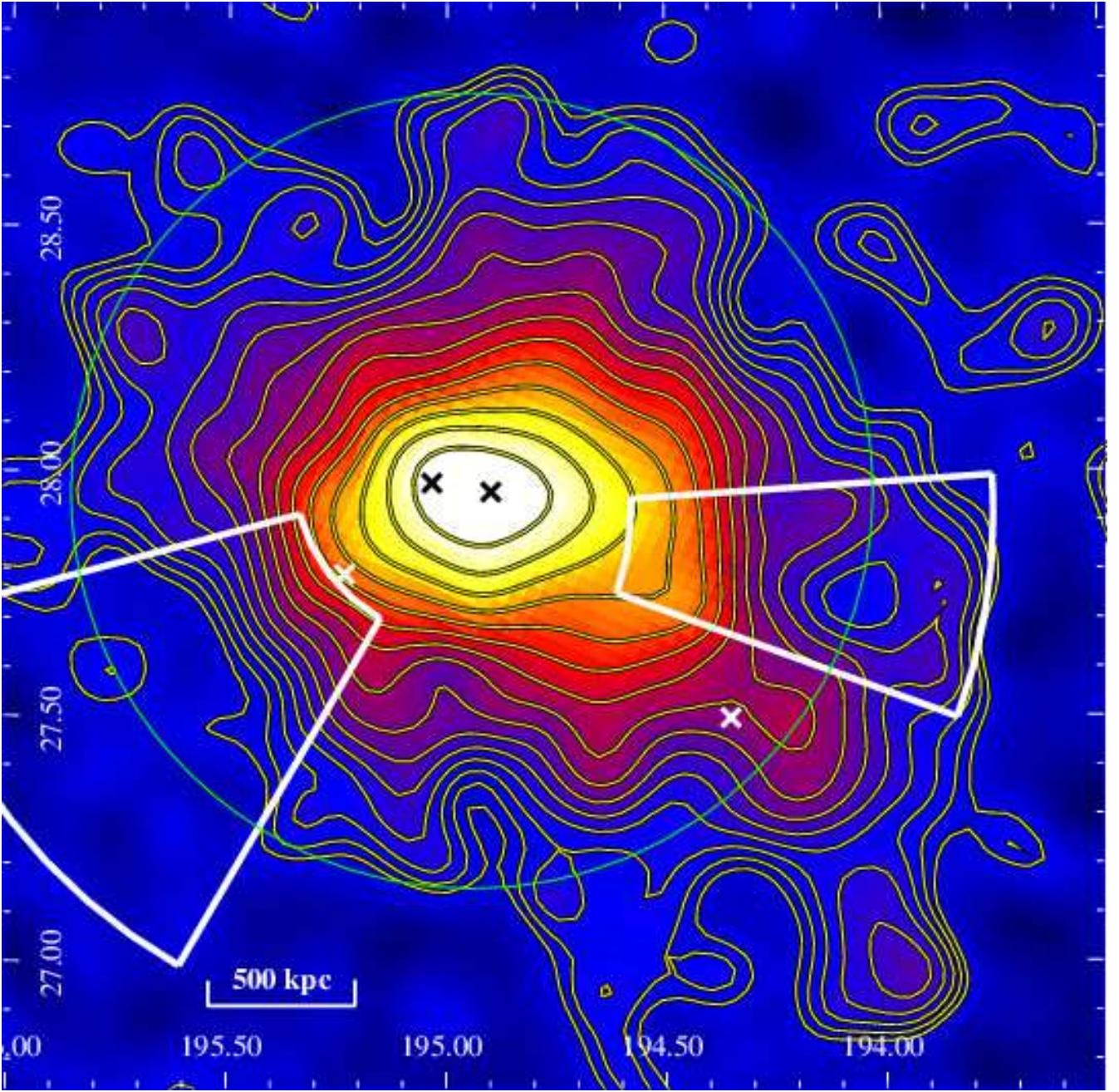}
}
\end{minipage}
\end{centering}
\caption{{\footnotesize 
The \Planck\ $y$ map of the Coma cluster obtained by combining the HFI channels from 100~GHz to 857~GHz.  North is up and west is to the right. The map is corrected for the additive constant $y_{\rm off}$.
The final map bin corresponds to  ${\rm FWHM}=10\arcm$. The image is about $130{\rm arcminute} \times 130 {\rm arcminute}$. 
The contour levels are logarithmically spaced by ${2}^{1/4}$~ (every 4 lines, $y$ increases by a factor 2). The outermost contour corresponds to $y=2\times \sigma_{\rm noise}=4.6\times 10^{-6}$. The green circle indicates $R_{500}$. White and black crosses indicate the position of the brightest galaxies in Coma. 
 The white sectors indicate two regions where the $y$ map shows a local steepening of the radial gradient (see Sect.~\ref{sec:shock} and Fig.~\ref{fig:shock}).}}
\label{fig:coma10}
\end{figure*}

\subsection{Zero level of the $y$ map and the maximum detection radius}

 As a result of the extraction algorithm, \planck\ $y$ maps contain 
an arbitrary additive constant $y_{\rm off}$ which is
 a free parameter in all our $y$-map models. 
This constant can be determined using the \planck\ patch by simply setting to zero the $y$ value measured at very large radii, where we expect to have small or no contribution to the signal 
from the Cluster itself. 
In particular, in the case of the $13.6\deg \times 13.6 \deg $
MILCA-based patch of the image centred on Coma, this constant is
negative, as illustrated in Fig.~\ref{fig:zeroth_lev}. The radial
profile of $y$ was extracted from the $y$ map in a set of circular
annuli centred at ($\rm{RA}$,~$\rm{DEC}$)=
($12^{\rm h}59^{\rm m}47^{\rm s},+27\degr 55\arcmin 53\arcsec$). The errors assigned to the points are crudely
estimated by calculating the variance of the $y$ map blocked to a
pixel size much larger than the size of the \planck\ PSF. The variance
is then rescaled for each annulus, assuming that the correlation of the
noise can be neglected on these spatial scales. For a model
consisting of a power law
plus a constant (over the radial range from 85~arcmin to 300~arcmin)
we find $y_{\rm off}=-6.3\times 10^{-7}\pm 0.9\times 10^{-7}$. 
We note that the precise
value of $y_{\rm off}$ depends weakly on the particular model
used, and on the range of radii involved in the fitting.

To determine the maximum radius at which \planck\ detects a significant
excess of $y$ compared to the rest of the image, we adopted the
following procedure. For every annulus around Coma with measured
$y=y_{\rm Coma}$ we have calculated the distribution of $y=y_{\rm random}$
measured in $300$ annuli of a similar size, but with the centres
randomly placed in any part of the image outside the $5 \times R_{500}$
circle around Coma, where $R_{500}$ is the radius at which the cluster density contrast is $\Delta=500$. When calculating $y_{\rm random}$ the parts of the
annuli within $5\times R_{500}$ were excluded. The comparison of $y_{\rm Coma}$
with the distribution of $y_{\rm random}$ is used to conservatively
estimate the probability of getting  $y>y_{\rm Coma}$ by chance in an
annulus of a given size at a random position in the image (see
Fig.~\ref{fig:zeroth_lev}, lower panel). For the
annulus between 2.6 and 3.1$~R_{500}$ (122~arcmin to 147~arcmin) the
probability of getting $y_{\rm Coma}$ by chance is $\approx 3\times 10^{-3}$
(a crude estimate, given $N=300$ random positions). For smaller radii the probability is much
lower, while at larger radii the probability of getting $y$ in excess
of $y_{\rm Coma}$ is $\sim$10\% or higher. We conclude that \planck\
detects the signal from Coma in narrow annuli $\displaystyle \Delta
R/R=0.2$ at least up to $R_{\rm max}\sim 3\times R_{500}$. This is a conservative
and model-independent estimate. In the rest of the paper we use
parametric models which cover the entire range of radii to fully exploit
\planck\ data even beyond $R_{\rm max}$.

\section{\textit{XMM-Newton} data analysis}\label{sec:X-ray}

 The \xmm\  results presented in this paper have been derived from analysis of the mosaic obtained by combining 
 27 \xmm\ pointings of the Coma cluster available in the archive.
The \xmm\ data have been prepared and analysed using the procedure described in detail in \cite{2011A&A...527A..21B}, and \cite{2008A&A...479..307B}. 
We estimated the $Y_{\rm X}=M_{\rm gas}\times T$ parameter of Coma iteratively using the $Y_{\rm X} - M_{500}$
scaling relation calibrated from hydrostatic mass estimates in a
nearby cluster sample observed with \xmm\ \citep{2010A&A...517A..92A};
we find  $R_{500}\approx (47\pm1){\rm arcmin}\approx (1.31\pm 0.03){~\rm Mpc}$
and we use this value throughout the paper. To study the surface brightness and temperature radial profiles we use the  forward approach described in \citet{2011A&A...527A..21B} taking care to project the temperature profile using  the formula  
appropriate for spectroscopy; i.e., we use the spectroscopic-like temperature introduced by \citealt{2004MNRAS.354...10M}.

\section{The Coma $y$ maps}\label{sec:imaging}

The main goal of this paper is to present the radial and sectoral properties of  the SZ signal from the Coma cluster. Here we  
describe some  general properties of the image;  the full image analysis will be presented in a forthcoming paper that will
make use of  all the \planck\ data, including the extended surveys.
 
Fig.~\ref{fig:coma10} shows the  \planck\ $y$ map of the Coma cluster obtained by combining the HFI channels from 100~GHz to 857~GHz.  
The effective point spread function (PSF) of this map corresponds to ${\rm FWHM}=10\arcm$ and its noise level is  $\sigma_{\rm noise}=2.3\times 10^{-6}$.

To highlight the spatial structure of the $y$ map, in Fig.~\ref{fig:coma10} we overlay the contour levels of the $y$ signal.  We notice that at this resolution, the $y$ signal observed by \planck\ traces the pressure distribution of the ICM up to $R_{500}$. As is already known from X-ray observations \cite[e.g.][]{1992A&A...259L..31B, 1993MNRAS.261L...8W, 2003A&A...400..811N}, the \planck\ $y$ map shows that the gas in Coma is elongated towards the west and extends in the south-west direction toward the NGC4839 subgroup.  Fig.~\ref{fig:coma10} shows that the SZ signal from this subgroup is clearly detected by \planck\ (see the white cross to the south-west).

Fig.~\ref{fig:coma10} also shows  clear compression of the isocontour lines in a number of cluster regions. 
We notice that, in most cases, the extent of the compression is
of the order of the $y$ map correlation length ($\approx 10\arcm$): it is likely  
that most of these are image artifacts  induced by  correlated noise in the y map.
Nevertheless, we also notice  at least 
two regions where the compression of the isocontour lines extends over angular scales significantly larger than the noise correlation length.
These two regions, located to the west and to the southeast of cluster centre, may indicate  real  steepenings of the radial gradient.
Such steepenings suggest the presence of a discontinuities in the 
cluster pressure profile,  which may be produced by a thermal shocks,  as we discuss in Sect.~\ref{sec:shock}.
For convenience,  in Fig.~\ref{fig:coma10} we outline the  regions from  which we extract the $y$ profiles used in Sect.~\ref{sec:shock} 
with white sectors. It is worth noting that the western steepening extends over a much larger angular scale 
than indicated by  the white sector. In Sect.~\ref{sec:shock} we explain why
we prefer a narrower sector for our quantitative analysis.

\begin{figure}[t]

\includegraphics[width=\linewidth,angle=0,keepaspectratio]{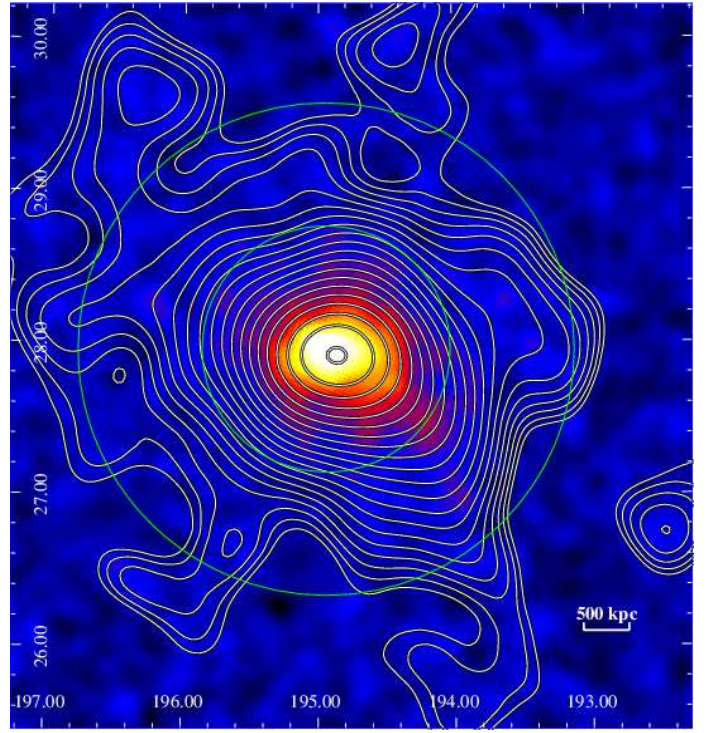}
\caption{{\footnotesize 
The \planck\ $y$ map of the Coma cluster obtained by combining the 70~GHz channel of LFI and the HFI channels from 100~GHz to 857~GHz.  The map has been smoothed to have a PSF with  ${\rm FWHM}=30\arcm$. The image is about $266 {\rm arcmin}\time 266 {\rm arcmin}$. 
The outermost contour corresponds to $y=2\times \sigma_{\rm noise30}=6.7\times 10^{-7}$. The green circles indicate $R_{500}$ and $2\times R_{500}\approx R_{200}$.
}}
\label{fig:coma30}
\end{figure}

In Fig.~\ref{fig:coma30} we show the \planck\ $y$ map of the Coma cluster obtained by adding the 70~GHz channel of LFI to the HFI channels and smoothing to a lower resolution. 
The PSF of  this map corresponds to ${\rm FWHM}=30\arcm$, which lowers the noise level by approximately one order of magnitude with respect to the 10$\arcm$ resolution map: $\sigma_{\rm noise30}=3.35\times 10^{-7}$.
 As for Fig.~\ref{fig:coma10}   the outermost contour level indicates $y=2\times \sigma_{\rm noise30}=6.7\times 10^{-7}$. 
 Due to the larger smoothing, this map shows less structure in the cluster centre, but clearly highlights  that \planck\ can  trace the pressure profile of the ICM well beyond  $R_{200}\approx 2\times R_{500}$ (see the outermost circle in Fig.~\ref{fig:coma30}).  

\section{Azimuthally averaged profile}\label{sec:profile}

\begin{figure}[t]

\includegraphics[width=\linewidth,angle=0,keepaspectratio]{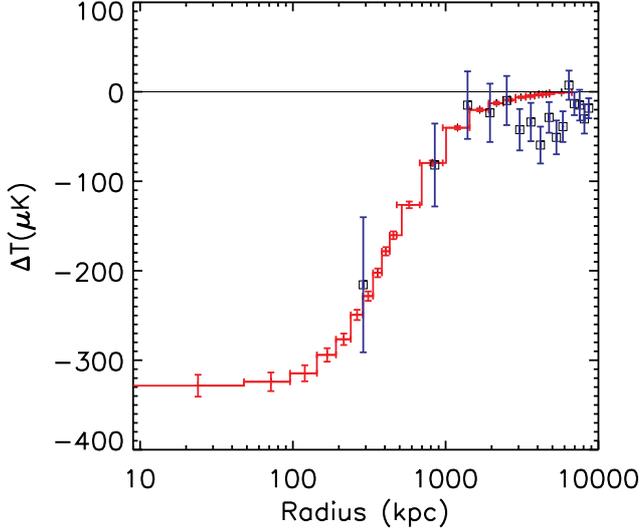}
\caption{{\footnotesize Comparison of the radial profile of the SZ effect towards the
Coma cluster, in units of the Rayleigh-Jeans equivalent temperature measured by \planck\ (crosses)  with the one obtained by {\it WMAP} (open squares) using the optimal V and W band data (from figure~14 of \citealt{2011ApJS..192...18K}). The plotted \planck\ errors are the square root of the diagonal elements of the covariance matrix. Notice that  profiles have been extracted from SZ maps with  $10\arcm $ and $30\arcm$ angular resolution from \planck\ and {\it WMAP}, respectively. 
}}
\label{fig:wmap_comparison}
\end{figure}

 Before studying the azimuthally averaged SZ profile of the Coma
cluster in detail, we first show a very simple performance test.
 In  Fig.~\ref{fig:wmap_comparison}  we compare the  SZ effect toward the Coma cluster, in units of the Rayleigh-Jeans equivalent temperature, measured by \planck\  and by {\it WMAP} 
using the optimal V and W bands  (from Fig.~14 of \citealt{2011ApJS..192...18K}). 
This figure  shows that, in addition to its greatly improved
angular resolution, \Planck\ frequency coverage  results in errors
on the profile which are $\approx 20$ times smaller than those from {\it WMAP}. 
Thanks to this higher sensitivity \planck\ allows us
to study, for the first time, the SZ signal of the Coma Cluster to its  very outermost regions. 
We do this by extracting the radial profile in concentric annuli centred on the cluster centroid  
(RA, Dec)=($12^{\rm h}59^{\rm m}47^{\rm s},+27\degr 55\arcmin 53\arcsec$).

We fit the  observed $y$ profile using the pressure formula proposed by \citet{2010A&A...517A..92A}:
\begin{equation}
P(x)=\frac{P_0}{(c_{500}x)^\gamma\left[1+(c_{500}x)^{\alpha}\right]^{(\beta-\gamma)/\alpha}},
\label{eq:arnaud}
\end{equation} 
where, $x=(R/R_{500})$.
This is done by fixing $R_{500}$ at the best-fit value obtained from the X-ray analysis ($R_{500}=1.31{\rm Mpc}$, see Sect.~\ref{sec:X-ray}) and  using three different combinations of parameters which we itemise below:
\begin{itemize}
\item  a ``universal'' pressure model (which we will refer to as Model A) for which we leave
 only $P_0$   as a free parameter and fix $c_{500}=1.177$, $\gamma= 0.3081$, $\alpha= 1.0510$, $\beta= 5.4905$ \citep{2010A&A...517A..92A};
\item  a pressure profile appropriate for clusters with disturbed X-ray morphology (Model B) for which we leave
 $P_0$   as a free parameter and fix $c_{500}=1.083$, $\gamma= 0.3798$, $\alpha= 1.406$, $\beta= 5.4905$  \citep{2010A&A...517A..92A};
\item a modified pressure profile (Model C) for which
 we let all the parameters vary (except $R_{500}$).   
\end {itemize}

\begin{figure*}[t]
\begin{centering}$
\begin{array}{cc}
\includegraphics[height=8.cm,angle=0,keepaspectratio]{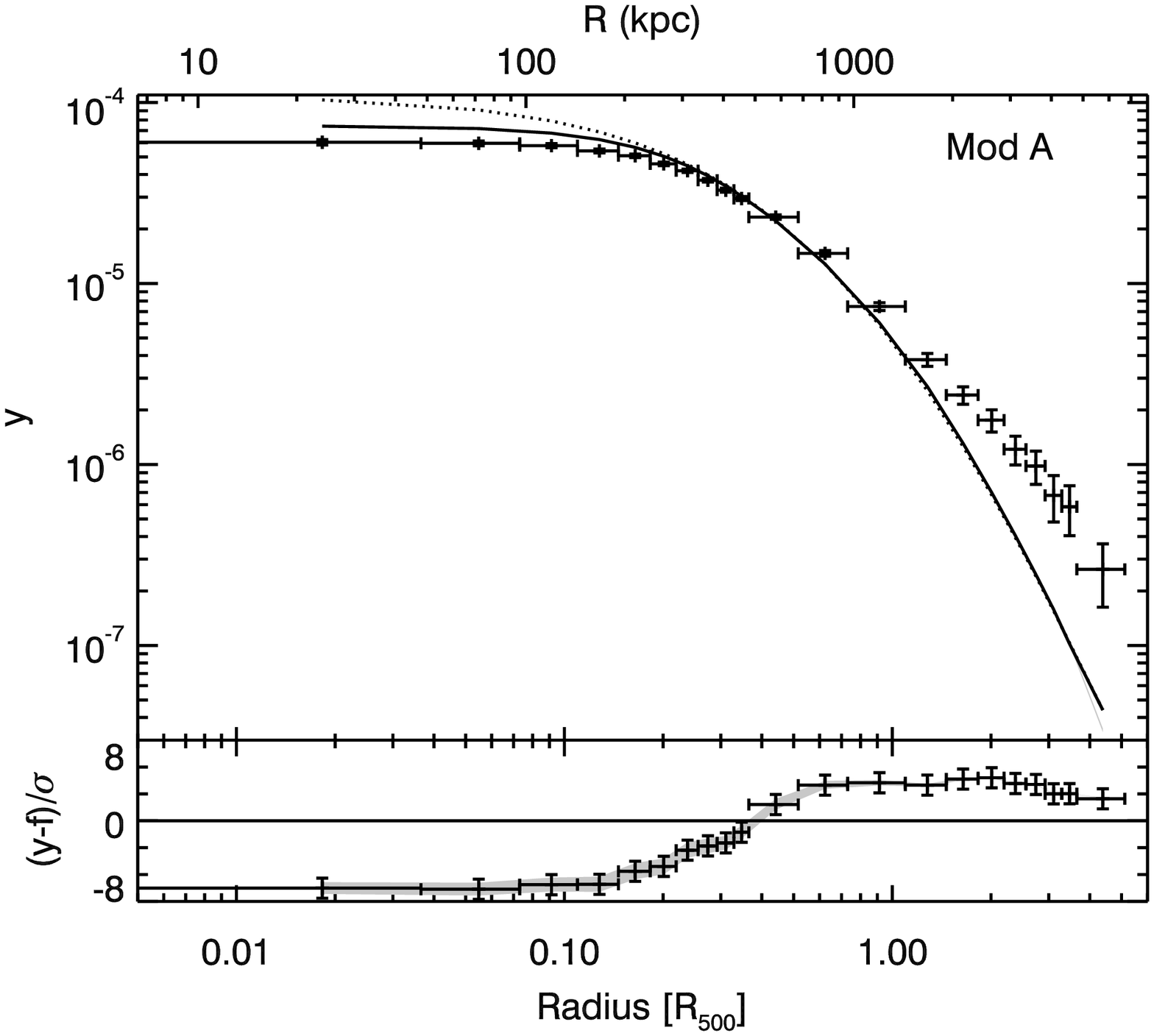}&
\includegraphics[height=8.cm,angle=0,keepaspectratio]{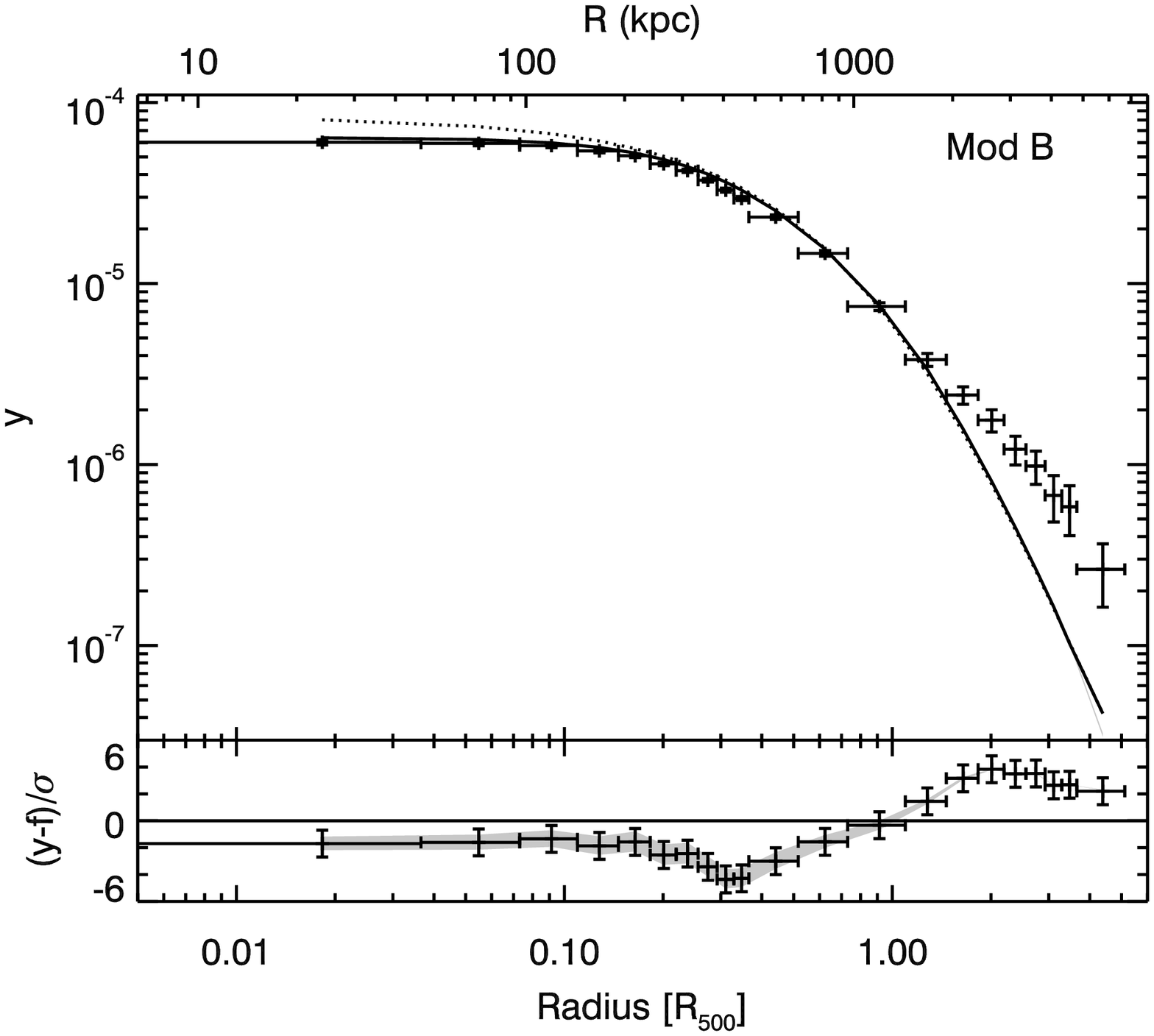}\\
\includegraphics[height=8.cm,angle=0,keepaspectratio]{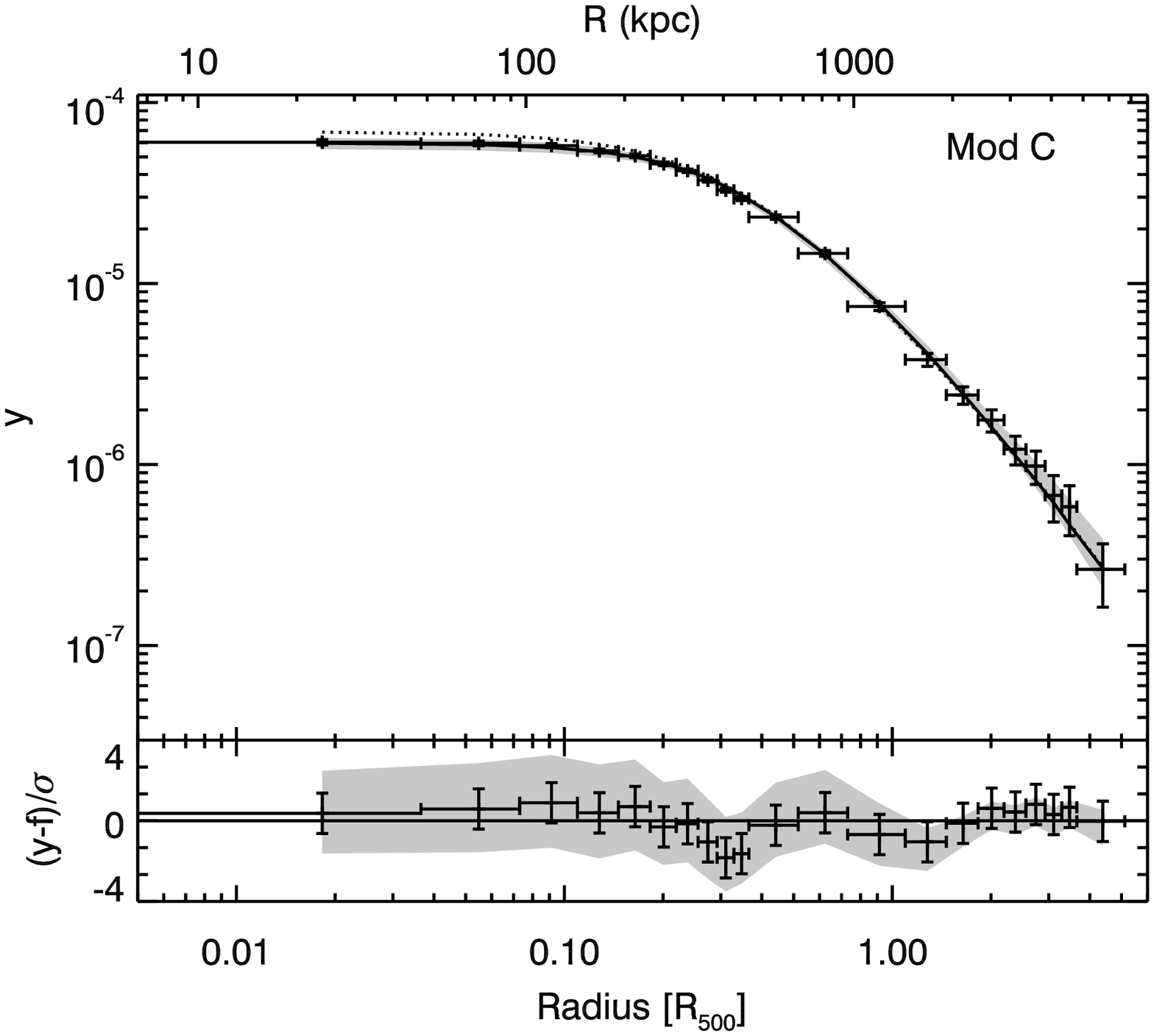}&
\includegraphics[height=8.cm,angle=0,keepaspectratio]{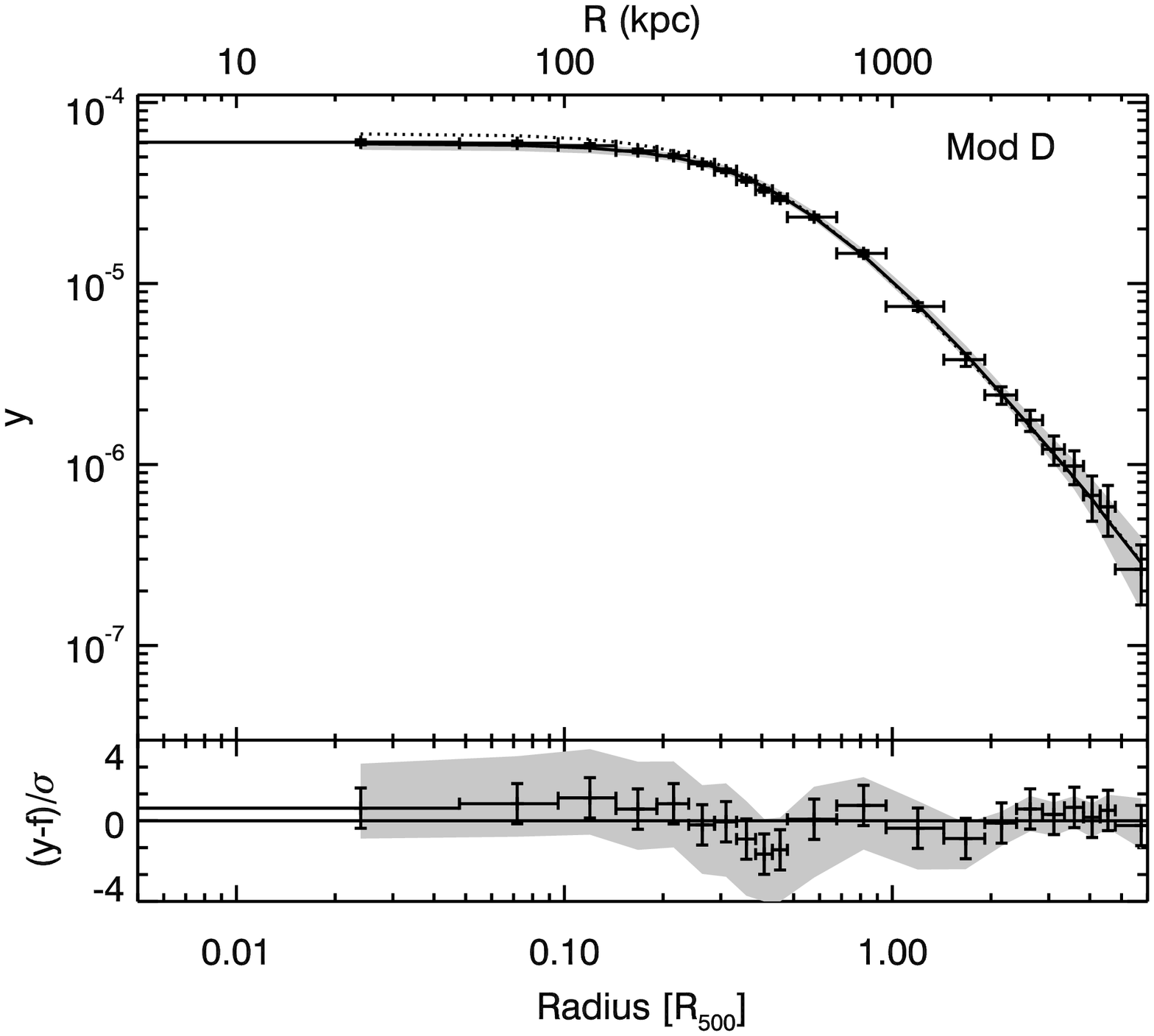}\\
\end{array}$
\end{centering}
\caption{{\footnotesize Comparison between  the azimuthally averaged $y$ profile of the Coma cluster
and various models. From  left to right, top to bottom, we show the best-fit $y$ models corresponding to the Arnaud et al. (2010) ``universal'' profile (A), the ``universal'' profile for merger systems (B), the modified ``universal'' profile (C, see \ref{tab:fit_par_y}), and the Vikhlinin et al. fitting formula (D, see. \ref{tab:fit_par_vic_y}). 
For each panel we show in the
 {\it Upper subpanel} the points indicating the Coma $y$ profile extracted in circular annuli centred at ($\rm{RA}$,$\rm{DEC}$)= ($12^{\rm h}59^{\rm m}47^{\rm s},+27\degr 55\arcmin 53\arcsec$). 
The plotted errors are the square root of the diagonal elements of the covariance matrix. Continuous and dotted lines  are
the best-fit projected $y$ model after and before the convolution with the \planck\ PSF, respectively. The gray shaded region indicates the envelope 
derived from the 68.4\% of models  with the lowest $\chi^2$.
 In the
 {\it Lower subpanel} we show the
ratio between the observed and the best-fit model of the projected $y$ profile in units of the relative error.  The gray shaded region indicates the envelope 
derived from the 68.4\% of models  with the lowest $\chi^2$.
}}
\label{fig:conbined_y}
\end{figure*}

The best-fit parameters, together with their 68.4\% confidence level errors, are reported in Table~\ref{tab:fit_par_y}.
The resulting best-fit models, together with the envelopes corresponding to the 68.4\% of models with the lowest $\chi^2$, 
are overlaid in the upper left, upper right and lower left panels of   Fig.~\ref{fig:conbined_y}, for models A, B, and C,  respectively.
We find that Eq.~(\ref{eq:arnaud}) fits the observed $y$ profile only if all the parameters (except $R_{500}$) are left free to vary (i.e., Model C).   

We also fit the observed radial $y$ profile using a
 fitting formula (Model D) derived from  the density and temperature functionals introduced by \citet{2006ApJ...640..691V}:  
\begin{equation}
  P=n_{\rm e} \times {\rm k}T,
  \label{p3d_vik}
\end{equation}

\noindent where,

\begin{eqnarray}
  n_{\rm e}^2(r) &=& n_{0}^2
  \frac{(r/r_{\rm c})^{-\alpha}}{[1+(r/r_{\rm c})^2]^{3\beta - \alpha/2}}
  \frac{1}{[1+(r/r_{\rm s})^3]^{\epsilon/3}} \nonumber \\ &&+ \frac
  {n_{02}^2}{[1+(r/r_{{\rm c}2})^2]^{3\beta_2}},
  \label{rho3d_equ}
\end{eqnarray}

\noindent and

\begin{equation}
  T(r) = T_0 \frac{(r/r_{\rm t})^{-a}}{[1+(r/r_{\rm t})^b]^{c/b}}.
  \label{t3d_equ}
\end{equation}

Notice that, for our purpose, Eq.~(\ref{p3d_vik}) is only used to fit the cluster  pressure profile.
For this reason, it is unlikely that, when considered separately, the best-fit parameters of 
 Eqs.~(\ref{rho3d_equ}) and ~(\ref{t3d_equ}) reproduce the actual cluster density and temperature profiles.
The best-fit parameters, together with their 68.4\% confidence level errors, are reported in Table~\ref{tab:fit_par_vic_y}.

The resulting model, with the 68.4\% envelope is overlaid in the lower-right panel of Fig.~\ref{fig:conbined_y}.
The above temperature and density functions contain many more free parameters than Eq.~(\ref{eq:arnaud}).
All these parameters have been specifically introduced to adequately fit all the observed surface brightness and temperature profiles of X-ray clusters of galaxies.  
This function, thus, is capable, in principle, of providing a better fit to any observed SZ profile.  Despite this,  we find that compared with Model C, Model D does not improve the quality of the fit. The reduced $\chi^2$ of  model D is
slightly higher ($\Delta \chi^2=0.3$) than for model C.

\begin{table*}[tmb]                 
\begingroup
\newdimen\tblskip \tblskip=5pt
\caption{ Best-fit parameters for the \citet{2010A&A...517A..92A} pressure model 
(Eqs. \ref{rho3d_equ} and \ref{t3d_equ}).}                          
\label{tab:fit_par_y}                            
\nointerlineskip
\vskip -3mm
\footnotesize
\setbox\tablebox=\vbox{
   \newdimen\digitwidth 
   \setbox0=\hbox{\rm 0} 
   \digitwidth=\wd0 
   \catcode`*=\active 
   \def*{\kern\digitwidth}
   \newdimen\signwidth 
   \setbox0=\hbox{+} 
   \signwidth=\wd0 
   \catcode`!=\active 
   \def!{\kern\signwidth}
{\tabskip=2em
\halign{#\hfill&\hfill#\hfill&\hfill#\hfill&\hfill#\hfill&\hfill#\hfill&\hfill#\hfill&\hfill#\hfill\cr                       
\noalign{\vskip 5pt}
\noalign{\doubleline}
Model          & $P_0$                & $ c_{500}$       & $\gamma$               & $\alpha$ & $\beta$                 & $R_{500}$\cr
           $({\rm Mpc})$    &  $(10^{-2} {\rm cm^{-3}}\, {\rm keV} )$         &                  &                        &          &                         & $({\rm Mpc})$\cr
\noalign{\vskip 3pt\hrule\vskip 5pt}
\hspace{.0cm} A (``Universal'')      & $2.57^{+0.04}_{-0.04}$& 1.17             &   0.308                &    1.051 & $ 5.4905 $              &  $1.31$                  \cr   
\vspace{.03in}
\noindent B (``Universal'' merger) & $1.08^{+0.02}_{-0.02}$& 1.083            &   $0.3798   $           &1.406     & $ 5.49$                 &  $1.31$                 \cr
\vspace{.03in}
C (``Universal'' all free)  & $2.2^{+0.3}_{-0.4}$& $2.9^{+0.3}_{-0.2}$ &   $<0.001$ &$ 1.8^{+0.5}_{-0.2}$    & $3.1^{+0.5}_{-0.2}$   &  $1.31$  \cr
\noalign{
\hrule\vskip 3pt}
}
}
}
\endPlancktablewide                 
\endgroup
\end{table*}                        

\begin{table*}[tmb]                 
\begingroup
\newdimen\tblskip \tblskip=5pt
\caption{Best-fit parameters for  pressure  model D \citep[Eq.~\ref{p3d_vik},][]{2006ApJ...640..691V}. 
As this model is  used  to fit the pressure, the best-fit density and temperature profiles are highly correlated 
and are unlikely to describe the actual cluster density and temperature profiles (see text).}                          
\label{tab:fit_par_vic_y}                           
\nointerlineskip
\vskip -3mm
\footnotesize
\setbox\tablebox=\vbox{
   \newdimen\digitwidth 
   \setbox0=\hbox{\rm 0} 
   \digitwidth=\wd0 
   \catcode`*=\active 
   \def*{\kern\digitwidth}
   \newdimen\signwidth 
   \setbox0=\hbox{+} 
   \signwidth=\wd0 
   \catcode`!=\active 
   \def!{\kern\signwidth}
{\tabskip=2em
\halign{#\hfill&#\hfill&#\hfill $|$ &#\hfill&#\hfill&#\hfill\cr                       
\noalign{\vskip 5pt}
\noalign{\doubleline}
\multispan3\hfill {Density}\hfill & \multispan3\hfill {Temperature}\hfill\cr
\noalign{\vskip 3pt\hrule\vskip 5pt}
$n_0$ &$(10^{-3}\rm cm^{-3})$ & $ 2.7^{+0.1}_{-0.3} $   & $ T_{0}$&$({\rm KeV})  $ & $6.9^{+0.1}_{-0.8} $ \cr    
\vspace{.03in}
$r_{\rm c}$ &$  (\rm Mpc)$        & $0.4^{+0.2}_{-0.02}$  & $r_{\rm t}$&$ ({\rm Mpc}) $   & $0.26^{+0.05}_{-0.07}  $\cr
\vspace{.03in}
$r_{\rm s}$&$ (\rm Mpc)$          & $0.7^{+0.2}_{-0.2}  $  & $a     $ &            & $0  $ \cr
\vspace{.03in}
$\alpha$   &              & $<10^{-6}     $  & $b     $     &        & $3.4^{+5.0}_{-0.2}     $  \cr
\vspace{.03in}
$\beta$    &              & $0.57^{+0.02}_{-0.3}$  & $c      $          &  & $ 0.6^{+0.7}_{-0.1} $ \cr
\vspace{.03in}
$\gamma$   &              & $3                 $ &    & &\cr
\vspace{.03in}
$\epsilon$  &             & $2.1^{+0.7}_{-0.7}$  &       &      &                              \cr
\vspace{.03in}
$n_{02}$&$ (\rm cm^{-3})$     &        $0^{a}$   &          &   &                              \cr
\noalign{
\hrule\vskip 3pt}
}
}
}
\endPlancktablewide                 
\tablenote  a The fit returns $n_{02}=0$ thus $r_{{\rm c}2}$ and $\beta_2$ are arbitrary.\par
\endgroup
\end{table*}                        

\begin{figure*}[t]
\begin{centering}$
\begin{array}{cc}
\includegraphics[height=7.cm,angle=0,keepaspectratio]{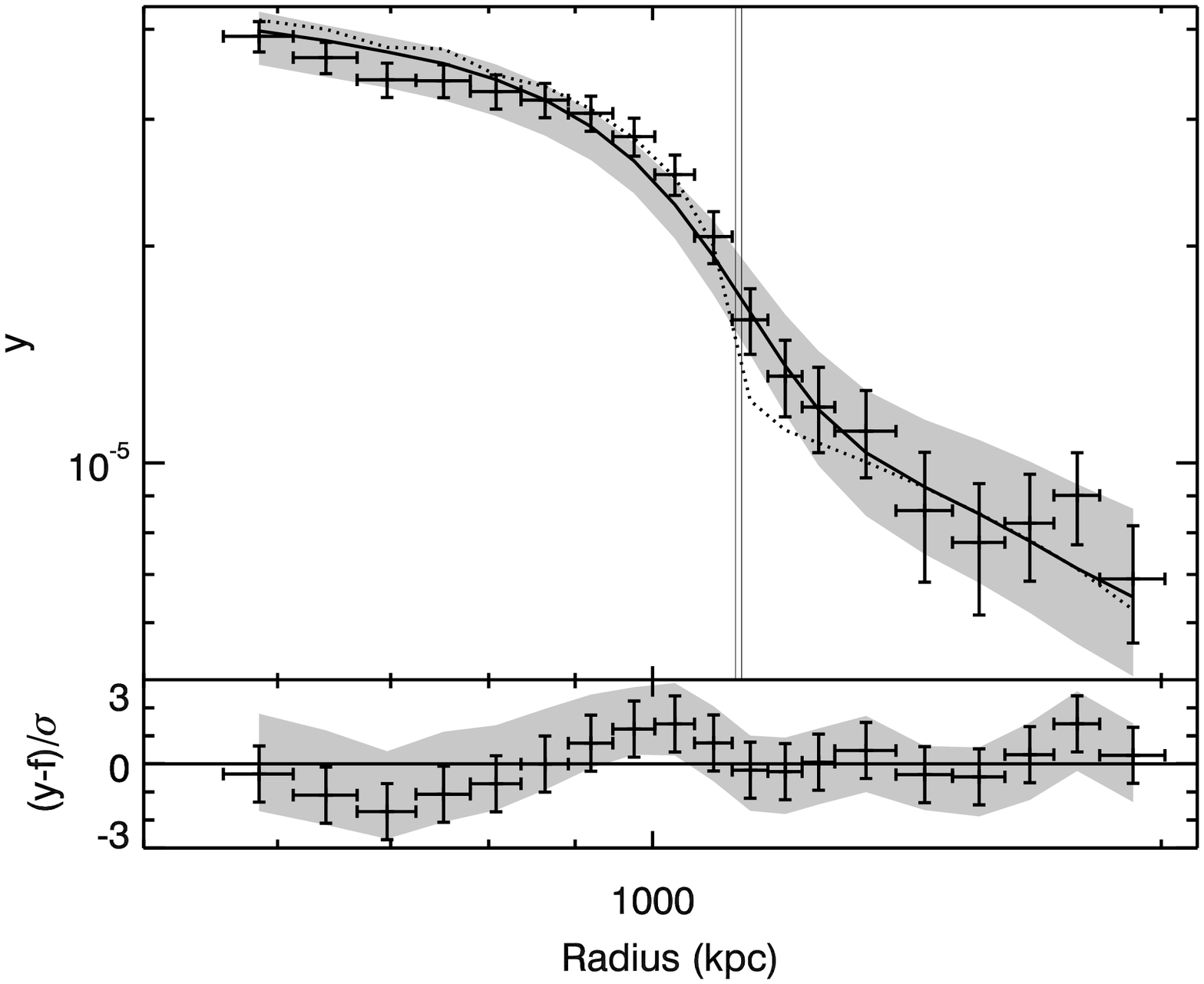}&
\includegraphics[height=7.cm,angle=0,keepaspectratio]{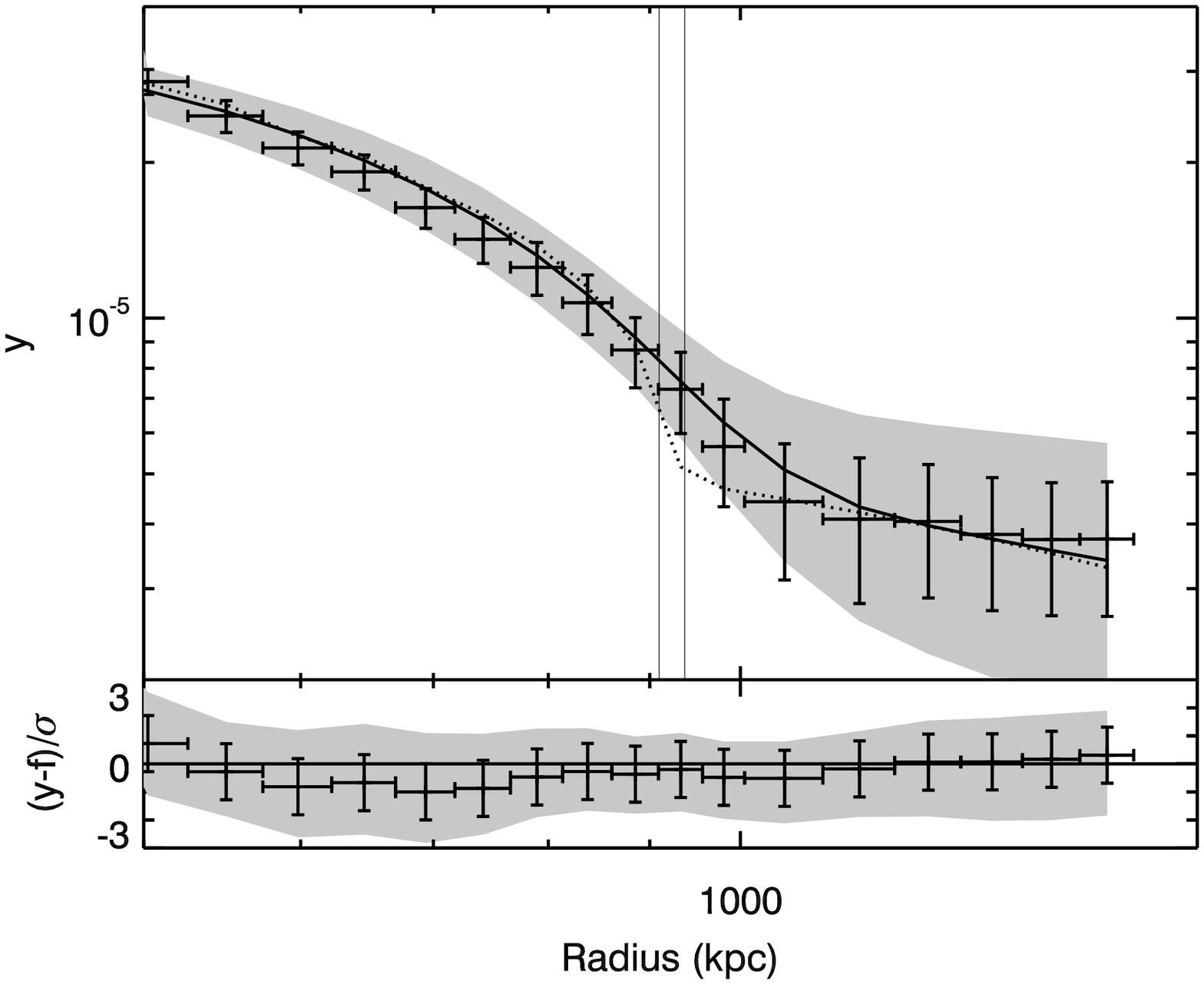}\\
\end{array}$
\end{centering}

\caption{{\footnotesize Comparison between the  projected $y$ radial profile and the best-fit shock model of the west ({\it left}) and south-east ({\it right}) pressure  jumps. 
 {\it Upper panels}: The points indicate the Coma $y$ profile extracted from the respective sectors, whose centres and position angles are reported in Table~\ref{tab:shock}. 
The plotted errors are the square root of the diagonal elements of the covariance matrix. 
Continuous and dotted lines  are
the best-fit projected $y$ model reported in  Table~\ref{tab:shock} after and before the convolution with the \planck\ PSF, respectively. The two vertical lines mark the $\pm 1\sigma$  position range of the jump.  The gray shaded region indicates the envelope 
derived from the 68.4\% of models  with the lowest $\chi^2$.   
 {\it Lower panels}:
Ratio between the observed and the best-fit model of the projected $y$ profile in units of the relative error.  The gray shaded region indicates the envelope 
derived from the 68.4\% of models  with the lowest $\chi^2$.
}}
\label{fig:shock}
\end{figure*}

\begin{figure*}[t]
\begin{centering}$
\begin{array}{cc}
\includegraphics[height=7.cm,angle=0,keepaspectratio]{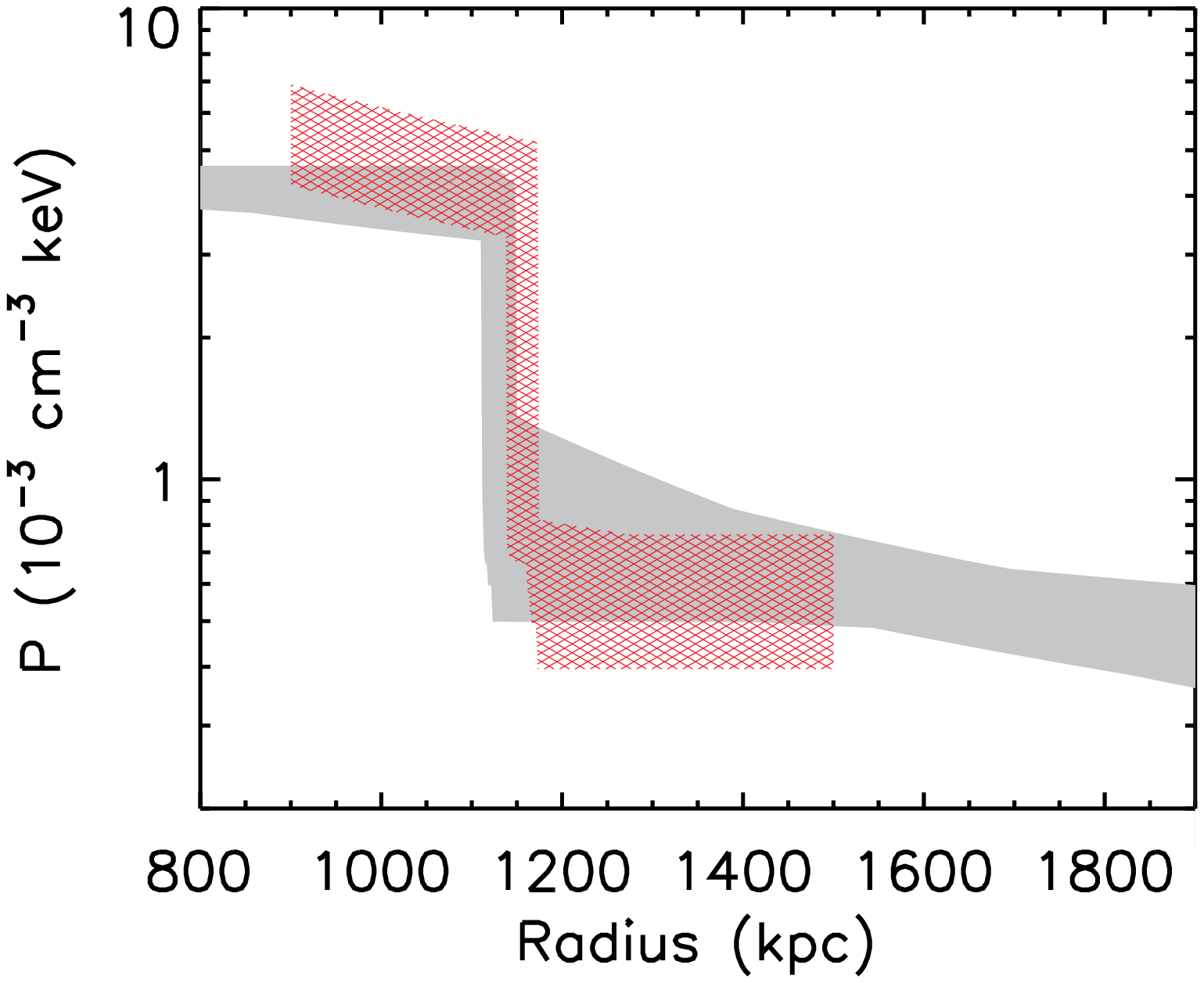}&
\includegraphics[height=7.cm,angle=0,keepaspectratio]{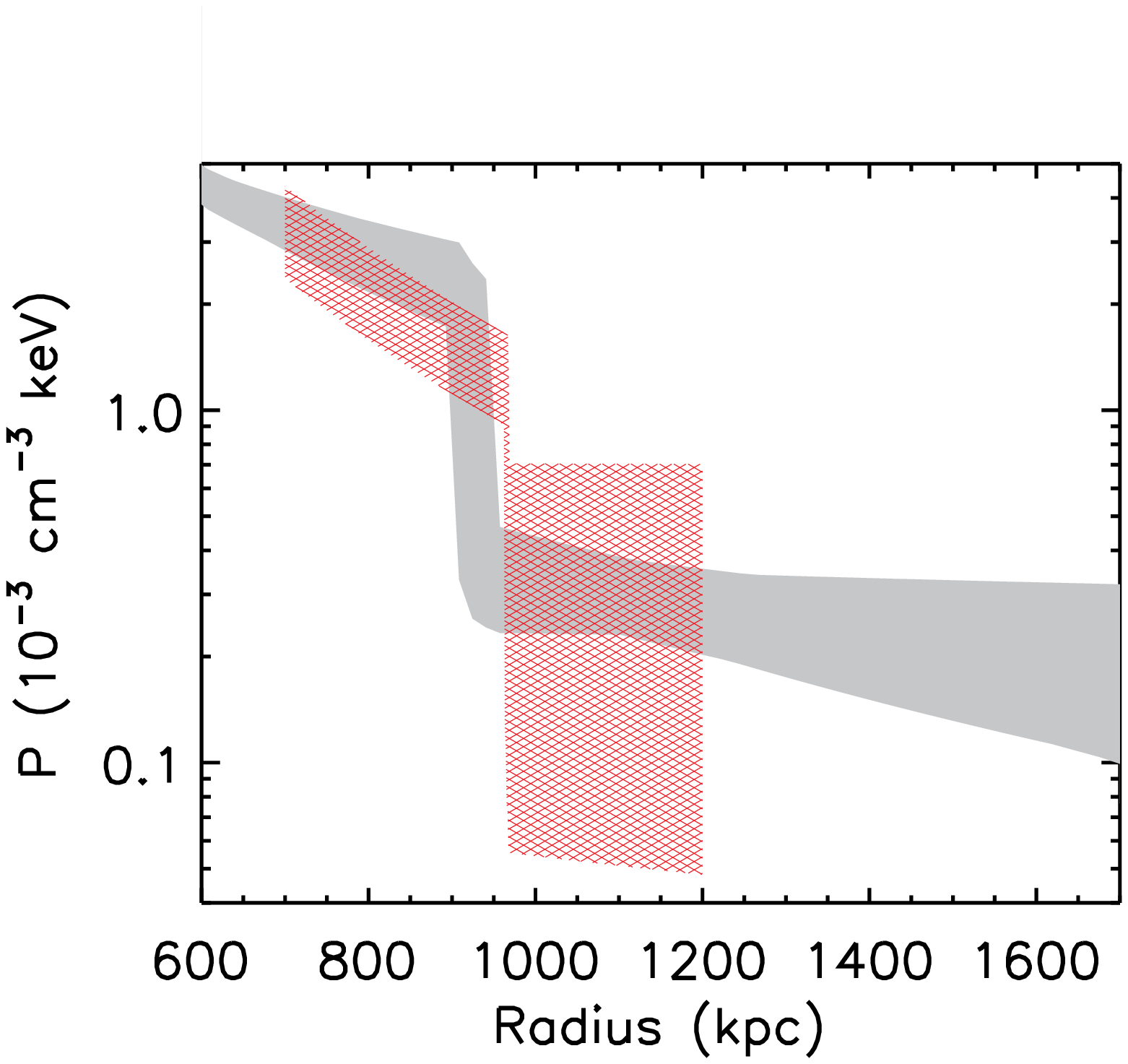}\\
\end{array}$
\end{centering}
\caption{{\footnotesize  68.4\% confidence level range of the 3D-pressure model for the west (left panel) and south-east (right panel) sectors in Fig.~\ref{fig:shock}.  Grey shaded regions are the profiles derived from the \Planck\ data. Red regions are the profiles derived from the \xmm\ data.
}}
\label{fig:p_shock}
\end{figure*}

\section{Pressure jumps}\label{sec:shock}

Fig.~\ref{fig:coma10}  shows at least two cluster regions where the $y$ isocontour lines appear to  be compressed 
on angular scales larger than the correlation length of the noise map.
This indicate a local steepening of  the $y$ signal. The most prominent  feature is located at about $0.5$~degrees from the cluster centre to the west.
Its position angle is quite large and extends from 340~deg to 45~deg. The second, less prominent feature, is located at  
$0.5$~degrees from the cluster center to the south-east.

Both features suggest the presence of  discontinuities in the underlying cluster pressure profiles. To test  this hypothesis and to try to estimate the  amplitude and the position of the pressure jumps we use the following simplified approach: i) we select two sectors; ii) we extract the $y$ profiles using circular annuli; and iii) assuming spherical symmetry, we fit them to a 3D pressure model with a pressure jump. 
This test requires that the extraction sectors are carefully  selected. Ideally one would like to follow, as close as possible, the curvature of the $y$ signal  around the  possible  pressure jumps. It is clear, however, that this procedure cannot be done exactly but it may be somewhat arbitrary.  The pressure jumps are unlikely to be perfectly 
spherically symmetric, thus, the sector selection  depends also on what is initially thought to be the leading edge of the underlying pressure jump. 
Despite of this arbitrariness, our approach remain valid for the purpose of testing for the presence of a shock. As matter of fact, even if we choose 
a sector that does not properly sample the pressure jump, our action  goes in the direction 
of mixing the signal from the pre- and post-pressure jump regions. This will simply result is a smoother profile which, when fitted with the 3D pressure model, will 
returns a smaller amplitude for the pressure jump itself. 
Thus, in the worst scenario, the measured pressure jumps would, in any case, represent a lower estimate of the jumps at the leading edges. 

 In order to minimise the mixing 
   of pre- and post-shock signals, one can reduce the width of
   the analysis sector to the limit allowed by signal statistics.
   Indeed, for very high signal-to-noise, one could, in principle,
   extract the y signal along a line perpendicular to the leading
   edge of the shock. This would limit mixing of pre- and post shock
   signals to line-of-sight and beam effects. 
 In the specific case of the Coma cluster we notice that the west feature is located in a higher signal-to-noise region than the south-east one. For this reason we decide to extract the west profile using a sector with an angular aperture smaller than the actual angular extent of this feature in the $y$ map.     

Following the above considerations, we set the  centres and orientations of the  west and the south-east sectors to the values reported in the first three columns of Table~\ref{tab:shock} and indicated in Fig.~\ref{fig:coma10}.  In Sect.~\ref{sec:shocksdisc} below we  demonstrate 
that, within the selected  sectors, the SZ and the X-ray analyses give consistent results. This indicates that,  despite the apparent arbitrariness in 
sector selection,: 
i) these SZ-selected sectors are representative of the  features under study and; ii) that the hypothesis of spherical symmetry is a good approximation, at least  within the selected sectors.

We fit the profiles using a 3D pressure model composed of  two power laws with index 
$\eta_1$ and $\eta_2$ and a jump by a factor $D_{\rm J}$ at  radius $r_{\rm J}$. It is important 
to note that, even if irrelevant for the estimate of the jump amplitude, the value of both the slope  $\eta_2$
and the absolute normalization of the 3D pressure at a given radius depends on   the slope and extension of 
the ICM along the line of sight. To take this into account we assume that outside the fitting region (i.e. at $r>r_{\rm s}$, with $r_{\rm s}=2 {\rm Mpc}$)  the slope of the pressure profile 
 follows the asymptotic average pressure profile corresponding to model C (i.e. $\eta_{3}=\beta=3.1$; see Sect.~\ref{sec:profile} and Table~\ref{tab:fit_par_y}).
The 3D pressure profile is thus given by:
\begin{equation}
P=P_0 \times 
\left\{
\begin{array}{ll}
 D_{\rm J}(r/r_{\rm J})^{-\eta_1}& r<r_{\rm J}; \\ 
 (r/r_{\rm J})^{-\eta_2}&r_{\rm J}<r<r_{\rm s};\\ 
    (r_{\rm s}/r_{\rm J})^{-\eta_2} (r/r_{\rm s})^{-\eta_3},              &       r>r_{\rm s}. \\
\end{array}
\right.
\label{eq:shock}
\end{equation}
We project the above 3D pressure model, 
integrating along the line of sight for $r<10{\rm Mpc}$.

The best-fit parameters, together with their 68.4\% errors, are reported in
Table~\ref{tab:shock}. 
 Note, that the error bars on $r_{\rm J}$ are smaller than the angular
resolution of \Planck. As explained in Appendix~\ref{sec:appendix}, this is not surprising and is simply due to projection effects.

In the left and right panels of Fig.~\ref{fig:p_shock} 
we show with a grey shadow the corresponding 3D pressure jump models with their errors for the west and south-east sectors, respectively. For convenience in Fig.~\ref{fig:shock}  we overlay  the data points with the best-fit projected $y$ models  after and before the convolution with the Planck PSF.  As shaded region, we report the envelope derived from the 68.4\% of models with the lowest $\chi^2$. In the lower panels we show  the ratio between the data and the best-fit model of the projected $y$ profile in units of the relative error.
This figure clearly shows that the pressure jump model provides a good fit to the observed  profiles for both  the west and south-east sectors.
 Furthermore the comparison of the projected model before and after the convolution with the PSF clearly shows that, for the Coma cluster, 
the effect of the \planck ~ PSF smoothing is secondary with respect to  projection effects.
This  indicates that there is only a modest 
gain, from the detection point of view, in observing this 
specific feature using an instrument with a much better angular resolution 
than \planck ~ (for a full discussion, see Appendix~\ref{sec:appendix}).

As reported  in Table~\ref{tab:shock} the pressure jumps corresponding to the observed profiles are
 $D_{\rm J}=4.9^{+0.4}_{-0.2}$ and  
$D_{\rm J}=5.0^{+1.3}_{-0.1}$ for the west and south-east sectors, respectively.


\begin{table*}[tmb]                 
\begingroup
\newdimen\tblskip \tblskip=5pt
\caption{ Best-fit parameters of the pressure jump model of Eq.~(\ref{eq:shock}).}                          
\label{tab:shock}                            
\nointerlineskip
\vskip -3mm
\footnotesize
\setbox\tablebox=\vbox{
   \newdimen\digitwidth 
   \setbox0=\hbox{\rm 0} 
   \digitwidth=\wd0 
   \catcode`*=\active 
   \def*{\kern\digitwidth}
   \newdimen\signwidth 
   \setbox0=\hbox{+} 
   \signwidth=\wd0 
   \catcode`!=\active 
   \def!{\kern\signwidth}
{\tabskip=2em
\halign{#\hfill&\hfill#\hfill&\hfill#\hfill&\hfill#\hfill&\hfill#\hfill&\hfill#\hfill&\hfill#\hfill&\hfill#\hfill&\hfill#\hfill\cr                     
\noalign{\vskip 5pt}
\noalign{\doubleline}
Sector&$^a$RA&$^a$Dec& $^a$Position angle & $P_{0}$                & $r_{\rm J}$ &$D_{\rm J}$&$\eta_{1}$&$\eta_{2}$\cr
     &(J2000)&(J2000)&(deg:deg)&$(10^{-4}\rm{cm^{-3}\, keV})$&$(\rm{Mpc})$&         &         &          \cr
\noalign{\vskip 3pt\hrule\vskip 5pt}
West &13 00 25.6&+27 54 44.00 &340:364 & $8.8^{+0.2}_{-0.5}$   &$1.13^{+0.03}_{-0.01}$    & $4.9^{+0.4}_{-0.2}$ &$0.0^{+0.2}_{-0.0}$ &$1.2^{+0.2}_{-0.2}$ \cr
\vspace{.03in}
South-east &12 59 48.9&+28 00 14.39 &195:240 & $3.6^{+0.1}_{-0.5}$   &$0.92^{+0.02}_{-0.01}$ &$ 5.0^{+1.3}_{-0.1}$   & $1.5^{+0.2}_{-0.2}$ &$1.00^{+0.3}_{-0.5}$  \cr
\noalign{
\hrule\vskip 3pt}
}
}
}
\endPlancktablewide                 
\tablenote a The RA and Dec indicate the centre of curvature of the sectors  from which the profiles have been extracted.\par
\tablenote b We fixed $r_{\rm s}=2~{\rm Mpc}$ and $\eta_3=3.1$ (see text).\par
\endgroup
\end{table*}                        

%

\begin{figure}[t]
\begin{centering}
\includegraphics[width=\linewidth,angle=0,keepaspectratio]{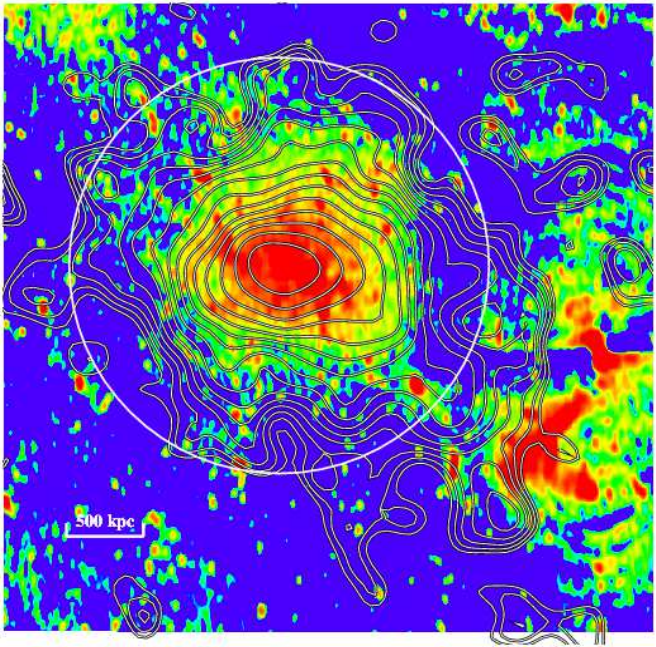}
\end{centering}
\caption{{\footnotesize
 Westerbork Synthesis Radio Telescope 352~MHz total intensity image of the Coma Cluster from figure 3 of
\citet{2011MNRAS.412....2B}  overlaid with the $y$ contour levels from Fig.~\ref{fig:coma10}. 
 Most of the radio flux from compact sources has been subtracted; the resolution is $133 {\rm arcsec} \times 68 {\rm arcsec}$ at $-1.5$ degrees (W of N). The white circle indicates $R_{500}$.
}}
\label{fig:diff_coma10}
\end{figure}

\begin{figure}[t]
\begin{centering}
\includegraphics[width=\linewidth,angle=0,keepaspectratio]{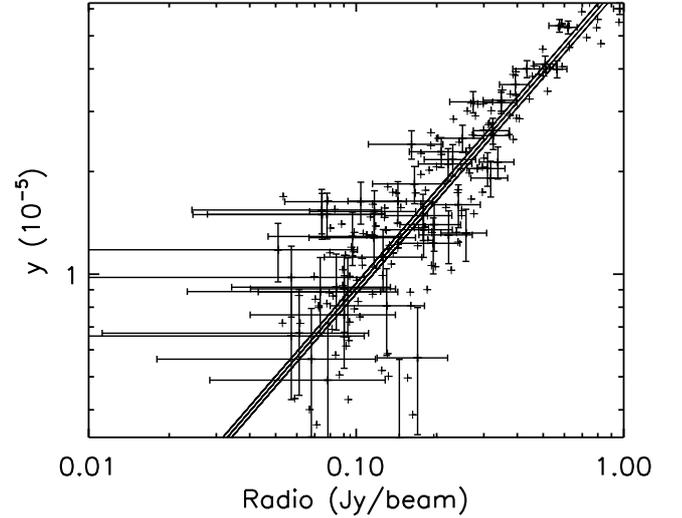}
\end{centering}
\caption{{\footnotesize  Scatter plot between the radio map after smoothing to ${\rm FWHM}=10{\arcm}$ 
and the $y$ signal for the Coma cluster.  To make the plot clearer,  we show errors only for some points.
}}
\label{fig:linreg}
\end{figure}

\begin{figure}[t]
\begin{centering}
\includegraphics[width=\linewidth,angle=0,keepaspectratio]{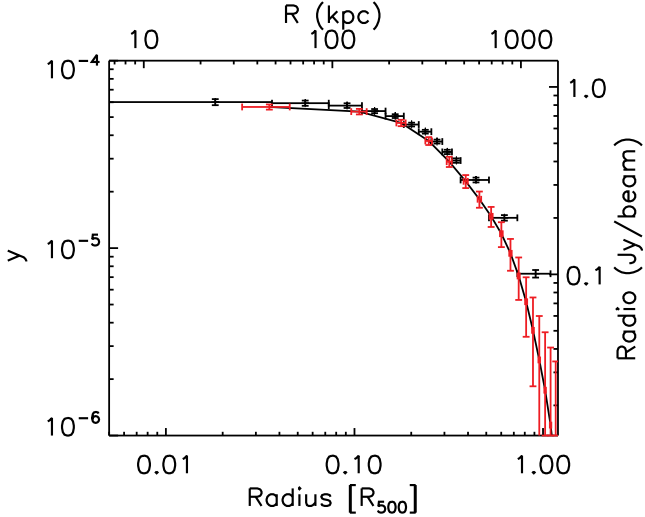}
\end{centering}
\caption{{\footnotesize
Comparison of the $y$ (black) and diffuse radio (red) global radio profiles in Coma. The radio profile has been convolved to 10~arcmin resolution to match the \Planck\ FWHM and simply rescaled by
the multiplication factor derived from the linear regression shown in Fig.~\ref{fig:linreg}. The radio errors are dominated by uncertainties in the
zero level due to a weak bowling effect resulting from the lack of short
interferometer spacings.
}}
\label{fig:radio_y_prof}
\end{figure}

\section {SZ-Radio comparison}\label{SZ-radio}

In Fig.~\ref{fig:diff_coma10} we overlay the $y$ contour levels from Fig.~\ref{fig:coma10}  with the 352~MHz Westerbork Synthesis Radio Telescope diffuse
total intensity image of the Coma cluster from figure 3 of \citet{2011MNRAS.412....2B}. Most of the emission from compact radio sources both in and behind the cluster has been automatically subtracted.
This image clearly shows a correlation between the diffuse radio emission and the $y$ signal.

 To provide a more quantitative comparison of the observed correlation, we first removed the remaining compact source emission in the radio image using the multiresolution filtering technique of \citet{2002PASP..114..427R}. This removed 99.9\% of the flux of unresolved sources, although residual emission likely associated with the  tailed radio galaxy NGC~4874 blends in to the halo emission and contributes to the observed brightness within the central $\sim 300\, {\rm kpc}$.  After filtering, 
we convolve the  
the diffuse radio emission to 10~arcmin resolution to match the \planck\ $y$ map. We then extract the radio and $y$ signals from the  $r<50~{\rm arcmin}$ region of the cluster and plot the results in Fig.~\ref{fig:linreg}.
 This is the first quantitative surface-brightness comparison of radio and SZ brightnesses\footnote{see e.g.  \citet{2011A&A...534L..12F} and \citet{2012JApA..tmp...20M}  for a morphological comparison between radio and SZ brightnesses}. 
We  fit the data in the log-log plane using the Bayesian  linear regression algorithm proposed by  \citet{2007ApJ...665.1489K}, which accounts for errors in both abscissa and ordinate. The radio errors of $50\, {\rm mJy}/10 \arcmin$ beam are estimated from the off-source scatter, which is dominated by emission over several degree scales which is incompletely sampled by the interferometer.
We find a quasi-linear relation between the radio emission and the $y$ signal:
\begin{equation}
\frac{ y}{10^{-5}}=10^{(0.86\pm 0.02)}F_{\rm R}^{(0.92\pm 0.04 )},
\end{equation}
\noindent where $F_{\rm R}$ is the radio brightness in ${\rm Jy\, beam^{-1}}$ (10~arcmin beam FWHM). 

Furthermore, using the same algorithm, we find that the intrinsic  scatter between the two observables is only $(9.6\pm 0.2)\%$.
The quasi-linear relation between the radio emission and $y$ signal, and its small scatter, are also clear from the good match of 
the radio and $y$  profiles shown in Fig.~\ref{fig:radio_y_prof},  obtained by simply rescaling the  10$\arcm$ FWHM convolved radio profile by $10^{0.86}\times 10^{5}$.
An approximate linear relationship between the radio halo and SZ total powers for a sample of clusters was also found by \citet{2012MNRAS.421L.112B}, for the case that the signals are calculated over the volume of the radio halos.

There are several sources of scatter contributing to the point-by-point correlation in Fig.~\ref{fig:linreg} and the radial radio profile in Fig.~\ref{fig:radio_y_prof}. First is the random noise in the measurements, which is $\sim$2--3${\rm mJy}/135 \arcsec$ beam. Even after convolving to a $10\, \arcmin$ beam, however, this is insignificant with respect to the other sources of scatter.  A second issue is the proper zero-level of the radio map, based on the  incomplete sampling of the largest scale structures by the interferometer. After making our best estimate of the zero-level correction, the remaining uncertainty is $\sim25\, {\rm mJy}/10 \arcmin$ beam, which is indicated  as error bars in Fig.~\ref{fig:radio_y_prof}.

Note that the radio profile is significantly flatter at large radii than presented by \citet{1997A&A...321...55D}.  However, their image, made with the Effelsberg 100m telescope at $1.4\,{\rm GHz}$, appears to have set the zero level too high;  they do not detect the faint Coma related emission mapped by 
\citet{2011MNRAS.412....2B} on the Green Bank Telescope, also at $1.4\, {\rm GHz}$, and by \citet{2007ApJ...659..267K}  at $0.4\, {\rm GHz}$ using Arecibo and DRAO. The addition of a zero level flux to the  \citet{1997A&A...321...55D} measurements at their lowest contour level flattens out their profile to be consistent with ours at their furthest radial sample at $900\, {\rm kpc}$.

Finally, there are azimuthal variations in the shape of the radial profile, both for the radio and Y images.  This is seen most clearly in Fig.~\ref{fig:shock}, comparing the west and southeast sectors.  In the radio, the radial profiles in 90 degree wide sectors differ by up to a factor of 1.6 from the average;  it is therefore important to understand Fig.~\ref{fig:radio_y_prof} as an average profile, not one that applies universally at all azimuths.   These azimuthal variations can also contribute to the scatter in the point-by-point correlation in Fig.~\ref{fig:linreg}, but only to the extent that the behavior differs between radio and Y.

\section {Discussion}\label{sec:discussion}

So far In this paper we have presented the data analysis of the Coma cluster observed in its SZ effect by the \planck\ satellite.
In Sect.~\ref{sec:imaging} and Sect.~\ref{sec:profile} we showed  that, thanks to its great sensitivity, \planck\ is capable of
 detecting significant SZ emission above the zero level of the $y$  map up to at least $4~\rm{Mpc}$ which corresponds to $R\approx 3\times R_{500}$. 
This allows, for the first time, the study of the ICM pressure distribution in the outermost 
cluster regions. Furthermore, we performed a comparison
 with radio synchrotron emission. 
 Here we discuss our results in more  detail.

\begin{figure}[t]
\begin{centering}
\includegraphics[width=\linewidth,angle=0,keepaspectratio]{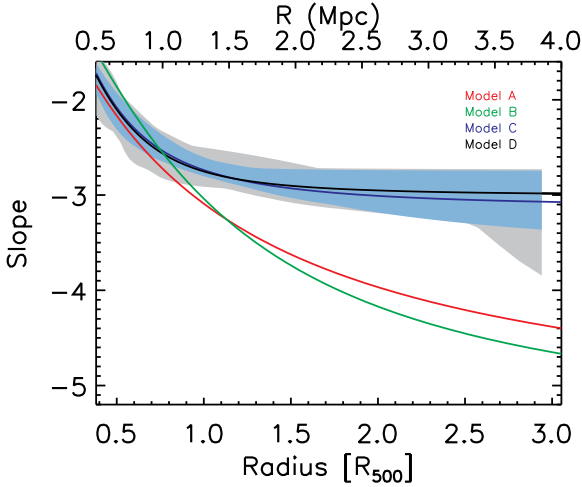}
\end{centering}
\caption{{\footnotesize Comparison  of the pressure slopes of the best-fit models shown in Fig.~\ref{fig:conbined_y}. The red, green, blue and grey lines correspond to  Models A, B, C, and D, respectively.  
}}
\label{fig:slopes}
\end{figure}

\begin{figure}[t]
\begin{centering}
\includegraphics[width=\linewidth,angle=0,keepaspectratio]{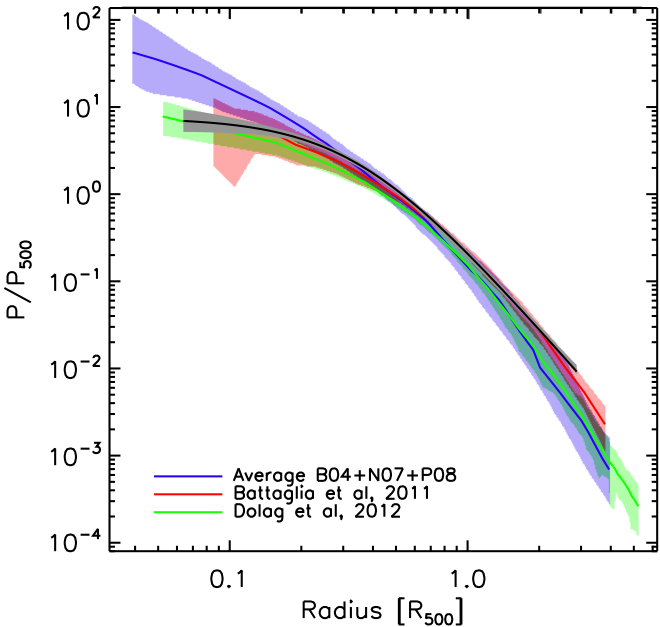}
\end{centering}
\vskip -2.2cm
\caption{{\footnotesize   Scaled Coma pressure profile with relative errors (black line and gray shaded region) overplotted on the  scaled pressure profiles derived from numerical simulations of B04+N07+P08 (blue line and violet shaded region), \citet{2011arXiv1109.3709B} (red line and shaded region), and Dolag et al. (in preparation, green line and light green shaded region).  
}}
\label{fig:comparison_p_prof}
\end{figure}

\subsection{Global pressure profile}

To study the 3D pressure distribution of the ICM up to $r=$3--4$\times R_{500}$, 
 we fit the observed $y$ profile using four analytic models
 summarised in Tables~\ref{tab:fit_par_y} and \ref{tab:fit_par_vic_y}    
(see Sect.~\ref{sec:profile}). 

From the ratio plot shown in Fig.~\ref{fig:conbined_y}  we immediately see that the ``universal'' pressure profile (Model A) is too steep both in the 
cluster centre and in the outskirts. The fit to the  data thus results in an overestimation and underestimation of the observed SZ signal at smaller and larger radii, respectively.  The overestimation of the observed profile at lower radii is consistent with {\it WMAP} \citep{2011ApJS..192...18K}.  This  is expected, 
since merging systems, such as Coma, have a flatter central pressure profile than the ``universal'' model \citep{2010A&A...517A..92A}.
For merging systems,  Model B should provide a better fit,  as it has been specifically calibrated, at $r<R_{500}$,   to reproduce the average X-ray profiles of
such systems 
\citep{2010A&A...517A..92A}. Fig.~\ref{fig:conbined_y} shows that this latter model indeed reproduces the data well at $r<R_{500}$. Nevertheless,  as for Model A, 
it still underestimates the observed $y$ signal at larger radii.
The observed profile clearly requires a shallower pressure profile in the cluster outskirts, as evident in Models C and D.
This is  important, as the  external pressure slopes of both Model A and B are tuned to reproduce the mean slope predicted by the hydrodynamic simulations of \citet{2004MNRAS.348.1078B}, 
\citet{2007ApJ...655...98N}, and \citet[from now on, B04+N07+P08]{2008A&A...491...71P}. The \planck\ 
observation shows that  the pressure slope for Coma is flatter than this value.
This is also illustrated in Fig.~\ref{fig:slopes} where we report the pressure slope as a function of the radius in our models:
we find  that while at $R=3\times R_{500}$ the  mean predicted pressure slope 
is  $>4.5$ for Models A and B,  
the observed pressure slope of Coma  is $\approx 3.1$  as seen in Model C and Model D. 
 
 In  Fig.~\ref{fig:comparison_p_prof} we compare 
the scaled  pressure profile of Coma  with the
pressure profiles derived from the  numerical simulations of B04+N07+P08 and with the numerical simulations of Dolag et al. (in preparation) and \citet{2011arXiv1109.3709B}. We note that the simulations
agree within their respective dispersions across the whole radial
range. The Dolag et al. (in preparation) and \citet{2011arXiv1109.3709B}  profiles
best agree within the central part, and are flatter than the
B04+N07+P08 profile. This is likely due to the implementation
of AGN feedback, which triggers energy injection at cluster
centre, balancing radiative cooling and thus stopping the
gas cooling. In the outer parts where cooling is negligible, the
B04+N07+P08 and Dolag et al. (in preparation) profiles are in perfect agreement.
The  \citet{2011arXiv1109.3709B} profile is slightly higher, but
still compatible within its dispersion with the two other sets.
Here again, differences are probably due to the  specific implementation of the simulations.

We find that the Coma pressure profile at $2\times R_{500}$ is already 
2 times higher than the average profile predicted by the B04+N07+P08 and  Dolag et al. (in preparation) simulations, 
although still  within the overall profile distribution which has  quite a large scatter. 
The pressure profile of \citet{2011arXiv1109.3709B}   appears to be more consistent with the Coma profile and, in general, with the \planck\ SZ pressure
profile obtained by stacking  62 nearby massive clusters \citep{planck2012-V}. Still Fig.~\ref{fig:comparison_p_prof} indicates that   
the Coma pressure profile  lies on the upper envelope of the pressure profile distribution derived from all the above simulations.

It is beyond the scope of this paper to discuss in detail the comparison between theoretical predictions.
Here we just stress  that,  at such large radii, there is the possibility that  the observed SZ signal  could be significantly contaminated by SZ sources  along the line of sight. This signal could  be generated by:
i) unresolved and undetected clusters; and ii) hot-warm gas filaments.
Contamination would produce an 
apparent flattening of the pressure profile. 
We  tested for possible contamination  by  unresolved clusters by re-extracting the $y$ profile, excluding  circular regions of 
$r=5\arcm$ centred on all NED identified clusters of galaxies present in the Coma cluster region. 
We  find that the new $y$ profile is consistent within the errors with 
the previous one, which implies that this kind of contamination is negligible in the Coma region.  Thus, if there is SZ contamination
it is probably related to the filamentary structures surrounding the cluster. 
We note that from the re-analysis of the \rosat~all-sky survey,
\citet{2009ApJ...696.1886B} and \citet{2003ApJ...585..722B} report 
the detection of extended soft X-ray emission in the Coma cluster region 
up to $5~{\rm Mpc}$ from the cluster centre. 
They propose that this emission is related to filaments that converge 
toward Coma and is generated either by non-thermal radiation caused by accretion shocks or by thermal emission from the filaments themselves.

\begin{figure}[t]
\includegraphics[width=\linewidth,angle=0,keepaspectratio]{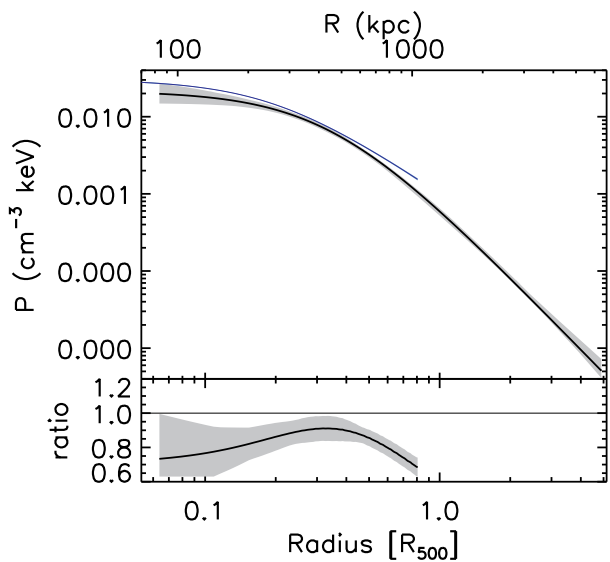}
\caption{{\footnotesize Comparison between  the \planck\ and \xmm\ derived deprojected total pressure profiles.
 {\it Upper panel}: Blue line and light blue shaded region are the deprojected  pressure profile, with its 68.4\% confidence level errors, obtained from the X-ray analysis of the \xmm\ data (see text). The black line and grey shaded regions are the best-fit and  68.4\% confidence level errors from  the Model C pressure profile resulting from the fit shown in Fig.~\ref{fig:conbined_y}. {\it Lower panel}: Ratio between the \xmm\  and \planck\ derived pressure profiles. The black line and the grey shading  indicate the best-fit and the 68.4\%  confidence level  errors, respectively.    
}}
\label{fig:total_pressure}
\end{figure}

\begin{figure}[t]
\includegraphics[width=\linewidth,angle=0,keepaspectratio]{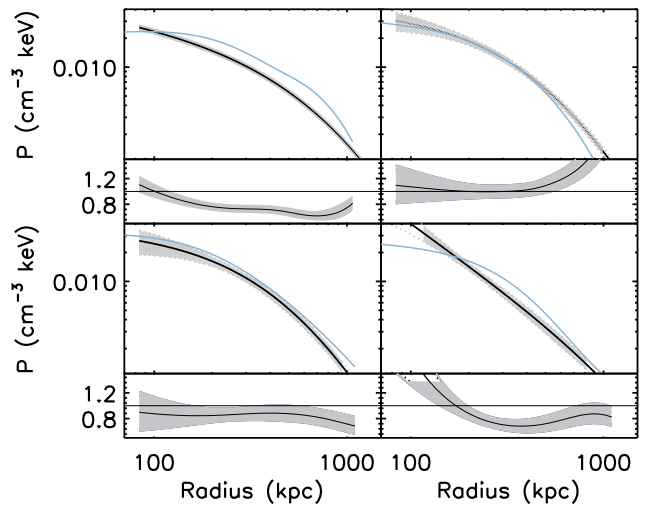}
\caption{{\footnotesize Same as Fig.~\ref{fig:total_pressure} but from profiles extracted in four $90\deg$ sectors. From left to right, top to bottom we report the west ($-45\deg,45\deg$), north ($45\deg,135\deg$), east ($135\deg,225\deg$) and south ($225\deg,3155\deg$) sectors, respectively. 
}}
\label{fig:sector_pressure}
\end{figure}

\subsection{X-ray and SZ pressure profile comparison}\label{sec:xray_sz}

 We can compare the 3D pressure profile derived from the SZ observations to that obtained by multiplying the 3D electron density and the gas temperature profiles derived from the data analysis of the \xmm\ mosaic of Coma. 

In Fig.~\ref{fig:total_pressure} we  compare  the 3D X-ray pressure profile with the 3D SZ profile of our reference Model C. 
We point the reader's attention to the very large dynamical range 
shown in the figure: the radius extends up to $r=4~{\rm Mpc}$, probing approximately four orders of magnitude in pressure. 
In contrast, due to a combination of relatively high background level and  available mosaic observations, \xmm\ can probe  the ICM  pressure profile of Coma only up to  $\sim 1\, {\rm Mpc}$. This is a four times smaller radius than \planck, probing only $\sim$ one order of magnitude in pressure.

Due to the good  statistics of both  \planck\ and \xmm\ data,   we see that the pressure profile derived from \planck\ appears  significantly  lower than that of \xmm,  even if they differ  by only $10-15\%$. 
This  discrepancy may be  related to the fact that we are applying  spherical models to a cluster that has a much more complex morphology, with a number of substructures. 
A detailed structural analysis  exploring these apparent pressure profile discrepancies is beyond  the scope of this paper and will be presented in a 
forthcoming study. 
Here we just show a comparison of the  3D  pressure profiles obtained from 
\planck\ and \xmm\ in four  $90\deg$ sectors centred on the cluster and oriented  towards  the four cardinal points (see Fig.~\ref{fig:sector_pressure}).
This shows that the pressure discrepancy depends strongly  on the sector considered. In particular, we find that while 
in the north sector the \planck\ and \xmm\ pressure profiles agree within 
the errors, in the west sector we find  discrepancies,  up to 25--30\%.
As known from X-ray observations \citep[see e.g.,][]{2003A&A...400..811N}
 the north sector is the one that is most regular, while the west sector is the one in which the ICM is strongly elongated, with the presence of major structures.

\begin{figure*}[t]
\begin{centering}$
\begin{array}{cc}
\includegraphics[width =9.cm,angle=0,keepaspectratio]{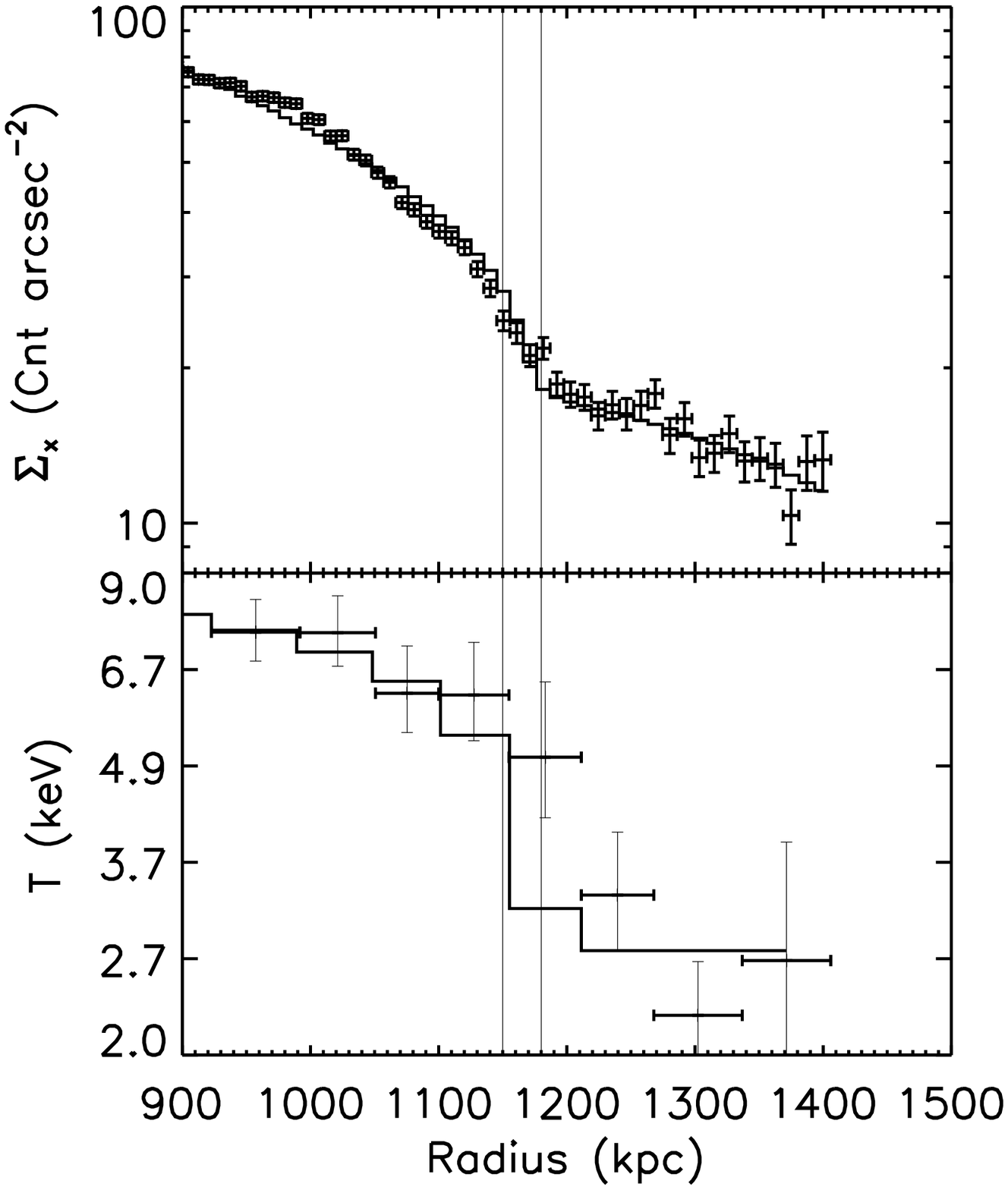}&
\includegraphics[width=9.cm,angle=0,keepaspectratio]{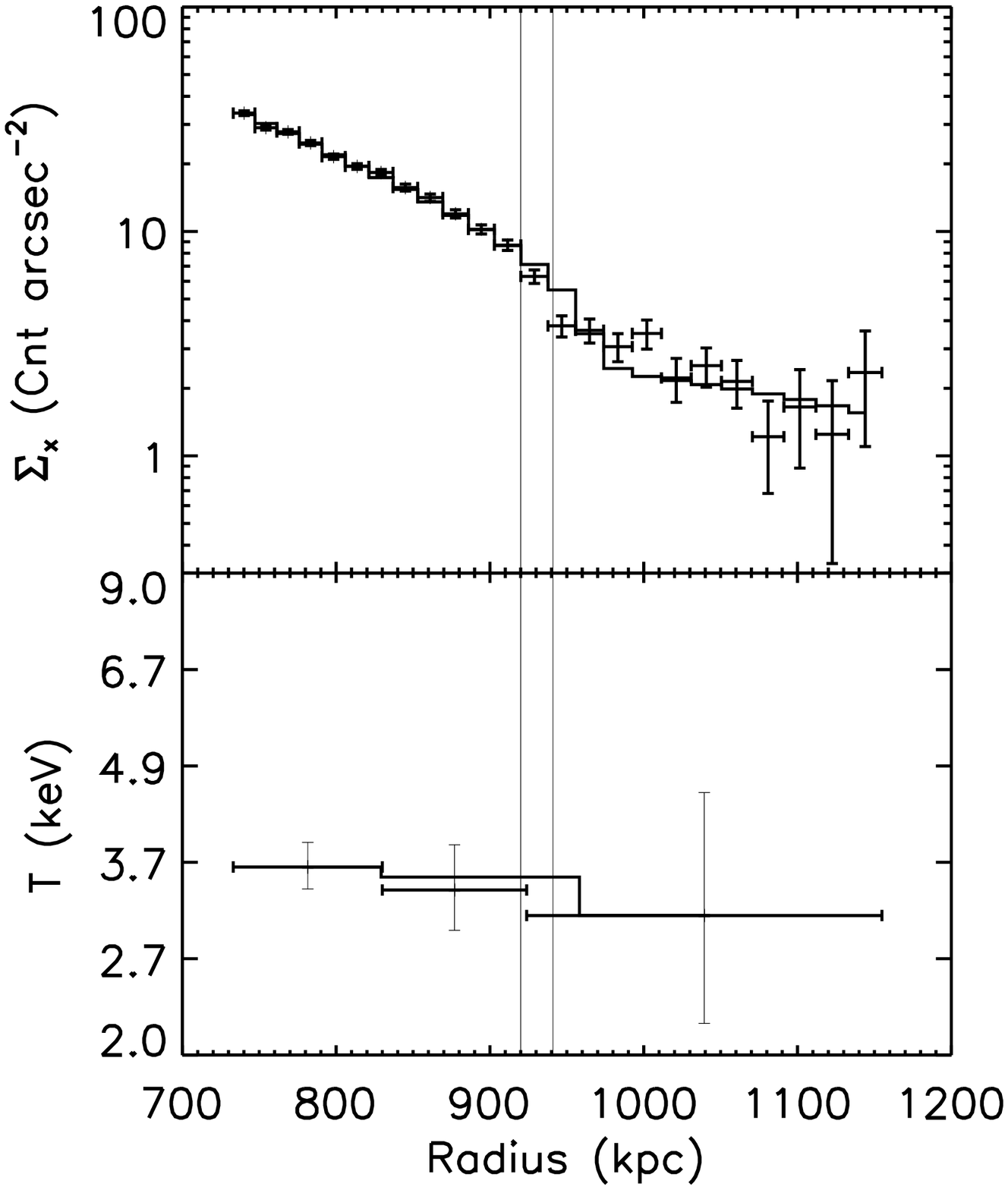}\\

\includegraphics[width=9.cm,angle=0,keepaspectratio]{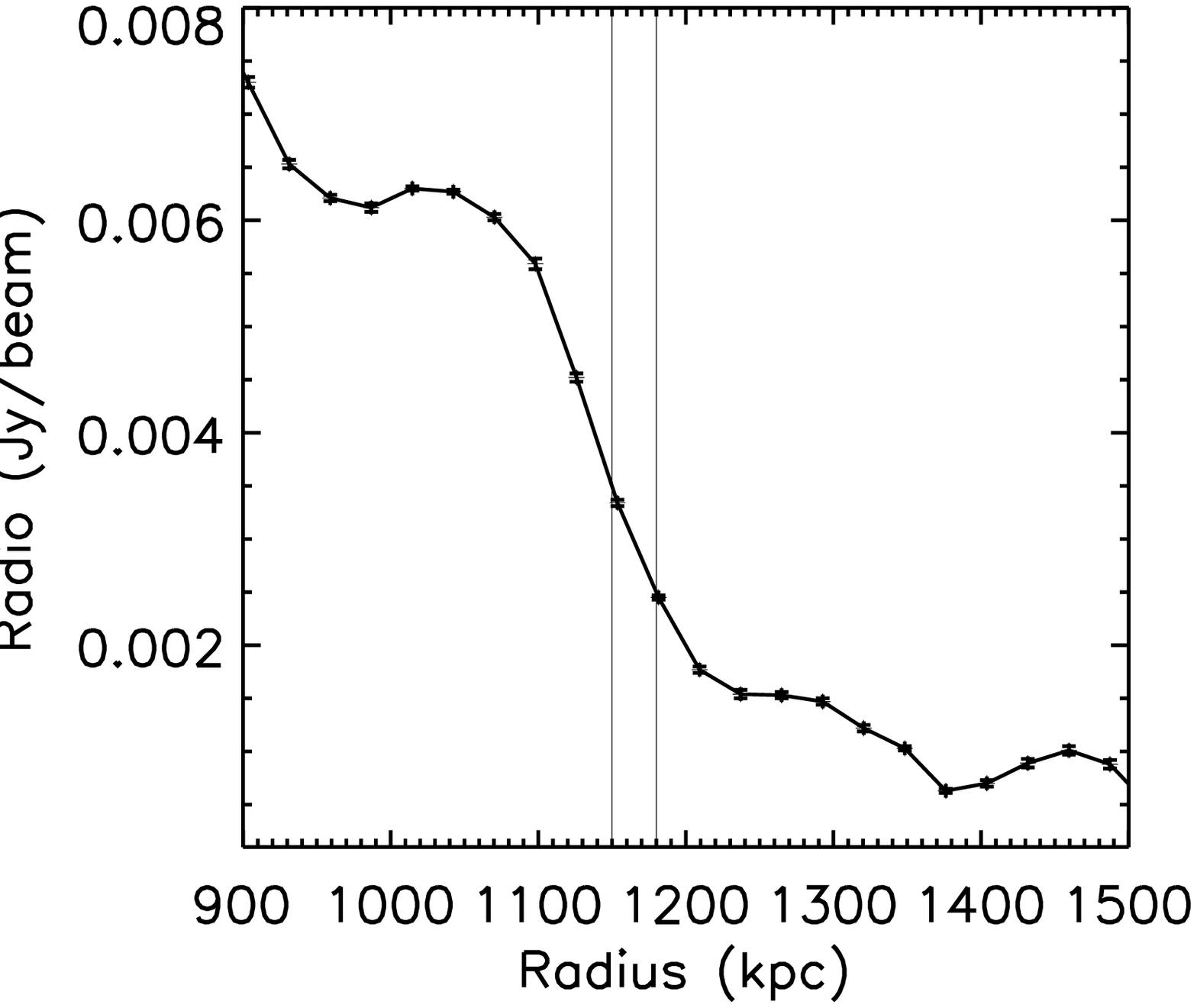}&
\includegraphics[width=9.cm,angle=0,keepaspectratio]{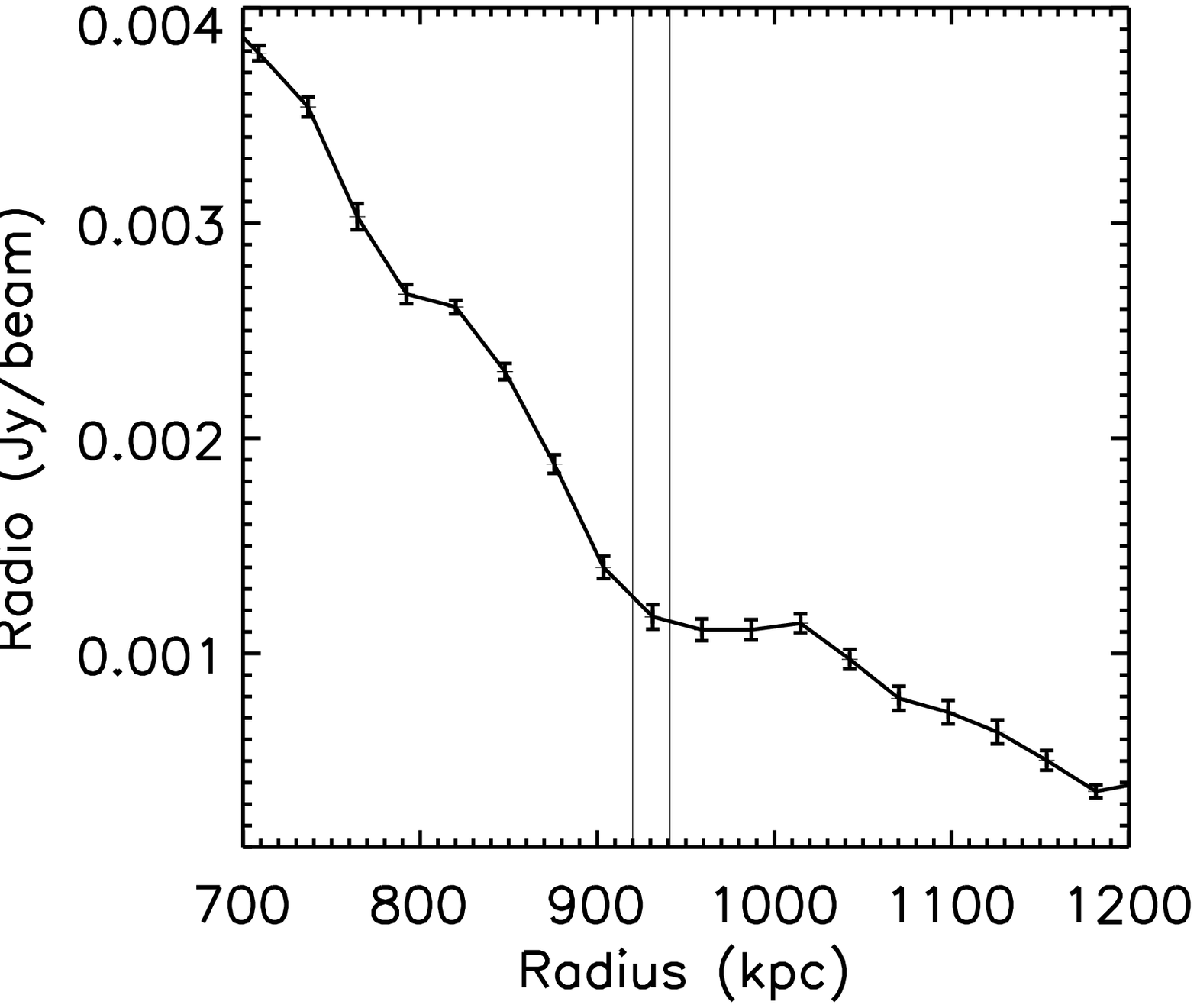}\\
\end{array}$
\end{centering}
\caption{{\footnotesize Comparison of the X-ray and radio properties  in the  west ({\it left panels}) and south-east ({\it right panels}) sectors. {\it Upper panels}: Surface brightness and temperature profiles 
 of the \xmm\ mosaic. The continuous histograms show  the best-fit  models. The 3D pressure model is overplotted in Fig.~\ref{fig:p_shock}.
{\it Lower panels}: Radial profiles of 352~MHz radio emission  at $2 {\rm arcmin}$ resolution in the west ({\it left}) and south-east ({\it right}) sectors after subtraction of radio emission from compact sources  \cite[see][]{2011MNRAS.412....2B}.  The two vertical lines mark the position range of the inferred jumps. 
}}
\label{fig:radio_shock}
\end{figure*}


\begin{table*}[tmb]                 
\begingroup
\newdimen\tblskip \tblskip=5pt
\caption{ Main parameters of the fit of the temperature and density models to the  \xmm\ data (see Eqs.~\ref{eq:shock_ne} and ~\ref{eq:shock_T}). The symbols $M_{\rm n}$, $M_{\rm T}$, $M_{\rm nT}$, and $M_{\rm SZ}$ represent the Mach numbers derived  from the X-ray density, temperature, and  pressure ($n\times T$), and SZ pressure jumps, respectively (see text).}                          
\label{tab:shock_xray}                            
\nointerlineskip
\vskip -3mm
\footnotesize
\setbox\tablebox=\vbox{
   \newdimen\digitwidth 
   \setbox0=\hbox{\rm 0} 
   \digitwidth=\wd0 
   \catcode`*=\active 
   \def*{\kern\digitwidth}
   \newdimen\signwidth 
   \setbox0=\hbox{+} 
   \signwidth=\wd0 
   \catcode`!=\active 
   \def!{\kern\signwidth}
{\tabskip=2em
\halign{\hfill#\hfill&\hfill#\hfill&\hfill#\hfill&\hfill#\hfill&\hfill#\hfill&\hfill#\hfill&\hfill#\hfill&\hfill#\hfill&\hfill#\hfill\cr                     
\noalign{\vskip 5pt}
\noalign{\doubleline}
Sector   & $r_{\rm x}$                 & $D_{\rm n}$                   & $M_{\rm n}$             & $D_{\rm T}$             & $M_{\rm T}$             & $D_{\rm n}\times D_{\rm T}$ & $ M_{\rm nT}$ &$M_{\rm SZ}$    \cr
         &$(\rm{Mpc})$&           &           &            &        &                &            &          \cr 
\noalign{\vskip 3pt\hrule\vskip 5pt}
West     &$1.173^{+0.0003}_{-0.003}$ & $2.00^{+0.03}_{-0.03}$ & $1.73^{+0.03}_{-0.03}$ &$3.0^{+0.7}_{-0.6}$  & $2.6^{+0.4}_{-0.4}$  & $6.0^{+1.4}_{-1.1}$           &$2.3^{+0.2}_{-0.2}$       &$2.03^{+0.09}_{-0.04}$ \cr
\vspace{.03in}
South-east&$0.9778^{+0.0002}_{-0}$    & $2.43^{+0.02}_{-0.02}$ & $2.10^{+0.01}_{-0.01}$ &$1.3^{+1.8}_{-0.6}$ & $1.3^{+1.3}_{-1.3}$  & $3.1^{+1.6}_{-1.1}$          &$1.6^{+0.3}_{-0.1} $      &$2.05^{+0.25}_{-0.02}$ \cr
\noalign{\vskip 5pt\hrule\vskip 3pt}
}
}
}
\endPlancktablewide                 
\endgroup
\end{table*}                        

\subsection{Shocks}\label{sec:shocksdisc}

In Sect.~\ref{sec:shock} we show
that Coma exhibits a localised steepening of its $y$ profile in at least
two directions, to the west and to the  south-east. 
 These suggest the presence of discontinuities in the underlying 3D pressure
profile of the cluster. Using two sectors designed to follow the curvature of the $y$ signal around the
 pressure jumps we  estimate their amplitudes. This represents the first attempt to identify and estimate the amplitude of possible pressure jumps in the cluster atmosphere directly from the SZ signal.
Interestingly, we find that similar features
are observed at the same locations in the X-ray and radio bands.

In Fig.~\ref{fig:radio_shock}  we compare the X-ray and radio cluster properties  from the  west  and south-east sectors selected from the SZ image.  The X-ray surface brightness and temperature profiles have been derived from the \xmm\ mosaic while the radio profile is extracted from the 352~MHz Westerbork observations at   $2 {\rm arcmin}$ resolution. To guide the reader's eye,  we mark, for each profile in the figure, 
the position of the pressure jump as derived from the analysis of the 
$y$ signal (See Table~\ref{tab:shock}).   For both sectors we find that the  X-ray surface brightness and radio profiles 
show relatively sharp features at the same position as the steepening of the \planck\ $y$ profiles. 
This is also the case for the temperature profile of the west sector. For the south-east sector, however,
 this evidence is less clear. Because it is located
 in a much lower signal-to-noise region of the cluster, the
 error of the outermost temperature bin is too large to be able to put a stringent constraint on a possible temperature jump.

 To check if  the X-ray features are also consistent with the hypothesized presence of a discontinuity in the 
cluster pressure profile we  simultaneously fit
the observed X-ray  surface brightness and temperature profiles using the following discontinuous 3D
density and temperature models:

\begin{equation}
n=n_0 \times 
\left\{
\begin{array}{ll}
 D_{\rm n}(r/r_{\rm X})^{-\xi_1}& r<r_{\rm X} \\ 
 (r/r_{\rm X})^{-\xi_2} &r>r_{\rm X};\\ 
\end{array}
\right.
\label{eq:shock_ne}
\end{equation}
and
\begin{equation}
T=T_0 \times 
\left\{
\begin{array}{ll}
 D_{\rm T}(r/r_{\rm X})^{-\zeta_1}& r<r_{\rm X} \\ 
 (r/r_{\rm X})^{-\zeta_2} &r>r_{\rm X}.\\ 
\end{array}
\right.
\label{eq:shock_T}
\end{equation}
\noindent Here $r_{\rm X}$ is the position of the X-ray jump and $D_{\rm n}$ and $D_{\rm T}$
are amplitudes of the density and temperature discontinuities, respectively.
The above models are projected along the line of sight for $r<10{\rm Mpc}$  using a temperature function appropriate for spectroscopic data  \citep{2004MNRAS.354...10M}.  Notice that due to the poor statistics of the temperature in the south-east sector, for this profile we fix ${\xi_1}={\xi_1}=0$. This choice does not affect the determination of the  jump  position  $r_{\rm X }$ which 
is mainly driven  by the surface brightness  rather than by the temperature profile. 

The best-fit position, density, and temperature jumps, together with their 68.4\%  confidence level errors are reported in Table~\ref{tab:shock_xray}. To make a direct comparison with the pressure jump measured from the SZ signal, in the same table we add the  amplitude of the X-ray pressure jump derived by multiplying the X-ray density and temperature  models (i.e., $P_{\rm x}=n_{\rm e}{\rm k}T$).

The best-fit  surface brightness and projected temperature models are shown as histograms in  Fig.~\ref{fig:radio_shock}. The best-fit 3D $P_{\rm x}$ model  and its 68.4\% confidence level errors are overlaid in Fig.~\ref{fig:p_shock}.

From Table~\ref{tab:shock_xray} we  see that the X-ray data from the west sector are consistent with the presence of a discontinuity, both in the 3D density and 3D temperature profiles. Both jumps are detected at $> 5\sigma$ confidence  and
the  pressure jumps derived from  X-ray and from SZ are consistent within the 68.4\% confidence level errors (Table~\ref{tab:shock} and Table~\ref{tab:shock_xray} ). 
This agreement near the discontinuity is also seen in  Fig.~\ref{fig:p_shock} which, in addition,  shows that the 3D pressure profiles for the west sector derived  from the SZ and the X-ray data are consistent not only near the jump, but also  over a much wider radial range.

These results indicate that the feature seen by \Planck\ is produced by a shock induced by
supersonic motions in the cluster's hot gas atmosphere.
 Assuming  Rankine-Hugoniot pressure jump conditions across the fronts
\citep[\S 85 of][]{landau1959fm}, the discontinuity in the density, temperature and pressure profiles are uniquely linked to the shock Mach number.

Table~\ref{tab:shock_xray} shows that the Mach number obtained from the SZ and X-ray pressure profiles are also consistent within the $\pm 1\sigma$ confidence level errors. Furthermore,  the Mach number derived from the X-ray density and temperature profiles agree  within the  $\pm 2 \sigma$ confidence level errors.
This agreement supports the hypothesis that the west feature observed by \planck\ is a shock front.

 For the south-east sector  Table~\ref{tab:shock_xray} shows that the X-ray surface brightness profile is consistent with the presence  of a significant discontinuity in the 3D density profile. Due to the modest statistics, the temperature model returns large errors and 
$D_{\rm T}$ is not constrained (see Table~\ref{tab:shock_xray}). Thus, although consistent, we cannot confirm the presence of a temperature jump.
Despite  this we find that, as for the west sector,  the  pressure jumps and the pressure profiles derived from  X-ray and from SZ are consistent within the 68.4\% confidence level errors (see Fig.~\ref{fig:p_shock} and Tables~\ref{tab:shock} and \ref{tab:shock_xray}). 
 Finally, Table~\ref{tab:shock_xray} shows  that the Mach numbers derived from the  amplitudes of the different 3D models are all consistent within the 68.4\% uncertainty levels. We would like to stress  that this is true not only for $M_{\rm T}$ and $M_{\rm nT}$ which, being directly connected to $D_{\rm T}$, have  relatively large errors,  but also for $M_{\rm n}$ and $M_{\rm SZ}$ which do not depends on $D_{\rm T}$ at all.
As for the west sector, this agreement supports the initial hypothesis that the south-east feature observed by \planck\ is also  a shock front. 

Notice that  the good agreement between the 3D pressure models derived from the X-ray and SZ data, both in the west and south-east sectors,  indicate that, within the selected regions, spherical symmetry is a good approximation to the underlying pressure distribution.

We conclude this section by  pointing  the reader's attention to the fact that, even though the 
radio and X-ray observations have a much better PSF than \planck,   Fig.~\ref{fig:radio_shock} shows that  
the respective  jumps in these observations appear smooth on a scale of $\approx 200 {\rm kpc}\approx 7{\arcm}$. 
As explained in detail in Appendix~\ref{sec:appendix} this is simply 
 a  projection effect  (see also Fig.~\ref{fig:p_shock} and Sect.~\ref{sec:shock}).  Despite its relatively large PSF,
\planck\ is able to measure pressure jumps in the 
atmosphere of the Coma cluster.

\subsection{Quasi-linear SZ-radio relation}
In Sect.~\ref{SZ-radio} we show that for the Coma cluster the radio  brightness and $y$ emission scale approximately linearly with a small scatter  between the radio emission and thermal pressure. Due to the near-linear correlation, where line-of-sight projection effects cancel out, we work here with volume-averaged emissivities. We first
express the monochromatic radio emissivity $[{\rm erg\, s^{-1}\, cm^{-3}\, sr^{-1}\, Hz^{-1}}]$ as:

\begin{equation}\epsilon_r \sim~n_{CRe}~B^{1+\alpha} \sim~\mathcal{Q}_{CRe}~\frac{B^{1+\alpha}}{B^2 + B_{\rm CMB}^2},
\end{equation}

\noindent  where $\alpha$ is the spectral index,$B$ is the magnetic field, $B_{\rm CMB} \approx 3 (1+z) ~{\rm \mu\, G}$ is the equivalent magnetic field of the CMB, and $n_{\rm CRe}$ and $\mathcal{Q}_{\rm CRe}$ are the density and injection rate of cosmic-ray electrons (CRe) , respectively.  In general, $\mathcal{Q}_{\rm CRe}$ can be a function of position and electron energy, and will depend on the model of cosmic-ray acceleration assumed. 
  In {\it secondary} (hadronic) acceleration models \citep{1980ApJ...239L..93D, 1982AJ.....87.1266V}, the relativistic electrons are produced in collisions of long-lived cosmic ray protons with the thermal electrons, resulting in $\mathcal{Q}_{\rm CRe}~\propto$~$n_{\rm e}$~$n_{\rm CRp}$, where $n_{\rm CRp}$ and $n_{\rm e}$ is the density of cosmic-ray protons and thermal electrons, respectively.  Recent models in this category  \citep{2010ApJ...722..737K, 2010ApJ...719L..74K} 
require that,  in contrast to $n_{\rm e}$, 
$n_{\rm CRp}$ should be constant over the cluster volume in order to match the cluster radio brightness profiles. 
\citet{2008MNRAS.385.1211P} show that there is strong cosmic-ray proton injection even in the cluster peripheries, due to the stronger
shock waves there. Strong radio cosmic-ray proton diffusion and streaming within the ICM could also lead to a completely flat cosmic-ray proton
profile
\citep{2011A&A...527A..99E}.  
In the limit where $B \gg B_{\rm CMB}$ and assuming $\alpha \approx 1$ \citep[e.g.][]{1993ApJ...406..399G, 1997A&A...321...55D},
this would lead to $\epsilon_r \propto$ n$_{\rm e} \propto y/T$. This is consistent with our observations\footnote{In the case that  $\alpha = 1 + \delta$, the relationship would be $\epsilon_r\propto (y/T)^{1+\delta/2}$, if we assume $B \propto \sqrt{n_{\rm e}}$. E.g., if we use $\alpha=1.2$ \citep{1993ApJ...406..399G}, we would expect $y \propto T \epsilon_{\rm r}^{0.91}$, which is approximately our measured value. However, we continue to use the term ``linear relationship", with 
the understanding that the difference between our measured slope and linearity 
is consistent for our simple assumption about the spectral index.},  
especially since n$_{\rm e}$(r)  varies much more than $T(r)$ in the Coma cluster   \cite[see e.g.][]{2001A&A...365L..67A, 2008A&A...478..615S}. \citet{2011ApJ...728...53J} 
derive a lower limit for the average field in Coma of $1.7 \mu$G, from limits on the {\it Fermi} $\gamma$-ray flux.
 The $\gamma$-ray analysis thus leaves open the question of whether Coma could be in the strong-field limit.
  
However, the rotation measure observations of \citet{2010A&A...513A..30B} 
provide  characteristic values of 4--5${\,\mu}$G  for the {\em combined} contributions of
the central diffuse cluster field and contributions local to each radio
source \citep[e.g.][]{2011MNRAS.413.2525G, 2003ApJ...588..143R}.   
The majority of Coma's volume,  which is outside of the cluster core, is thus in the weak-field limit, which leads to $\epsilon_r \propto y~B^2 /T $.  To remain consistent with the linear correlation found here, the magnetic field would thus need to be nearly independent of thermal density.  The non-ideal MHD simulations of 
\citet{2011MNRAS.418.2234B} show a typical scaling of $B \propto n_{\rm e}^{0.6}$, which would yield $\epsilon_r \propto y^{2.2}/T^{2.2}$.  This could make the secondary model inconsistent with the observations in the weak-field limit.\\

{\it Primary} (re-)acceleration models assume that relativistic electrons are accelerated directly from shocks and/or turbulence generated in cluster mergers.  The turbulent re-acceleration model  \citep{1987A&A...182...21S, 2001MNRAS.320..365B, 2001ApJ...557..560P}
leads to a scaling of $\epsilon_r \propto$ n$_{\rm e}$~$T^{1.5}\propto y\sqrt{T}$ in the $B<B_{\rm CMB}$ limit \citep{2007MNRAS.378.1565C} if one assumes $B^{2} \propto n_{\rm e}$ which is close to the simulation scaling results of \citet{2011MNRAS.418.2234B}.  
Such a scaling relation
 is consistent with the observed 
correlation. However, in order to connect the cosmic-ray electron density to n$_{\rm e}$,  primary models depend on a large number of free parameters, which are generally fit to match the observations. 
Recent attempts to reduce the number of assumptions by introducing secondary cosmic ray electrons and protons as seed particles \citep[][see above]{2011MNRAS.410..127B} 
fail to reproduce the linear correlation in the weak-field limit. This is another manifestation of the problem all simple models have in explaining the large extent of cluster radio profiles when compared to the X-rays and inferred magnetic fields  
\citep[e.g.,][]{2000A&A...362..151D, 2001A&A...369..441G, 2010MNRAS.407.1565D, 2011MNRAS.412....2B}. In  future, robust measurements of the cluster's magnetic field profile, coupled with high-resolution radio/X-ray/SZ correlations, will be needed to rule out these naive models.  

\subsection{Pressure jumps and radio emission}

Shocks play an important role in the production of radio emission.  We expect that shocks created during cluster mergers will compress magnetic fields and accelerate relativistic particles.  However, the radiating electrons will quickly lose their energy post-shock, and may not be visible for more than
 $\sim 100 {\rm kpc}$ behind the shock \citep[e.g.,][]{2005ApJ...627..733M},
given characteristic shock velocities and magnetic fields at ${\rm \mu\, G}$ levels.  
These shock-accelerated electrons, in shock-compressed magnetic fields, have been proposed as the explanation
for the observed polarised radio synchrotron radiation from cluster peripheral relic sources \citep{1998A&A...332..395E}.
Lower fields do not increase the electron lifetimes, and can even decrease them at fixed observing frequency, because of inverse Compton losses against the CMB.
 Recent simulations show that the presence of  cluster-wide turbulence following a major merger is maintained for a few Gyr at a few percent thermal pressure \citep[e.g.,][]{2005MNRAS.364..753D, 2006MNRAS.369L..14V, 2007ApJ...669..729K, 2011ApJ...726...17P}. 
This turbulence can re-accelerate mildly relativistic seed electrons, and is potentially responsible for the large-scale halo emission (see above).    In addition, an extensive population of low Mach number shocks is also seen in simulations \citep[e.g.,][]{2000ApJ...542..608M, 2006MNRAS.367..113P} and could play an important role in particle re-acceleration.

Shocks will also induce turbulence in the post-shock region ($\sim$200--300$\, {\rm kpc}$).
 There are hints from the small-scale  X-ray residuals (figure 3 of \citealt{2004A&A...426..387S}) 
that such turbulence may exist interior to the possible shocks seen in the west and south-east.
 For the western region, the combination of the  SZ/X-ray pressure jump, X-ray suggested turbulence, and excess synchrotron emission, points toward a connection between turbulence and diffuse synchrotron emission. The details of that connection, however, are not
clear. In addition to direct acceleration by turbulence, the post-shock
synchrotron emission could be a result of a population of weaker, as yet undetected shocks, or freshly accelerated cosmic-ray protons interacting with the ICM in a region where turbulence
has amplified the magnetic field
 \citep[e.g.,][]{2005MNRAS.364..753D, 2008Sci...320..909R, 2009JCAP...09..024K, 2010arXiv1011.0729K}.   Synchrotron spectral indices and magnetic field measurements,   in combination with reliable measurements of weaker shocks and turbulence, would be needed to discriminate between potential models.

The expected rapid loss of radio emissivity post-shock can also help us understand why shocks are sometimes easily detected in the radio, but other times are not.  In the clearest cases, radio shocks are seen beyond any central halo as relatively thin structures known as ``peripheral relics'' \citep{2010Sci...330..347V},  where they can accelerate relativistic electrons. 
Radio shocks may also be found at or near the edge of the halo, and would be characterised by a sharp, but low contrast, rise in brightness, while the post-shock emission blends in with the halo instead of falling off.  The western shock described here in Coma, as well as suggested shocks at the edges of halos in Abell clusters 521 and 754 \citep{2008A&A...486..347G, 2010arXiv1010.3660M, 2011ApJ...728...82M} 
are likely examples of this case.  Contrast effects will camouflage the appearance of shocks that are projected against any radio halo emission.  This is probably the case for the $y$ shock in the south-east, where the radio halo extends far beyond the shock.  
The Coma cluster thus hosts all three types  of  ``radio shocks":  a)  the classic {\it peripheral relic } at  a distance of $1.7{\rm Mpc}$ from the centre (which \citealt{1998A&A...332..395E} and \citealt{2011MNRAS.412....2B} suggest is an ``infall" shock); 
b) the western shock at the {\it edge} of the halo; and c) the south-east 
shock {\it projected} against the fading radio halo. \\

\section {Conclusions}\label{sec:conclusion}

We present the SZ observations of the  Coma cluster  based on the \planck\ nominal survey of 14 months.   
The excellent sensitivity of \planck\ allows, for the first time, the detection of SZ emission out to at least $R\approx 3 \times R_{500}$.
We limit our investigation to the radial  and sectoral properties of the intracluster medium, 
and we study the pressure distribution  to the outermost cluster regions.
Our three main results can be summarised as follows:
\begin{itemize}
\item the Coma pressure profile is flatter than the mean 
pressure profile predicted by the  B04+N07+P08 numerical simulations and lies on the upper envelope 
of  the simulated profile distribution. This effect has also been found in the pressure profile derived by  stacking 62 nearby clusters of galaxies observed with \planck\ \citep{planck2012-V}. 

\item  \planck\  detects a localised steepening of the $y$ profile
about half a degree to the west and also to the south-east of the cluster
centre. Features in the X-ray and radio synchrotron profiles at
similar locations suggest the presence of shock waves that
propagate with Mach number $M_{\rm w}=2.03^{+0.09}_{-0.04}$ and  $M_{\rm se}=2.05^{+0.25}_{-0.02}$ in  the west and south-east directions, respectively. 

\item the $y$ and radio-synchrotron signals are quasi-linearly correlated on Mpc-scales with only small intrinsic scatter. 
This implies either that,  unlike the thermal plasma,  the energy density of cosmic-ray electrons is relatively constant throughout the cluster, or that the magnetic fields fall off much more slowly with radius than previously thought. We detect a correspondence between the western $y$ shock and a previously reported radio/X-ray edge,  and we argue that either the magnetic fields are strong in the cluster outskirts, which would permit the hadronic model to explain the radio emission,  or some sort of re-acceleration by turbulence or additional shock waves must operate in the region behind the detected outer shock structures.
\end{itemize}
Even though this analysis is based on only about half of the data collected by \planck , 
our results represent a substantial step forward in the study of the physics of the Coma cluster.  The full set of data collected by \Planck,  will not only  improve the signal-to-noise
by another factor $\sim\sqrt(2)$
but also significantly improve our understanding of instrumental effects. Thus, we will be able to generate more accurate $y$ maps, and more thoroughly
unveil Coma's  two-dimensional SZ 
structure and its  filamentary environment.

\begin{acknowledgements}
A description of the Planck Collaboration and a
list of its members, indicating which technical or scientific
activities they have been involved in, can be found at  http://www.rssd.esa.int/Planck\_Collaboration.
The Planck Collaboration acknowledges the support of: ESA; CNES and CNRS/INSU-IN2P3-INP (France); ASI, CNR, and INAF (Italy); NASA and DoE (USA); STFC and UKSA (UK); CSIC, MICINN and JA (Spain); Tekes, AoF and CSC (Finland); DLR and MPG (Germany); CSA (Canada); DTU Space (Denmark); SER/SSO (Switzerland); RCN (Norway); SFI (Ireland); FCT/MCTES (Portugal); and DEISA (EU).
Partial support for this work for L. Rudnick comes from U.S. NSF Grant 09-08668 to the University of Minnesota. We would also like to acknowledge useful conversations with G. Brunetti.

\end{acknowledgements}

\bibliographystyle{aa}
\bibliography{Planck_coma.bib,Planck_bib.bib}

\appendix

\renewcommand{\figurename}{Fig. A}
\setcounter{figure}{0}

\begin{figure*}[t]
\begin{centering}$
\begin{array}{cccc}
\includegraphics[height=7.7cm,angle=0,keepaspectratio]{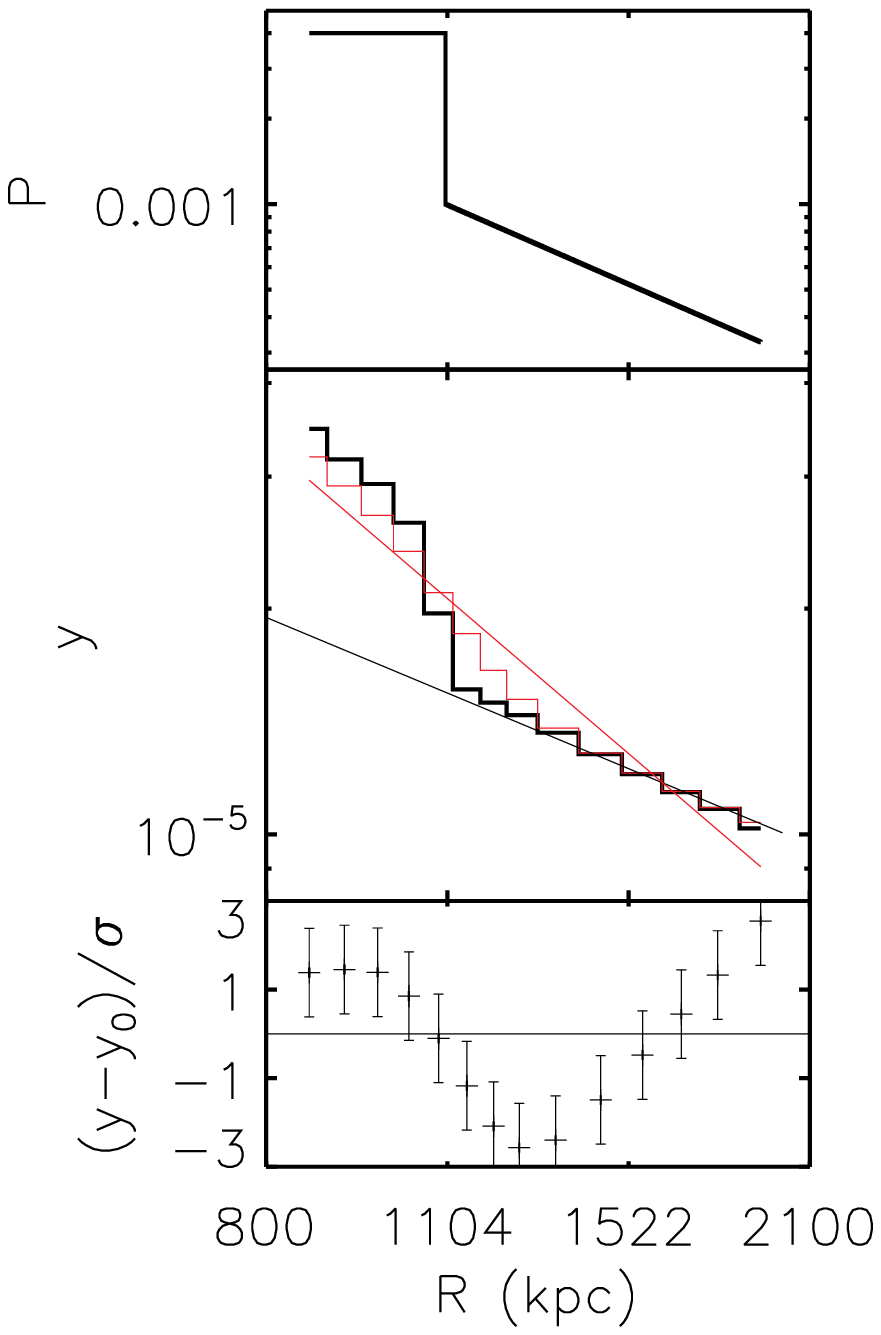}\hspace{-.3in}&
\includegraphics[height=7.7cm,angle=0,keepaspectratio]{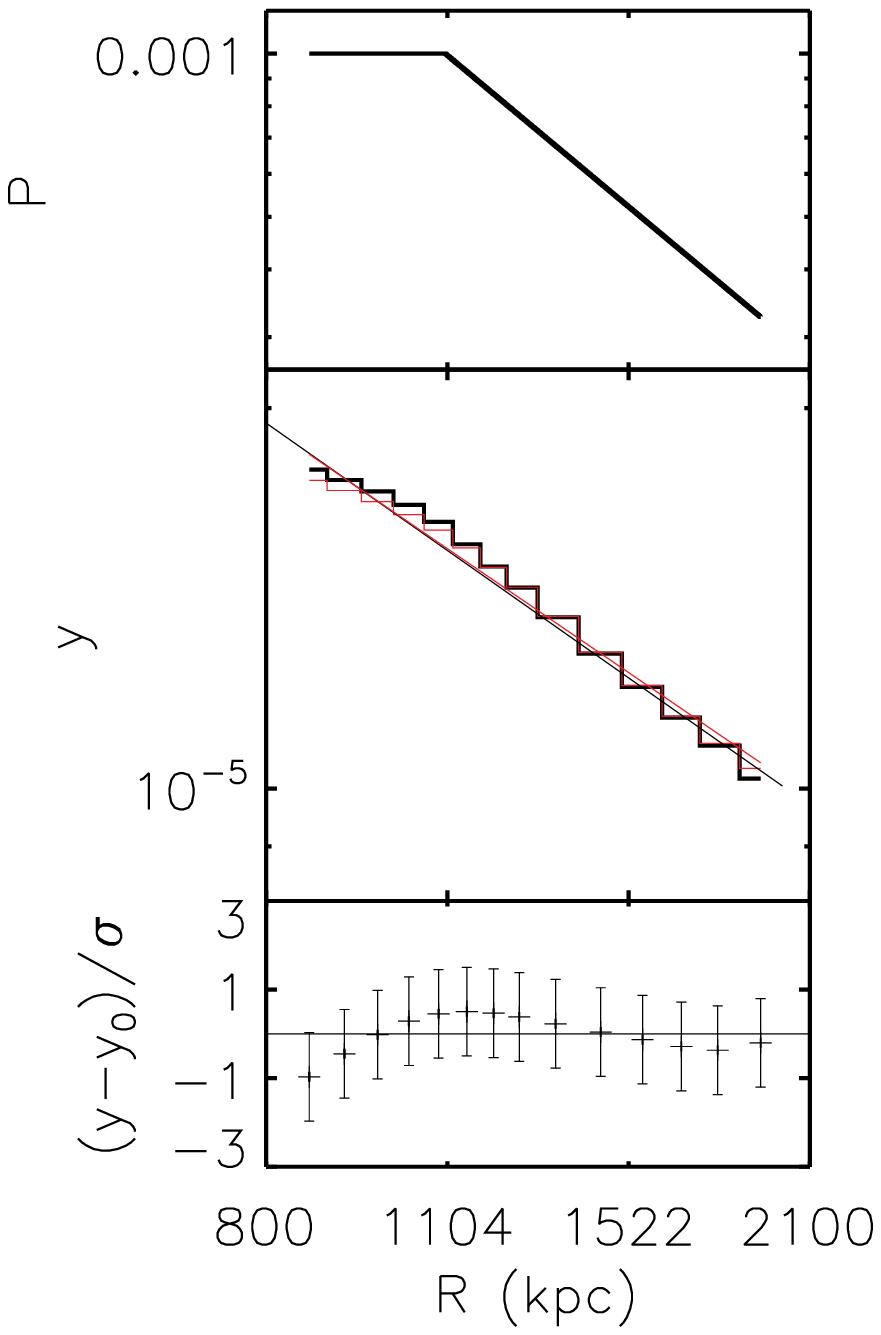}\hspace{-.3in}&
\includegraphics[height=7.7cm,angle=0,keepaspectratio]{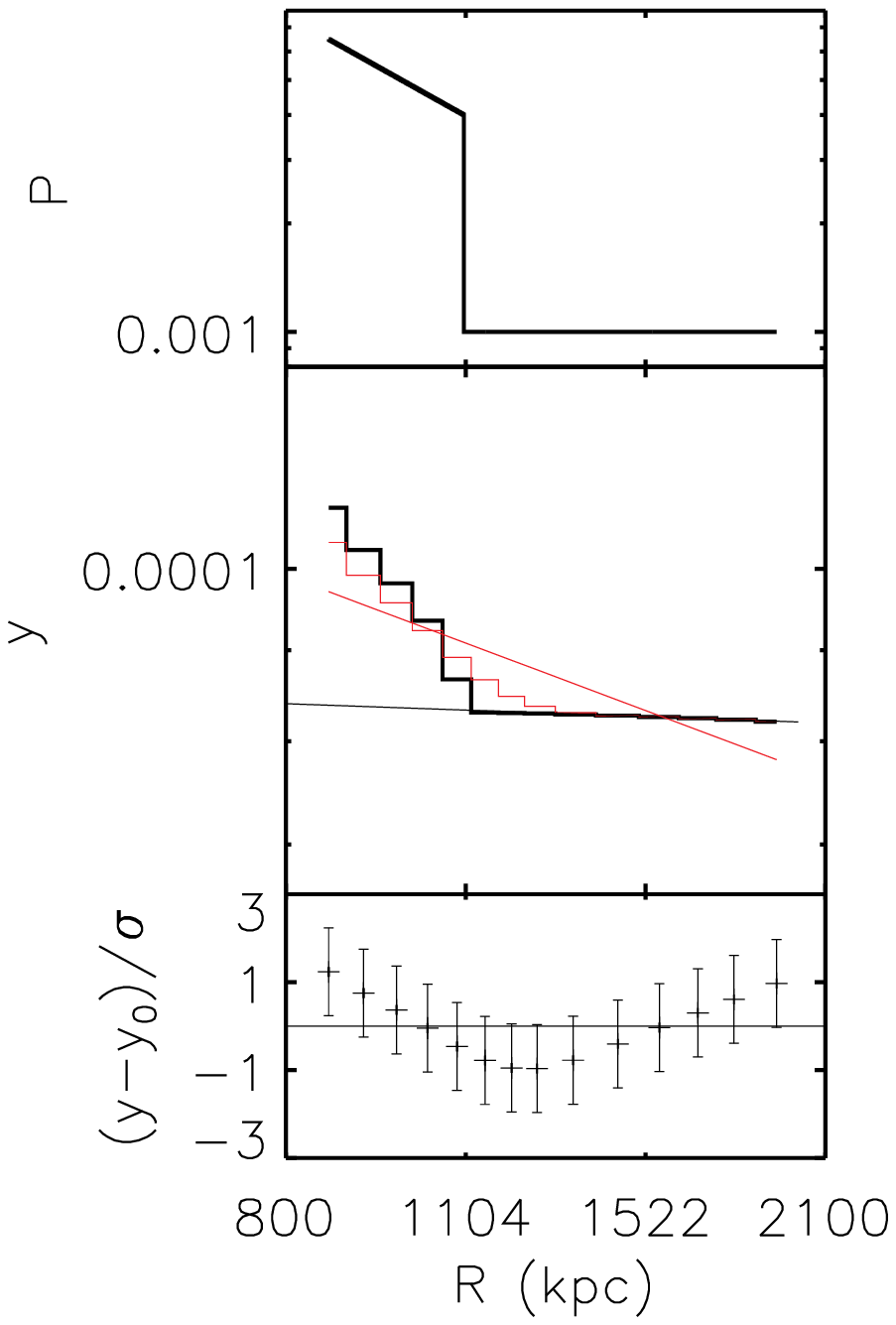}\hspace{-.3in}&
\includegraphics[height=7.7cm,angle=0,keepaspectratio]{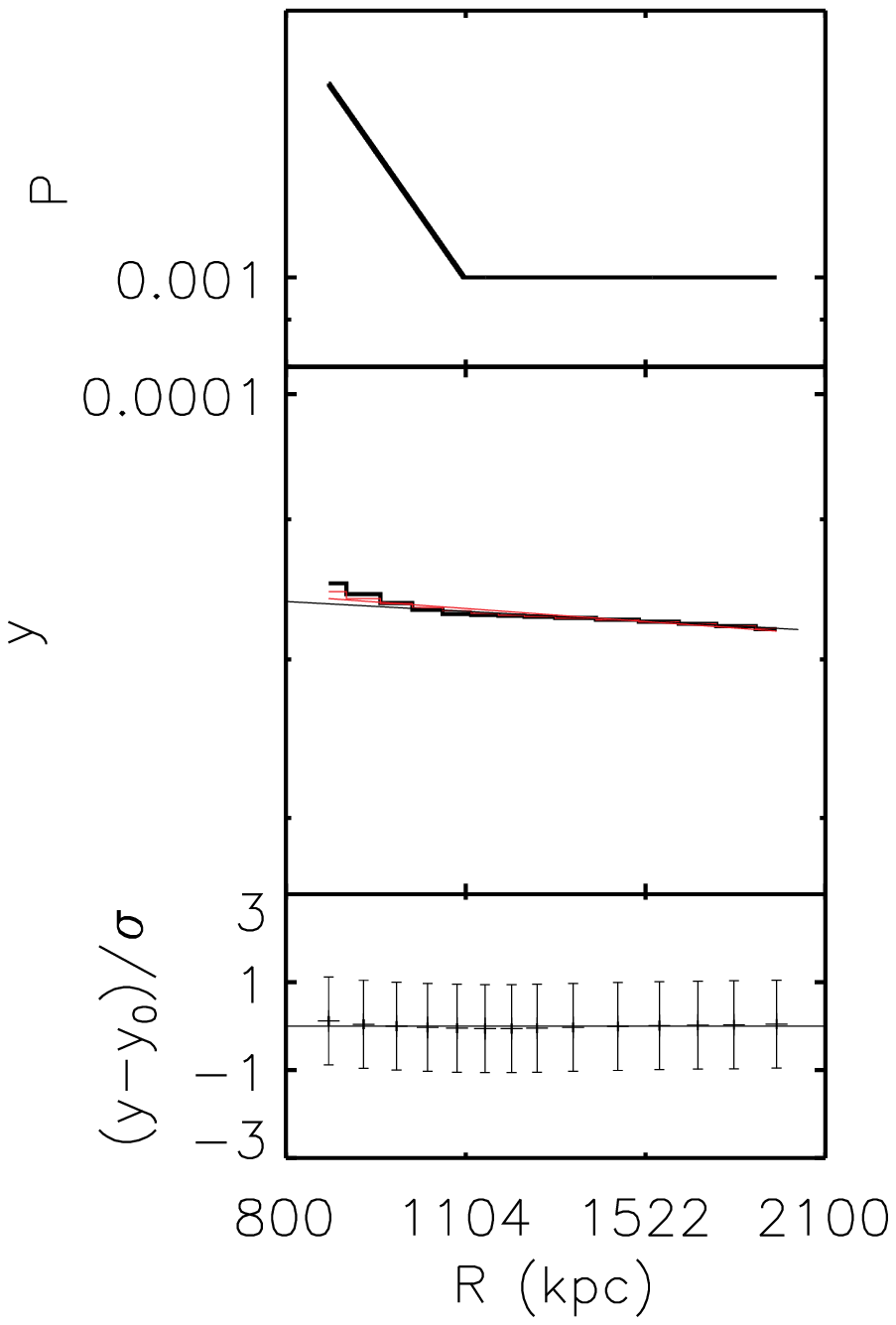}\\
\end{array}$
\end{centering}
\caption{{\footnotesize  Histogram of the effects of  projection and \planck\ PSF on the $y$ radial profile produced by an underlying broken power law pressure profile with and without a pressure discontinuity. In this figure we fix $P_0=10.\times 10^{-4}{\rm cm^{-3}\, keV}$, $r_{\rm J}=1.1~{\rm Mpc}$ and, from left to right we consider 
four different cases: 
i) $\eta_1=0$ , $\eta_2=2$, $D_{\rm J}=4$; ii) $\eta_1=0$ , $\eta_2=2$, $D_{\rm J}=1$;
iii) $\eta_1=2$ , $\eta_2=0$, $D_{\rm J}=4$; and iv) $\eta_1=2$ , $\eta_2=0$, $D_{\rm J}=1$.
{\it Upper Panels:} The underlying 3D pressure profile. {\it Middle Panels:}  The black and red  histograms are the projected $y$ profiles observed by an instrument with infinite angular resolution and with a PSF of $10~{\rm arcmin}$, 
respectively. The red line represents the best-fit of a simple power law to the entire convolved profile (red histogram). The black line is the same as the red line, but 
considering only the three outermost projected profile bins. {\it Lower Panels:} Ratio between the PSF-convolved and projected $y$ profile and its  best-fit power law model (red histogram and lines in middle panel) in units of a relative error, which, for this illustration, we set to 10\%. 
}}\label{fig:model_pwl} 
\end{figure*}

\begin{figure}[t]
\begin{centering}
\includegraphics[width=\linewidth,angle=0,keepaspectratio]{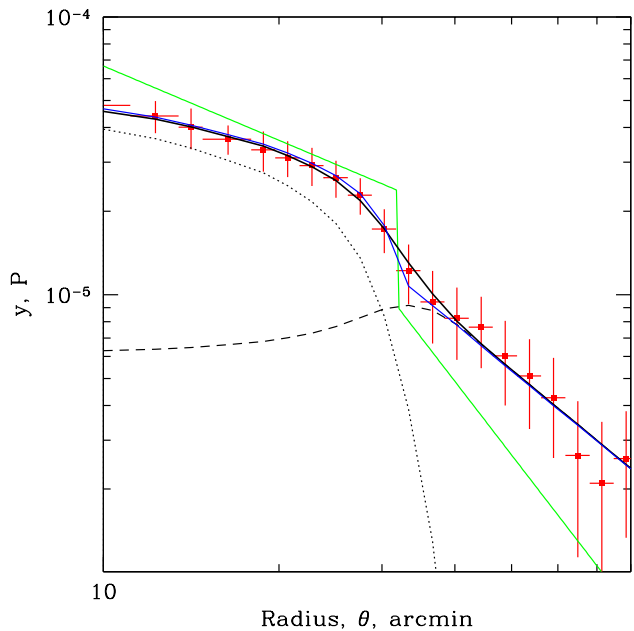}
\end{centering}
\caption{{\footnotesize 
Comparison of the 3D pressure model (green line) and the
corresponding projected $y$ profile, smoothed with the 10$\arcm$
beam. The dotted black line shows the expected $y$ profile due to the inner
power component, while the dashed line shows the contribution of the outer
power law component. The black solid line is the sum of these two
components. For comparison, the blue line shows the same model, but not
convolved with the 10$\arcm$ beam. In this plot $r_{\rm J}\sim 30\arcm$. Due to
projection effects, the range of radii affected by the value of $r_{\rm J}$
is of order $r_{\rm J}$ itself. Since $r_{\rm J}$ exceeds $10\arcm$, many independent data
points with large signal-to-noise ratio  contributes to the
determination of $r_{\rm J}$, allowing $r_{\rm J}$ to be estimated with an uncertainty
below the nominal angular resolution of the telescope. 
}}\label{fig:eugenemodel} 
\end{figure}

\begin{figure*}[t]
\begin{centering}$
\begin{array}{ccc}
\includegraphics[height=5.cm,angle=0,keepaspectratio]{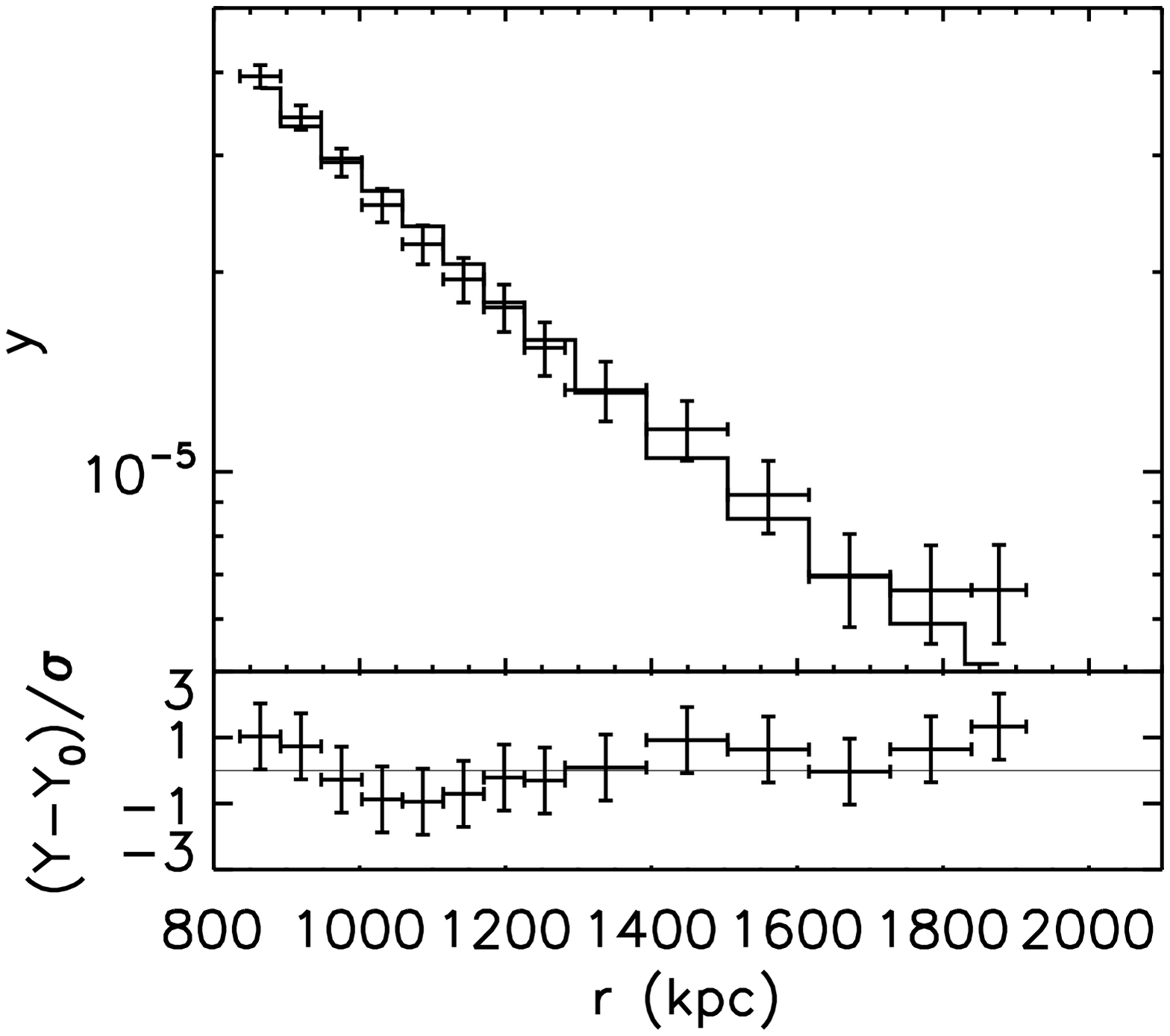}&
\includegraphics[height=5.cm,angle=0,keepaspectratio]{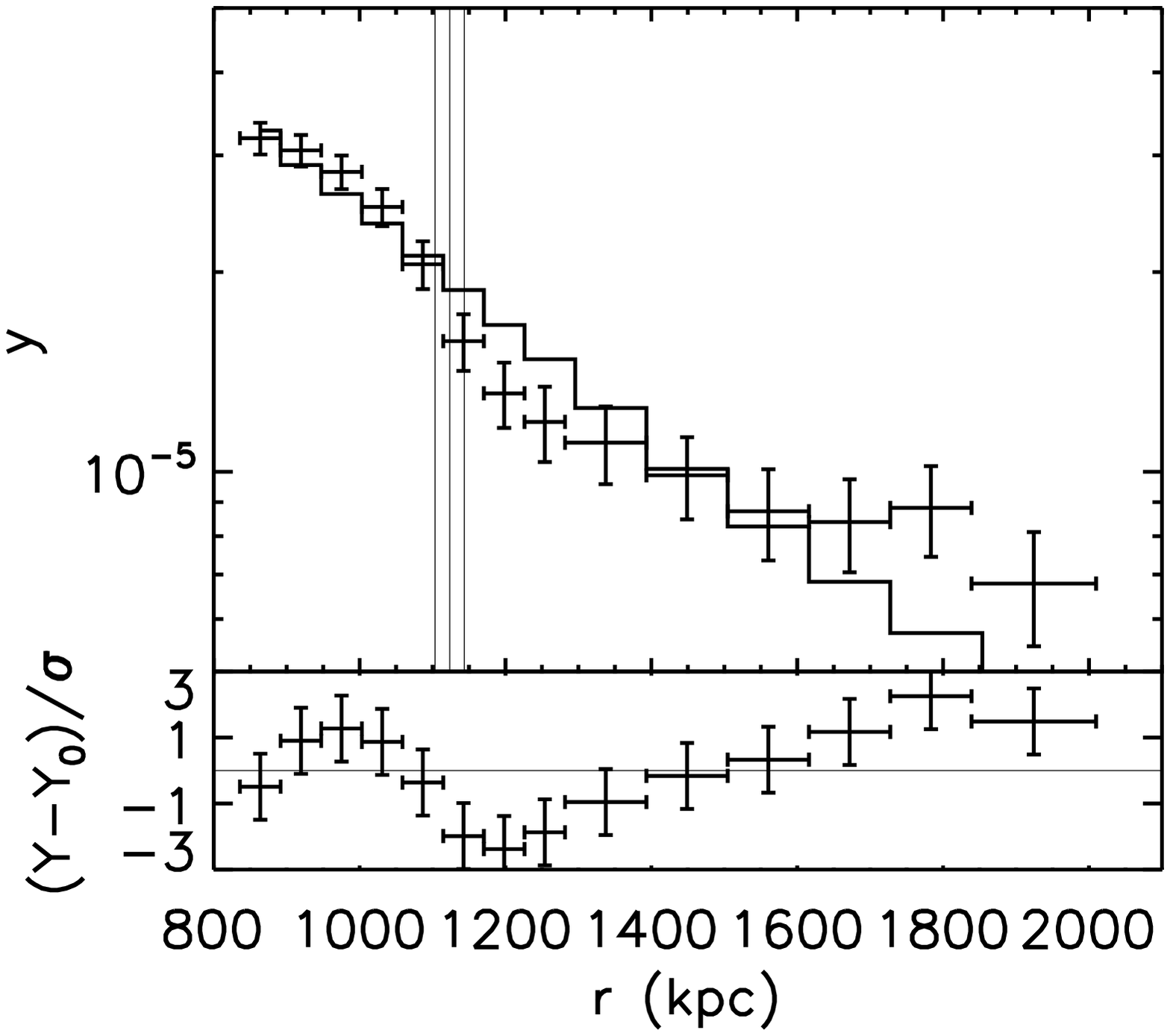}&
\includegraphics[height=5.cm,angle=0,keepaspectratio]{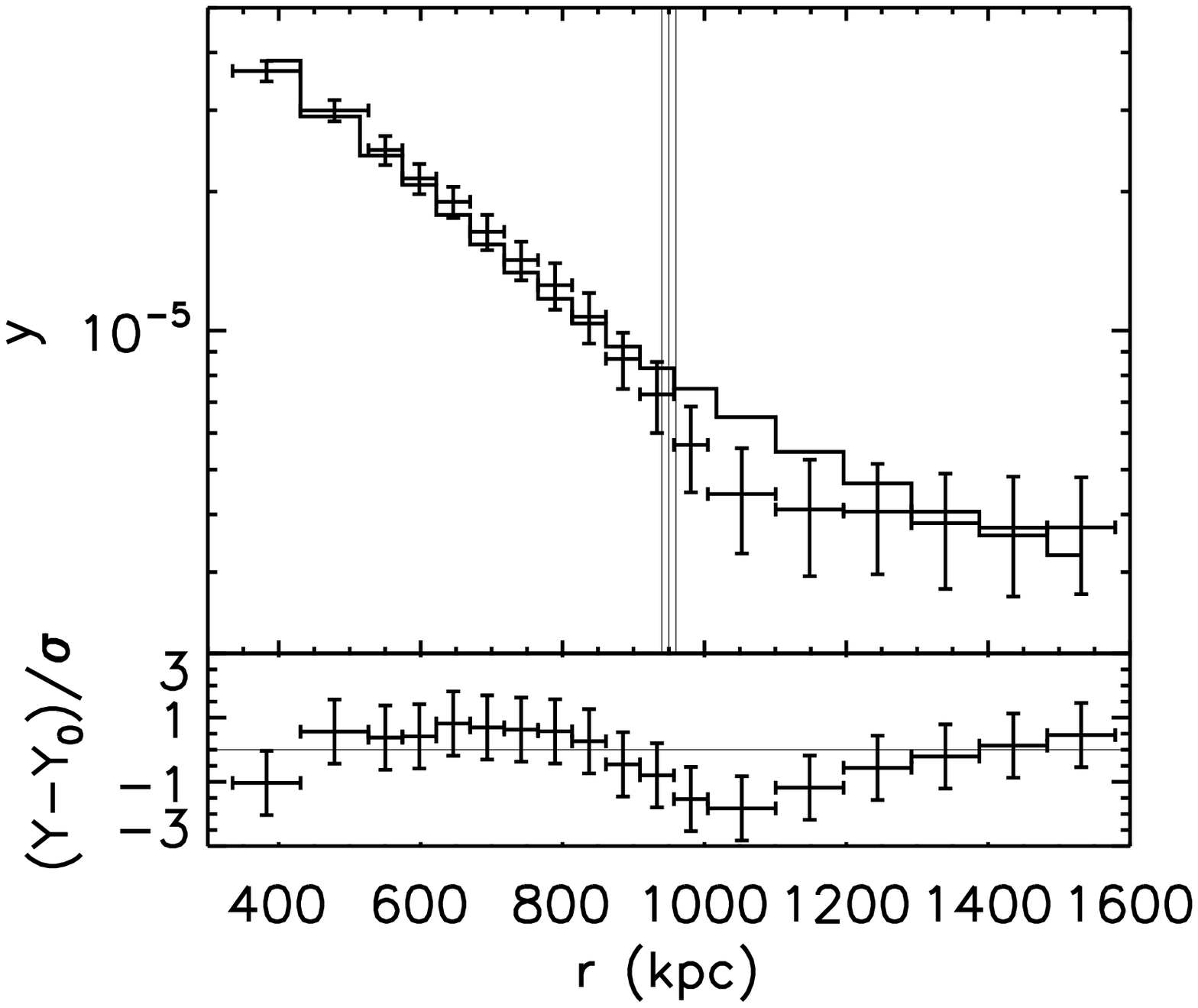}
\end{array}$
\end{centering}
\caption{{\footnotesize 
Results of the fit of the  $y$ profile extracted from three cluster sectors with a simple power law. 
{\it Left Panel:} sector with the same angular size and extension as the west shock but pointing to the north,  where there are no visible shock features. {\it Middle Panel:}  profile of the west shock. {\it Right Panel:} profile of the south-east shock.
In the lower panel of each figure we report the ratio between the observed and the best-fit model of the projected $y$ profile in units of the relative error.
The figure clearly illustrates that while the power law gives a good fit for the north sector where no shock is present, it returns a poor fit in the west and south-east sectors. These two cluster regions require a discontinuity in the pressure jump, as shown in Fig.~\ref{fig:shock}. 
}}\label{fig:powerlaw} 
\end{figure*}

\section{Pressure profile discontinuities as seen by \Planck}\label{sec:appendix}

In this section we show that   the 10~arcmin angular resolution of 
\planck\ is still sufficient to detect and measure 3D pressure jump features in the Coma cluster. This is because  at 
the cluster redshift,  $10~{\rm arcmin}\approx 280~{\rm kpc}$, which is of the
  order of the smoothing induced by projection.
To show this,  we assume  a discontinuous 3D pressure profile described by Eq.~(\ref{eq:shock}).
We  project this profile along the line of sight and we calculate the corresponding $y$ profile as observed with an instrument with: 
\begin{description}
\item i) infinite angular resolution;
\item ii) a  10$\arcm$ FWHM angular resolution,  as for   \planck .
\end{description}

The results of this exercise are  illustrated in Fig.~A\ref{fig:model_pwl},
 where we compare  four different cases  with and without  pressure jumps. 
In  the upper section of each  panel of Fig.~A\ref{fig:model_pwl}
 we show the input 3D pressure profiles.
In the middle sections  we show  the projected pressure profiles without 
any smoothing (as black histograms). The lower sections show deviations from a single power law fit.
The  panel on the left shows that, due to  simple projection effects, the $y$ profile appears 
smoothed  with an equivalent  smoothing  scale of $\approx $200--250$\, {\rm kpc}$.  For Coma, this corresponds to an angular scale of 7--9$\, {\rm arcmin}$.
In the same panel we overplot, as a red histogram, the $y$ profile convolved with the \planck\ PSF. This illustrates that the effect of the PSF smoothing is secondary with respect to  projection effects.
This  indicates that there is only a modest 
gain, from the detection point of view, in observing this 
specific feature using an instrument with a much better angular resolution 
than \planck .
 Notice that the fact that the projection of the 3D pressure
distribution onto a plane converts the sharp jump into a curved surface
brightness profile allows us to recover the position of $r_{\rm J}$ with an accuracy higher than
the angular resolution of \Planck. 

Fig.~A\ref{fig:eugenemodel} clearly shows
that the range of radii affected by the
$r_{\rm J}$ value is of the order of $r_{\rm J}$ itself. Since $r_{\rm J} > 10\arcm$, many
independent data points with large signal-to-noise ratio are
contributing to the determination of $r_{\rm J}$, driving the uncertainty
well below the nominal angular resolution of the telescope. Of course
the value of $r_{\rm J}$ is still subject to systematic uncertainties, 
e.g. from our assumption of spherical symmetry.

We use the above exercise  to illustrate two practical ways to 
identify the presence of a possible underlying 3D discontinuous pressure 
profile, hidden behind  some observed projected profile extracted in a 
specific cluster sector.
The first way is to search for the actual pressure jump in the observed pressure profile. As  projection smooths the profile, this needs to be done by looking at the profile extremes. We first notice that
 the outermost bins of the profile are practically unaffected by  PSF smoothing. 
This is clear  in the middle sections of  Fig.~A\ref{fig:model_pwl},
where  the red and back histograms are similar in  the  outermost 3 or 4 bins.
At this point, if we fit a line to either the 3 or 4   outermost bins, this will give an indication of the un-convolved $y$ profile slope at large radii. 
If we  extrapolate  this line to the centre, we 
can easily highlight the presence of a pressure jump.
This procedure is shown in the middle section of  Fig.~A\ref{fig:model_pwl},
using black straight lines. From  Fig.~A\ref{fig:model_pwl} we can see that
this procedure  highlights  
the intensity variation due to a pressure jump (see first and third panels).
In the case where we have no pressure jump  (second and forth panels) the best-fit 
line to the outermost bins tends to closely  follow the entire profile.

The second way to highlight the presence of a pressure jump is to fit a line to the entire observed profile and to examine the residuals. This procedure
is illustrated in the middle and lower sections of Fig.~A\ref{fig:model_pwl}.
The red lines in the middle section are the best-fit power law relations to the entire observed profile (red histogram). The crosses in the lower panels
 indicate the differences, in units of the relative errors, between the PSF 
convolved projected $y$ profile and its best-fit power law model. 
This figure shows that a 3D pressure jump  induces a characteristic  
signature in the residuals. 

 In Fig.~A\ref{fig:powerlaw} we  apply this second technique to the Coma cluster
by  showing  the fit of the $y$ profile extracted from three cluster sectors with a simple power law. The figure clearly illustrates that while the power law gives a good fit for the north sector where no shock is present, it returns a poor fit in the west and south-east sectors. These two cluster regions require a discontinuity in the pressure jump, as shown in Fig.~\ref{fig:shock}. 

\raggedright

\end{document}

%% file: PIP_10_authors_and_institutes.tex
\author{\small
Planck Collaboration:
P.~A.~R.~Ade\inst{92}
\and
N.~Aghanim\inst{63}
\and
M.~Arnaud\inst{78}
\and
M.~Ashdown\inst{75, 6}
\and
F.~Atrio-Barandela\inst{19}
\and
J.~Aumont\inst{63}
\and
C.~Baccigalupi\inst{91}
\and
A.~Balbi\inst{38}
\and
A.~J.~Banday\inst{101, 9}
\and
R.~B.~Barreiro\inst{71}
% ...
}
\institute{\small
APC, AstroParticule et Cosmologie, Universit\'{e} Paris Diderot, CNRS/IN2P3, CEA/lrfu, Observatoire de Paris, Sorbonne Paris Cit\'{e}, 10, rue Alice Domon et L\'{e}onie Duquet, 75205 Paris Cedex 13, France\\
\and
Aalto University Mets\"{a}hovi Radio Observatory, Mets\"{a}hovintie 114, FIN-02540 Kylm\"{a}l\"{a}, Finland\\
\and
Academy of Sciences of Tatarstan, Bauman Str., 20, Kazan, 420111, Republic of Tatarstan, Russia\\
\and
Agenzia Spaziale Italiana Science Data Center, c/o ESRIN, via Galileo Galilei, Frascati, Italy\\
\and
Agenzia Spaziale Italiana, Viale Liegi 26, Roma, Italy\\
\and
Astrophysics Group, Cavendish Laboratory, University of Cambridge, J J Thomson Avenue, Cambridge CB3 0HE, U.K.\\
\and
Atacama Large Millimeter/submillimeter Array, ALMA Santiago Central Offices, Alonso de Cordova 3107, Vitacura, Casilla 763 0355, Santiago, Chile\\
\and
CITA, University of Toronto, 60 St. George St., Toronto, ON M5S 3H8, Canada\\
\and
CNRS, IRAP, 9 Av. colonel Roche, BP 44346, F-31028 Toulouse cedex 4, France\\
\and
California Institute of Technology, Pasadena, California, U.S.A.\\
\and
Centre of Mathematics for Applications, University of Oslo, Blindern, Oslo, Norway\\
\and
Centro de Astrof\'{\i}sica, Universidade do Porto, Rua das Estrelas, 4150-762 Porto, Portugal\\
\and
Centro de Estudios de F\'{i}sica del Cosmos de Arag\'{o}n (CEFCA), Plaza San Juan, 1, planta 2, E-44001, Teruel, Spain\\
\and
Computational Cosmology Center, Lawrence Berkeley National Laboratory, Berkeley, California, U.S.A.\\
\and
Consejo Superior de Investigaciones Cient\'{\i}ficas (CSIC), Madrid, Spain\\
\and
DSM/Irfu/SPP, CEA-Saclay, F-91191 Gif-sur-Yvette Cedex, France\\
\and
DTU Space, National Space Institute, Technical University of Denmark, Elektrovej 327, DK-2800 Kgs. Lyngby, Denmark\\
\and
D\'{e}partement de Physique Th\'{e}orique, Universit\'{e} de Gen\`{e}ve, 24, Quai E. Ansermet,1211 Gen\`{e}ve 4, Switzerland\\
\and
Departamento de F\'{\i}sica Fundamental, Facultad de Ciencias, Universidad de Salamanca, 37008 Salamanca, Spain\\
\and
Departamento de F\'{\i}sica, Universidad de Oviedo, Avda. Calvo Sotelo s/n, Oviedo, Spain\\
\and
Department of Astronomy and Geodesy, Kazan Federal University,  Kremlevskaya Str., 18, Kazan, 420008, Russia\\
\and
Department of Astrophysics, IMAPP, Radboud University, P.O. Box 9010, 6500 GL Nijmegen,  The Netherlands\\
\and
Department of Physics \& Astronomy, University of British Columbia, 6224 Agricultural Road, Vancouver, British Columbia, Canada\\
\and
Department of Physics and Astronomy, Dana and David Dornsife College of Letter, Arts and Sciences, University of Southern California, Los Angeles, CA 90089, U.S.A.\\
\and
Department of Physics and Astronomy, University of Iowa, 203 Van Allen Hall, Iowa City, IA 52242, U.S.A.\\
\and
Department of Physics, Gustaf H\"{a}llstr\"{o}min katu 2a, University of Helsinki, Helsinki, Finland\\
\and
Department of Physics, Princeton University, Princeton, New Jersey, U.S.A.\\
\and
Department of Physics, University of California, Berkeley, California, U.S.A.\\
\and
Department of Physics, University of California, One Shields Avenue, Davis, California, U.S.A.\\
\and
Department of Physics, University of California, Santa Barbara, California, U.S.A.\\
\and
Department of Physics, University of Illinois at Urbana-Champaign, 1110 West Green Street, Urbana, Illinois, U.S.A.\\
\and
Department of Statistics, Purdue University, 250 N. University Street, West Lafayette, Indiana, U.S.A.\\
\and
Dipartimento di Fisica e Astronomia G. Galilei, Universit\`{a} degli Studi di Padova, via Marzolo 8, 35131 Padova, Italy\\
\and
Dipartimento di Fisica, Universit\`{a} La Sapienza, P. le A. Moro 2, Roma, Italy\\
\and
Dipartimento di Fisica, Universit\`{a} degli Studi di Milano, Via Celoria, 16, Milano, Italy\\
\and
Dipartimento di Fisica, Universit\`{a} degli Studi di Trieste, via A. Valerio 2, Trieste, Italy\\
\and
Dipartimento di Fisica, Universit\`{a} di Ferrara, Via Saragat 1, 44122 Ferrara, Italy\\
\and
Dipartimento di Fisica, Universit\`{a} di Roma Tor Vergata, Via della Ricerca Scientifica, 1, Roma, Italy\\
\and
Dipartimento di Matematica, Universit\`{a} di Roma Tor Vergata, Via della Ricerca Scientifica, 1, Roma, Italy\\
\and
Discovery Center, Niels Bohr Institute, Blegdamsvej 17, Copenhagen, Denmark\\
\and
Dpto. Astrof\'{i}sica, Universidad de La Laguna (ULL), E-38206 La Laguna, Tenerife, Spain\\
\and
European Southern Observatory, ESO Vitacura, Alonso de Cordova 3107, Vitacura, Casilla 19001, Santiago, Chile\\
\and
European Space Agency, ESAC, Planck Science Office, Camino bajo del Castillo, s/n, Urbanizaci\'{o}n Villafranca del Castillo, Villanueva de la Ca\~{n}ada, Madrid, Spain\\
\and
European Space Agency, ESTEC, Keplerlaan 1, 2201 AZ Noordwijk, The Netherlands\\
\and
GEPI, Observatoire de Paris, Section de Meudon, 5 Place J. Janssen, 92195 Meudon Cedex, France\\
\and
Helsinki Institute of Physics, Gustaf H\"{a}llstr\"{o}min katu 2, University of Helsinki, Helsinki, Finland\\
\and
INAF - Osservatorio Astrofisico di Catania, Via S. Sofia 78, Catania, Italy\\
\and
INAF - Osservatorio Astronomico di Padova, Vicolo dell'Osservatorio 5, Padova, Italy\\
\and
INAF - Osservatorio Astronomico di Roma, via di Frascati 33, Monte Porzio Catone, Italy\\
\and
INAF - Osservatorio Astronomico di Trieste, Via G.B. Tiepolo 11, Trieste, Italy\\
\and
INAF Istituto di Radioastronomia, Via P. Gobetti 101, 40129 Bologna, Italy\\
\and
INAF/IASF Bologna, Via Gobetti 101, Bologna, Italy\\
\and
INAF/IASF Milano, Via E. Bassini 15, Milano, Italy\\
\and
INFN, Sezione di Roma 1, Universit`{a} di Roma Sapienza, Piazzale Aldo Moro 2, 00185, Roma, Italy\\
\and
INRIA, Laboratoire de Recherche en Informatique, Universit\'{e} Paris-Sud 11, B\^{a}timent 490, 91405 Orsay Cedex, France\\
\and
IPAG: Institut de Plan\'{e}tologie et d'Astrophysique de Grenoble, Universit\'{e} Joseph Fourier, Grenoble 1 / CNRS-INSU, UMR 5274, Grenoble, F-38041, France\\
\and
ISDC Data Centre for Astrophysics, University of Geneva, ch. d'Ecogia 16, Versoix, Switzerland\\
\and
IUCAA, Post Bag 4, Ganeshkhind, Pune University Campus, Pune 411 007, India\\
\and
Imperial College London, Astrophysics group, Blackett Laboratory, Prince Consort Road, London, SW7 2AZ, U.K.\\
\and
Infrared Processing and Analysis Center, California Institute of Technology, Pasadena, CA 91125, U.S.A.\\
\and
Institut N\'{e}el, CNRS, Universit\'{e} Joseph Fourier Grenoble I, 25 rue des Martyrs, Grenoble, France\\
\and
Institut Universitaire de France, 103, bd Saint-Michel, 75005, Paris, France\\
\and
Institut d'Astrophysique Spatiale, CNRS (UMR8617) Universit\'{e} Paris-Sud 11, B\^{a}timent 121, Orsay, France\\
\and
Institut d'Astrophysique de Paris, CNRS (UMR7095), 98 bis Boulevard Arago, F-75014, Paris, France\\
\and
Institute for Space Sciences, Bucharest-Magurale, Romania\\
\and
Institute of Astro and Particle Physics, Technikerstrasse 25/8, University of Innsbruck, A-6020, Innsbruck, Austria\\
\and
Institute of Astronomy and Astrophysics, Academia Sinica, Taipei, Taiwan\\
\and
Institute of Astronomy, University of Cambridge, Madingley Road, Cambridge CB3 0HA, U.K.\\
\and
Institute of Theoretical Astrophysics, University of Oslo, Blindern, Oslo, Norway\\
\and
Instituto de Astrof\'{\i}sica de Canarias, C/V\'{\i}a L\'{a}ctea s/n, La Laguna, Tenerife, Spain\\
\and
Instituto de F\'{\i}sica de Cantabria (CSIC-Universidad de Cantabria), Avda. de los Castros s/n, Santander, Spain\\
\and
Jet Propulsion Laboratory, California Institute of Technology, 4800 Oak Grove Drive, Pasadena, California, U.S.A.\\
\and
Jodrell Bank Centre for Astrophysics, Alan Turing Building, School of Physics and Astronomy, The University of Manchester, Oxford Road, Manchester, M13 9PL, U.K.\\
\and
Kapteyn Astronomical Institute, University of Groningen, Landleven 12, 9747 AD Groningen, The Netherlands\\
\and
Kavli Institute for Cosmology Cambridge, Madingley Road, Cambridge, CB3 0HA, U.K.\\
\and
LAL, Universit\'{e} Paris-Sud, CNRS/IN2P3, Orsay, France\\
\and
LERMA, CNRS, Observatoire de Paris, 61 Avenue de l'Observatoire, Paris, France\\
\and
Laboratoire AIM, IRFU/Service d'Astrophysique - CEA/DSM - CNRS - Universit\'{e} Paris Diderot, B\^{a}t. 709, CEA-Saclay, F-91191 Gif-sur-Yvette Cedex, France\\
\and
Laboratoire Traitement et Communication de l'Information, CNRS (UMR 5141) and T\'{e}l\'{e}com ParisTech, 46 rue Barrault F-75634 Paris Cedex 13, France\\
\and
Laboratoire de Physique Subatomique et de Cosmologie, Universit\'{e} Joseph Fourier Grenoble I, CNRS/IN2P3, Institut National Polytechnique de Grenoble, 53 rue des Martyrs, 38026 Grenoble cedex, France\\
\and
Laboratoire de Physique Th\'{e}orique, Universit\'{e} Paris-Sud 11 \& CNRS, B\^{a}timent 210, 91405 Orsay, France\\
\and
Lawrence Berkeley National Laboratory, Berkeley, California, U.S.A.\\
\and
Max-Planck-Institut f\"{u}r Astrophysik, Karl-Schwarzschild-Str. 1, 85741 Garching, Germany\\
\and
Max-Planck-Institut f\"{u}r Extraterrestrische Physik, Giessenbachstra{\ss}e, 85748 Garching, Germany\\
\and
MilliLab, VTT Technical Research Centre of Finland, Tietotie 3, Espoo, Finland\\
\and
Minnesota Institute for Astrophysics, School of Physics and Astronomy, University of Minnesota, 116 Church St. SE, Minneapolis, MN 55455, U.S.A.\\
\and
National University of Ireland, Department of Experimental Physics, Maynooth, Co. Kildare, Ireland\\
\and
Niels Bohr Institute, Blegdamsvej 17, Copenhagen, Denmark\\
\and
Observational Cosmology, Mail Stop 367-17, California Institute of Technology, Pasadena, CA, 91125, U.S.A.\\
\and
Optical Science Laboratory, University College London, Gower Street, London, U.K.\\
\and
SISSA, Astrophysics Sector, via Bonomea 265, 34136, Trieste, Italy\\
\and
School of Physics and Astronomy, Cardiff University, Queens Buildings, The Parade, Cardiff, CF24 3AA, U.K.\\
\and
Space Research Institute (IKI), Profsoyuznaya 84/32, Moscow, Russia\\
\and
Space Research Institute (IKI), Russian Academy of Sciences, Profsoyuznaya Str, 84/32, Moscow, 117997, Russia\\
\and
Space Sciences Laboratory, University of California, Berkeley, California, U.S.A.\\
\and
Stanford University, Dept of Physics, Varian Physics Bldg, 382 Via Pueblo Mall, Stanford, California, U.S.A.\\
\and
T\"{U}B\.{I}TAK National Observatory, Akdeniz University Campus, 07058, Antalya, Turkey\\
\and
UPMC Univ Paris 06, UMR7095, 98 bis Boulevard Arago, F-75014, Paris, France\\
\and
Universit\"{a}t Heidelberg, Institut f\"{u}r Theoretische Astrophysik, Albert-\"{U}berle-Str. 2, 69120, Heidelberg, Germany\\
\and
Universit\'{e} Denis Diderot (Paris 7), 75205 Paris Cedex 13, France\\
\and
Universit\'{e} de Toulouse, UPS-OMP, IRAP, F-31028 Toulouse cedex 4, France\\
% ...
}

%% file: Planck.tex
\def\setsymbol#1#2{\expandafter\def\csname #1\endcsname{#2}}
\def\getsymbol#1{\csname #1\endcsname}

\def\Planck{\textit{Planck}}

\def\HeJT{$^4$He-JT}

\def\allearlypapers{\nocite{planck2011-1.1, planck2011-1.3, planck2011-1.4, planck2011-1.5, planck2011-1.6, planck2011-1.7, planck2011-1.10, planck2011-1.10sup, planck2011-5.1a, planck2011-5.1b, planck2011-5.2a, planck2011-5.2b, planck2011-5.2c, planck2011-6.1, planck2011-6.2, planck2011-6.3a, planck2011-6.4a, planck2011-6.4b, planck2011-6.6, planck2011-7.0, planck2011-7.2, planck2011-7.3, planck2011-7.7a, planck2011-7.7b, planck2011-7.12, planck2011-7.13}}

\newbox\tablebox    \newdimen\tablewidth
\def\leaderfil{\leaders\hbox to 5pt{\hss.\hss}\hfil}
%
%
\def\endPlancktable{\tablewidth=\columnwidth 
    $$\hss\copy\tablebox\hss$$
    \vskip-\lastskip\vskip -2pt}
\def\endPlancktablewide{\tablewidth=\textwidth 
    $$\hss\copy\tablebox\hss$$
    \vskip-\lastskip\vskip -2pt}
\def\tablenote#1 #2\par{\begingroup \parindent=0.8em
    \abovedisplayshortskip=0pt\belowdisplayshortskip=0pt
    \noindent
    $$\hss\vbox{\hsize\tablewidth \hangindent=\parindent \hangafter=1 \noindent
    \hbox to \parindent{$^#1$\hss}\strut#2\strut\par}\hss$$
    \endgroup}
\def\doubleline{\vskip 3pt\hrule \vskip 1.5pt \hrule \vskip 5pt}

%
\def\L2{\ifmmode L_2\else $L_2$\fi}
\def\dtt{\Delta T/T}
\def\DeltaT{\ifmmode \Delta T\else $\Delta T$\fi}
\def\deltat{\ifmmode \Delta t\else $\Delta t$\fi}
\def\fknee{\ifmmode f_{\rm knee}\else $f_{\rm knee}$\fi}
\def\Fmax{\ifmmode F_{\rm max}\else $F_{\rm max}$\fi}
\def\solar{\ifmmode{\rm M}_{\mathord\odot}\else${\rm M}_{\mathord\odot}$\fi}
\def\Msolar{\ifmmode{\rm M}_{\mathord\odot}\else${\rm M}_{\mathord\odot}$\fi}
\def\Lsolar{\ifmmode{\rm L}_{\mathord\odot}\else${\rm L}_{\mathord\odot}$\fi}
\def\mag{\sup{m}}
\def\inv{\ifmmode^{-1}\else$^{-1}$\fi}
\def\mo{\ifmmode^{-1}\else$^{-1}$\fi}
\def\sup#1{\ifmmode ^{\rm #1}\else $^{\rm #1}$\fi}
\def\expo#1{\ifmmode \times 10^{#1}\else $\times 10^{#1}$\fi}
\def\,{\thinspace}
\def\lsim{\mathrel{\raise .4ex\hbox{\rlap{$<$}\lower 1.2ex\hbox{$\sim$}}}}
\def\gsim{\mathrel{\raise .4ex\hbox{\rlap{$>$}\lower 1.2ex\hbox{$\sim$}}}}
\let\lea=\lsim
\let\gea=\gsim
\def\simprop{\mathrel{\raise .4ex\hbox{\rlap{$\propto$}\lower 1.2ex\hbox{$\sim$}}}}
\def\deg{\ifmmode^\circ\else$^\circ$\fi}
\def\pdeg{\ifmmode $\setbox0=\hbox{$^{\circ}$}\rlap{\hskip.11\wd0 .}$^{\circ}
          \else \setbox0=\hbox{$^{\circ}$}\rlap{\hskip.11\wd0 .}$^{\circ}$\fi}
\def\arcs{\ifmmode {^{\scriptstyle\prime\prime}}
          \else $^{\scriptstyle\prime\prime}$\fi}
\def\arcm{\ifmmode {^{\scriptstyle\prime}}
          \else $^{\scriptstyle\prime}$\fi}
\newdimen\sa  \newdimen\sb
\def\parcs{\sa=.07em \sb=.03em
     \ifmmode \hbox{\rlap{.}}^{\scriptstyle\prime\kern -\sb\prime}\hbox{\kern -\sa}
     \else \rlap{.}$^{\scriptstyle\prime\kern -\sb\prime}$\kern -\sa\fi}
\def\parcm{\sa=.08em \sb=.03em
     \ifmmode \hbox{\rlap{.}\kern\sa}^{\scriptstyle\prime}\hbox{\kern-\sb}
     \else \rlap{.}\kern\sa$^{\scriptstyle\prime}$\kern-\sb\fi}
\def\ra[#1 #2 #3.#4]{#1\sup{h}#2\sup{m}#3\sup{s}\llap.#4}
\def\dec[#1 #2 #3.#4]{#1\deg#2\arcm#3\arcs\llap.#4}
\def\deco[#1 #2 #3]{#1\deg#2\arcm#3\arcs}
\def\rra[#1 #2]{#1\sup{h}#2\sup{m}}
\def\page{\vfill\eject}
\def\dots{\relax\ifmmode \ldots\else $\ldots$\fi}
%
%
\def\WHzsr{\ifmmode $W\,Hz\mo\,sr\mo$\else W\,Hz\mo\,sr\mo\fi}
\def\mHz{\ifmmode $\,mHz$\else \,mHz\fi}
\def\GHz{\ifmmode $\,GHz$\else \,GHz\fi}
\def\mKs{\ifmmode $\,mK\,s$^{1/2}\else \,mK\,s$^{1/2}$\fi}
\def\muKs{\ifmmode \,\mu$K\,s$^{1/2}\else \,$\mu$K\,s$^{1/2}$\fi}
\def\muKRJs{\ifmmode \,\mu$K$_{\rm RJ}$\,s$^{1/2}\else \,$\mu$K$_{\rm RJ}$\,s$^{1/2}$\fi}
\def\muKHz{\ifmmode \,\mu$K\,Hz$^{-1/2}\else \,$\mu$K\,Hz$^{-1/2}$\fi}
\def\MJysr{\ifmmode \,$MJy\,sr\mo$\else \,MJy\,sr\mo\fi}
\def\MJysrmK{\ifmmode \,$MJy\,sr\mo$\,mK$_{\rm CMB}\mo\else \,MJy\,sr\mo\,mK$_{\rm CMB}\mo$\fi}
\def\microns{\ifmmode \,\mu$m$\else \,$\mu$m\fi}
\def\micron{\microns}
\def\muK{\ifmmode \,\mu$K$\else \,$\mu$\hbox{K}\fi}
\def\microK{\ifmmode \,\mu$K$\else \,$\mu$\hbox{K}\fi}
\def\muW{\ifmmode \,\mu$W$\else \,$\mu$\hbox{W}\fi}
\def\kms{\ifmmode $\,km\,s$^{-1}\else \,km\,s$^{-1}$\fi}
\def\kmsMpc{\ifmmode $\,\kms\,Mpc\mo$\else \,\kms\,Mpc\mo\fi}
%
%


\setsymbol{LFI:center:frequency:70GHz:units}{70.3\,GHz}
\setsymbol{LFI:center:frequency:44GHz:units}{44.1\,GHz}
\setsymbol{LFI:center:frequency:30GHz:units}{28.5\,GHz}

\setsymbol{LFI:center:frequency:70GHz}{70.3}
\setsymbol{LFI:center:frequency:44GHz}{44.1}
\setsymbol{LFI:center:frequency:30GHz}{28.5}

\setsymbol{LFI:center:frequency:LFI18:Rad:M:units}{71.7\GHz}
\setsymbol{LFI:center:frequency:LFI19:Rad:M:units}{67.5\GHz}
\setsymbol{LFI:center:frequency:LFI20:Rad:M:units}{69.2\GHz}
\setsymbol{LFI:center:frequency:LFI21:Rad:M:units}{70.4\GHz}
\setsymbol{LFI:center:frequency:LFI22:Rad:M:units}{71.5\GHz}
\setsymbol{LFI:center:frequency:LFI23:Rad:M:units}{70.8\GHz}
\setsymbol{LFI:center:frequency:LFI24:Rad:M:units}{44.4\GHz}
\setsymbol{LFI:center:frequency:LFI25:Rad:M:units}{44.0\GHz}
\setsymbol{LFI:center:frequency:LFI26:Rad:M:units}{43.9\GHz}
\setsymbol{LFI:center:frequency:LFI27:Rad:M:units}{28.3\GHz}
\setsymbol{LFI:center:frequency:LFI28:Rad:M:units}{28.8\GHz}
\setsymbol{LFI:center:frequency:LFI18:Rad:S:units}{70.1\GHz}
\setsymbol{LFI:center:frequency:LFI19:Rad:S:units}{69.6\GHz}
\setsymbol{LFI:center:frequency:LFI20:Rad:S:units}{69.5\GHz}
\setsymbol{LFI:center:frequency:LFI21:Rad:S:units}{69.5\GHz}
\setsymbol{LFI:center:frequency:LFI22:Rad:S:units}{72.8\GHz}
\setsymbol{LFI:center:frequency:LFI23:Rad:S:units}{71.3\GHz}
\setsymbol{LFI:center:frequency:LFI24:Rad:S:units}{44.1\GHz}
\setsymbol{LFI:center:frequency:LFI25:Rad:S:units}{44.1\GHz}
\setsymbol{LFI:center:frequency:LFI26:Rad:S:units}{44.1\GHz}
\setsymbol{LFI:center:frequency:LFI27:Rad:S:units}{28.5\GHz}
\setsymbol{LFI:center:frequency:LFI28:Rad:S:units}{28.2\GHz}

\setsymbol{LFI:center:frequency:LFI18:Rad:M}{71.7}
\setsymbol{LFI:center:frequency:LFI19:Rad:M}{67.5}
\setsymbol{LFI:center:frequency:LFI20:Rad:M}{69.2}
\setsymbol{LFI:center:frequency:LFI21:Rad:M}{70.4}
\setsymbol{LFI:center:frequency:LFI22:Rad:M}{71.5}
\setsymbol{LFI:center:frequency:LFI23:Rad:M}{70.8}
\setsymbol{LFI:center:frequency:LFI24:Rad:M}{44.4}
\setsymbol{LFI:center:frequency:LFI25:Rad:M}{44.0}
\setsymbol{LFI:center:frequency:LFI26:Rad:M}{43.9}
\setsymbol{LFI:center:frequency:LFI27:Rad:M}{28.3}
\setsymbol{LFI:center:frequency:LFI28:Rad:M}{28.8}
\setsymbol{LFI:center:frequency:LFI18:Rad:S}{70.1}
\setsymbol{LFI:center:frequency:LFI19:Rad:S}{69.6}
\setsymbol{LFI:center:frequency:LFI20:Rad:S}{69.5}
\setsymbol{LFI:center:frequency:LFI21:Rad:S}{69.5}
\setsymbol{LFI:center:frequency:LFI22:Rad:S}{72.8}
\setsymbol{LFI:center:frequency:LFI23:Rad:S}{71.3}
\setsymbol{LFI:center:frequency:LFI24:Rad:S}{44.1}
\setsymbol{LFI:center:frequency:LFI25:Rad:S}{44.1}
\setsymbol{LFI:center:frequency:LFI26:Rad:S}{44.1}
\setsymbol{LFI:center:frequency:LFI27:Rad:S}{28.5}
\setsymbol{LFI:center:frequency:LFI28:Rad:S}{28.2}


\setsymbol{LFI:white:noise:sensitivity:70GHz:units}{134.7\muKs}
\setsymbol{LFI:white:noise:sensitivity:44GHz:units}{164.7\muKs}
\setsymbol{LFI:white:noise:sensitivity:30GHz:units}{143.4\muKs}

\setsymbol{LFI:white:noise:sensitivity:70GHz}{134.7}
\setsymbol{LFI:white:noise:sensitivity:44GHz}{164.7}
\setsymbol{LFI:white:noise:sensitivity:30GHz}{143.4}


\setsymbol{LFI:white:noise:sensitivity:LFI18:Rad:M:units}{512.0\muKs}
\setsymbol{LFI:white:noise:sensitivity:LFI19:Rad:M:units}{581.4\muKs}
\setsymbol{LFI:white:noise:sensitivity:LFI20:Rad:M:units}{590.8\muKs}
\setsymbol{LFI:white:noise:sensitivity:LFI21:Rad:M:units}{455.2\muKs}
\setsymbol{LFI:white:noise:sensitivity:LFI22:Rad:M:units}{492.0\muKs}
\setsymbol{LFI:white:noise:sensitivity:LFI23:Rad:M:units}{507.7\muKs}
\setsymbol{LFI:white:noise:sensitivity:LFI24:Rad:M:units}{462.2\muKs}
\setsymbol{LFI:white:noise:sensitivity:LFI25:Rad:M:units}{413.6\muKs}
\setsymbol{LFI:white:noise:sensitivity:LFI26:Rad:M:units}{478.6\muKs}
\setsymbol{LFI:white:noise:sensitivity:LFI27:Rad:M:units}{277.7\muKs}
\setsymbol{LFI:white:noise:sensitivity:LFI28:Rad:M:units}{312.3\muKs}
\setsymbol{LFI:white:noise:sensitivity:LFI18:Rad:S:units}{465.7\muKs}
\setsymbol{LFI:white:noise:sensitivity:LFI19:Rad:S:units}{555.6\muKs}
\setsymbol{LFI:white:noise:sensitivity:LFI20:Rad:S:units}{623.2\muKs}
\setsymbol{LFI:white:noise:sensitivity:LFI21:Rad:S:units}{564.1\muKs}
\setsymbol{LFI:white:noise:sensitivity:LFI22:Rad:S:units}{534.4\muKs}
\setsymbol{LFI:white:noise:sensitivity:LFI23:Rad:S:units}{542.4\muKs}
\setsymbol{LFI:white:noise:sensitivity:LFI24:Rad:S:units}{399.2\muKs}
\setsymbol{LFI:white:noise:sensitivity:LFI25:Rad:S:units}{392.6\muKs}
\setsymbol{LFI:white:noise:sensitivity:LFI26:Rad:S:units}{418.6\muKs}
\setsymbol{LFI:white:noise:sensitivity:LFI27:Rad:S:units}{302.9\muKs}
\setsymbol{LFI:white:noise:sensitivity:LFI28:Rad:S:units}{285.3\muKs}

\setsymbol{LFI:white:noise:sensitivity:LFI18:Rad:M}{512.0}
\setsymbol{LFI:white:noise:sensitivity:LFI19:Rad:M}{581.4}
\setsymbol{LFI:white:noise:sensitivity:LFI20:Rad:M}{590.8}
\setsymbol{LFI:white:noise:sensitivity:LFI21:Rad:M}{455.2}
\setsymbol{LFI:white:noise:sensitivity:LFI22:Rad:M}{492.0}
\setsymbol{LFI:white:noise:sensitivity:LFI23:Rad:M}{507.7}
\setsymbol{LFI:white:noise:sensitivity:LFI24:Rad:M}{462.2}
\setsymbol{LFI:white:noise:sensitivity:LFI25:Rad:M}{413.6}
\setsymbol{LFI:white:noise:sensitivity:LFI26:Rad:M}{478.6}
\setsymbol{LFI:white:noise:sensitivity:LFI27:Rad:M}{277.7}
\setsymbol{LFI:white:noise:sensitivity:LFI28:Rad:M}{312.3}
\setsymbol{LFI:white:noise:sensitivity:LFI18:Rad:S}{465.7}
\setsymbol{LFI:white:noise:sensitivity:LFI19:Rad:S}{555.6}
\setsymbol{LFI:white:noise:sensitivity:LFI20:Rad:S}{623.2}
\setsymbol{LFI:white:noise:sensitivity:LFI21:Rad:S}{564.1}
\setsymbol{LFI:white:noise:sensitivity:LFI22:Rad:S}{534.4}
\setsymbol{LFI:white:noise:sensitivity:LFI23:Rad:S}{542.4}
\setsymbol{LFI:white:noise:sensitivity:LFI24:Rad:S}{399.2}
\setsymbol{LFI:white:noise:sensitivity:LFI25:Rad:S}{392.6}
\setsymbol{LFI:white:noise:sensitivity:LFI26:Rad:S}{418.6}
\setsymbol{LFI:white:noise:sensitivity:LFI27:Rad:S}{302.9}
\setsymbol{LFI:white:noise:sensitivity:LFI28:Rad:S}{285.3}


\setsymbol{LFI:knee:frequency:70GHz:units}{29.5\mHz}
\setsymbol{LFI:knee:frequency:44GHz:units}{56.2\mHz}
\setsymbol{LFI:knee:frequency:30GHz:units}{113.7\mHz}

\setsymbol{LFI:knee:frequency:70GHz}{29.5}
\setsymbol{LFI:knee:frequency:44GHz}{56.2}
\setsymbol{LFI:knee:frequency:30GHz}{113.7}

\setsymbol{LFI:knee:frequency:LFI18:Rad:M:units}{16.3\mHz}
\setsymbol{LFI:knee:frequency:LFI19:Rad:M:units}{15.1\mHz}
\setsymbol{LFI:knee:frequency:LFI20:Rad:M:units}{18.7\mHz}
\setsymbol{LFI:knee:frequency:LFI21:Rad:M:units}{37.2\mHz}
\setsymbol{LFI:knee:frequency:LFI22:Rad:M:units}{12.7\mHz}
\setsymbol{LFI:knee:frequency:LFI23:Rad:M:units}{34.6\mHz}
\setsymbol{LFI:knee:frequency:LFI24:Rad:M:units}{46.2\mHz}
\setsymbol{LFI:knee:frequency:LFI25:Rad:M:units}{24.9\mHz}
\setsymbol{LFI:knee:frequency:LFI26:Rad:M:units}{67.6\mHz}
\setsymbol{LFI:knee:frequency:LFI27:Rad:M:units}{187.4\mHz}
\setsymbol{LFI:knee:frequency:LFI28:Rad:M:units}{122.2\mHz}
\setsymbol{LFI:knee:frequency:LFI18:Rad:S:units}{17.7\mHz}
\setsymbol{LFI:knee:frequency:LFI19:Rad:S:units}{22.0\mHz}
\setsymbol{LFI:knee:frequency:LFI20:Rad:S:units}{8.7\mHz}
\setsymbol{LFI:knee:frequency:LFI21:Rad:S:units}{25.9\mHz}
\setsymbol{LFI:knee:frequency:LFI22:Rad:S:units}{15.8\mHz}
\setsymbol{LFI:knee:frequency:LFI23:Rad:S:units}{129.8\mHz}
\setsymbol{LFI:knee:frequency:LFI24:Rad:S:units}{100.9\mHz}
\setsymbol{LFI:knee:frequency:LFI25:Rad:S:units}{38.9\mHz}
\setsymbol{LFI:knee:frequency:LFI26:Rad:S:units}{58.9\mHz}
\setsymbol{LFI:knee:frequency:LFI27:Rad:S:units}{104.4\mHz}
\setsymbol{LFI:knee:frequency:LFI28:Rad:S:units}{40.7\mHz}

\setsymbol{LFI:knee:frequency:LFI18:Rad:M}{16.3}
\setsymbol{LFI:knee:frequency:LFI19:Rad:M}{15.1}
\setsymbol{LFI:knee:frequency:LFI20:Rad:M}{18.7}
\setsymbol{LFI:knee:frequency:LFI21:Rad:M}{37.2}
\setsymbol{LFI:knee:frequency:LFI22:Rad:M}{12.7}
\setsymbol{LFI:knee:frequency:LFI23:Rad:M}{34.6}
\setsymbol{LFI:knee:frequency:LFI24:Rad:M}{46.2}
\setsymbol{LFI:knee:frequency:LFI25:Rad:M}{24.9}
\setsymbol{LFI:knee:frequency:LFI26:Rad:M}{67.6}
\setsymbol{LFI:knee:frequency:LFI27:Rad:M}{187.4}
\setsymbol{LFI:knee:frequency:LFI28:Rad:M}{122.2}
\setsymbol{LFI:knee:frequency:LFI18:Rad:S}{17.7}
\setsymbol{LFI:knee:frequency:LFI19:Rad:S}{22.0}
\setsymbol{LFI:knee:frequency:LFI20:Rad:S}{8.7}
\setsymbol{LFI:knee:frequency:LFI21:Rad:S}{25.9}
\setsymbol{LFI:knee:frequency:LFI22:Rad:S}{15.8}
\setsymbol{LFI:knee:frequency:LFI23:Rad:S}{129.8}
\setsymbol{LFI:knee:frequency:LFI24:Rad:S}{100.9}
\setsymbol{LFI:knee:frequency:LFI25:Rad:S}{38.9}
\setsymbol{LFI:knee:frequency:LFI26:Rad:S}{58.9}
\setsymbol{LFI:knee:frequency:LFI27:Rad:S}{104.4}
\setsymbol{LFI:knee:frequency:LFI28:Rad:S}{40.7}


\setsymbol{LFI:slope:70GHz:units}{$-1.03$\mHz}
\setsymbol{LFI:slope:44GHz:units}{$-0.89$\mHz}
\setsymbol{LFI:slope:30GHz:units}{$-0.87$\mHz}

\setsymbol{LFI:slope:70GHz}{$-1.03$}
\setsymbol{LFI:slope:44GHz}{$-0.89$}
\setsymbol{LFI:slope:30GHz}{$-0.87$}

\setsymbol{LFI:slope:LFI18:Rad:M:units}{$-1.04$\mHz}
\setsymbol{LFI:slope:LFI19:Rad:M:units}{$-1.09$\mHz}
\setsymbol{LFI:slope:LFI20:Rad:M:units}{$-0.69$\mHz}
\setsymbol{LFI:slope:LFI21:Rad:M:units}{$-1.56$\mHz}
\setsymbol{LFI:slope:LFI22:Rad:M:units}{$-1.01$\mHz}
\setsymbol{LFI:slope:LFI23:Rad:M:units}{$-0.96$\mHz}
\setsymbol{LFI:slope:LFI24:Rad:M:units}{$-0.83$\mHz}
\setsymbol{LFI:slope:LFI25:Rad:M:units}{$-0.91$\mHz}
\setsymbol{LFI:slope:LFI26:Rad:M:units}{$-0.95$\mHz}
\setsymbol{LFI:slope:LFI27:Rad:M:units}{$-0.87$\mHz}
\setsymbol{LFI:slope:LFI28:Rad:M:units}{$-0.88$\mHz}
\setsymbol{LFI:slope:LFI18:Rad:S:units}{$-1.15$\mHz}
\setsymbol{LFI:slope:LFI19:Rad:S:units}{$-1.00$\mHz}
\setsymbol{LFI:slope:LFI20:Rad:S:units}{$-0.95$\mHz}
\setsymbol{LFI:slope:LFI21:Rad:S:units}{$-0.92$\mHz}
\setsymbol{LFI:slope:LFI22:Rad:S:units}{$-1.01$\mHz}
\setsymbol{LFI:slope:LFI23:Rad:S:units}{$-0.95$\mHz}
\setsymbol{LFI:slope:LFI24:Rad:S:units}{$-0.73$\mHz}
\setsymbol{LFI:slope:LFI25:Rad:S:units}{$-1.16$\mHz}
\setsymbol{LFI:slope:LFI26:Rad:S:units}{$-0.79$\mHz}
\setsymbol{LFI:slope:LFI27:Rad:S:units}{$-0.82$\mHz}
\setsymbol{LFI:slope:LFI28:Rad:S:units}{$-0.91$\mHz}

\setsymbol{LFI:slope:LFI18:Rad:M}{$-1.04$}
\setsymbol{LFI:slope:LFI19:Rad:M}{$-1.09$}
\setsymbol{LFI:slope:LFI20:Rad:M}{$-0.69$}
\setsymbol{LFI:slope:LFI21:Rad:M}{$-1.56$}
\setsymbol{LFI:slope:LFI22:Rad:M}{$-1.01$}
\setsymbol{LFI:slope:LFI23:Rad:M}{$-0.96$}
\setsymbol{LFI:slope:LFI24:Rad:M}{$-0.83$}
\setsymbol{LFI:slope:LFI25:Rad:M}{$-0.91$}
\setsymbol{LFI:slope:LFI26:Rad:M}{$-0.95$}
\setsymbol{LFI:slope:LFI27:Rad:M}{$-0.87$}
\setsymbol{LFI:slope:LFI28:Rad:M}{$-0.88$}
\setsymbol{LFI:slope:LFI18:Rad:S}{$-1.15$}
\setsymbol{LFI:slope:LFI19:Rad:S}{$-1.00$}
\setsymbol{LFI:slope:LFI20:Rad:S}{$-0.95$}
\setsymbol{LFI:slope:LFI21:Rad:S}{$-0.92$}
\setsymbol{LFI:slope:LFI22:Rad:S}{$-1.01$}
\setsymbol{LFI:slope:LFI23:Rad:S}{$-0.95$}
\setsymbol{LFI:slope:LFI24:Rad:S}{$-0.73$}
\setsymbol{LFI:slope:LFI25:Rad:S}{$-1.16$}
\setsymbol{LFI:slope:LFI26:Rad:S}{$-0.79$}
\setsymbol{LFI:slope:LFI27:Rad:S}{$-0.82$}
\setsymbol{LFI:slope:LFI28:Rad:S}{$-0.91$}


\setsymbol{LFI:FWHM:70GHz:units}{13\parcm01}
\setsymbol{LFI:FWHM:44GHz:units}{27\parcm92}
\setsymbol{LFI:FWHM:30GHz:units}{32\parcm65}

\setsymbol{LFI:FWHM:70GHz}{13.01}
\setsymbol{LFI:FWHM:44GHz}{27.92}
\setsymbol{LFI:FWHM:30GHz}{32.65}

\setsymbol{LFI:FWHM:LFI18:units}{13\parcm39}
\setsymbol{LFI:FWHM:LFI19:units}{13\parcm01}
\setsymbol{LFI:FWHM:LFI20:units}{12\parcm75}
\setsymbol{LFI:FWHM:LFI21:units}{12\parcm74}
\setsymbol{LFI:FWHM:LFI22:units}{12\parcm87}
\setsymbol{LFI:FWHM:LFI23:units}{13\parcm27}
\setsymbol{LFI:FWHM:LFI24:units}{22\parcm98}
\setsymbol{LFI:FWHM:LFI25:units}{30\parcm46}
\setsymbol{LFI:FWHM:LFI26:units}{30\parcm31}
\setsymbol{LFI:FWHM:LFI27:units}{32\parcm65}
\setsymbol{LFI:FWHM:LFI28:units}{32\parcm66}

\setsymbol{LFI:FWHM:LFI18}{13.39}
\setsymbol{LFI:FWHM:LFI19}{13.01}
\setsymbol{LFI:FWHM:LFI20}{12.75}
\setsymbol{LFI:FWHM:LFI21}{12.74}
\setsymbol{LFI:FWHM:LFI22}{12.87}
\setsymbol{LFI:FWHM:LFI23}{13.27}
\setsymbol{LFI:FWHM:LFI24}{22.98}
\setsymbol{LFI:FWHM:LFI25}{30.46}
\setsymbol{LFI:FWHM:LFI26}{30.31}
\setsymbol{LFI:FWHM:LFI27}{32.65}
\setsymbol{LFI:FWHM:LFI28}{32.66}



\setsymbol{LFI:FWHM:uncertainty:LFI18:units}{0.170\arcm}
\setsymbol{LFI:FWHM:uncertainty:LFI19:units}{0.174\arcm}
\setsymbol{LFI:FWHM:uncertainty:LFI20:units}{0.170\arcm}
\setsymbol{LFI:FWHM:uncertainty:LFI21:units}{0.156\arcm}
\setsymbol{LFI:FWHM:uncertainty:LFI22:units}{0.164\arcm}
\setsymbol{LFI:FWHM:uncertainty:LFI23:units}{0.171\arcm}
\setsymbol{LFI:FWHM:uncertainty:LFI24:units}{0.652\arcm}
\setsymbol{LFI:FWHM:uncertainty:LFI25:units}{1.075\arcm}
\setsymbol{LFI:FWHM:uncertainty:LFI26:units}{1.131\arcm}
\setsymbol{LFI:FWHM:uncertainty:LFI27:units}{1.266\arcm}
\setsymbol{LFI:FWHM:uncertainty:LFI28:units}{1.287\arcm}

\setsymbol{LFI:FWHM:uncertainty:LFI18}{0.170}
\setsymbol{LFI:FWHM:uncertainty:LFI19}{0.174}
\setsymbol{LFI:FWHM:uncertainty:LFI20}{0.170}
\setsymbol{LFI:FWHM:uncertainty:LFI21}{0.156}
\setsymbol{LFI:FWHM:uncertainty:LFI22}{0.164}
\setsymbol{LFI:FWHM:uncertainty:LFI23}{0.171}
\setsymbol{LFI:FWHM:uncertainty:LFI24}{0.652}
\setsymbol{LFI:FWHM:uncertainty:LFI25}{1.075}
\setsymbol{LFI:FWHM:uncertainty:LFI26}{1.131}
\setsymbol{LFI:FWHM:uncertainty:LFI27}{1.266}
\setsymbol{LFI:FWHM:uncertainty:LFI28}{1.287}


\setsymbol{HFI:center:frequency:100GHz:units}{100\,GHz}
\setsymbol{HFI:center:frequency:143GHz:units}{143\,GHz}
\setsymbol{HFI:center:frequency:217GHz:units}{217\,GHz}
\setsymbol{HFI:center:frequency:353GHz:units}{353\,GHz}
\setsymbol{HFI:center:frequency:545GHz:units}{545\,GHz}
\setsymbol{HFI:center:frequency:857GHz:units}{857\,GHz}

\setsymbol{HFI:center:frequency:100GHz}{100}
\setsymbol{HFI:center:frequency:143GHz}{143}
\setsymbol{HFI:center:frequency:217GHz}{217}
\setsymbol{HFI:center:frequency:353GHz}{353}
\setsymbol{HFI:center:frequency:545GHz}{545}
\setsymbol{HFI:center:frequency:857GHz}{857}


\setsymbol{HFI:Ndetectors:100GHz}{8}
\setsymbol{HFI:Ndetectors:143GHz}{11}
\setsymbol{HFI:Ndetectors:217GHz}{12}
\setsymbol{HFI:Ndetectors:353GHz}{12}
\setsymbol{HFI:Ndetectors:545GHz}{3}
\setsymbol{HFI:Ndetectors:857GHz}{4}


\setsymbol{HFI:FWHM:Maps:100GHz:units}{9\parcm88}
\setsymbol{HFI:FWHM:Maps:143GHz:units}{7\parcm18}
\setsymbol{HFI:FWHM:Maps:217GHz:units}{4\parcm87}
\setsymbol{HFI:FWHM:Maps:353GHz:units}{4\parcm65}
\setsymbol{HFI:FWHM:Maps:545GHz:units}{4\parcm72}
\setsymbol{HFI:FWHM:Maps:857GHz:units}{4\parcm39}
\setsymbol{HFI:FWHM:Maps:100GHz}{9.88}
\setsymbol{HFI:FWHM:Maps:143GHz}{7.18}
\setsymbol{HFI:FWHM:Maps:217GHz}{4.87}
\setsymbol{HFI:FWHM:Maps:353GHz}{4.65}
\setsymbol{HFI:FWHM:Maps:545GHz}{4.72}
\setsymbol{HFI:FWHM:Maps:857GHz}{4.39}


\setsymbol{HFI:beam:ellipticity:Maps:100GHz}{1.15}
\setsymbol{HFI:beam:ellipticity:Maps:143GHz}{1.01}
\setsymbol{HFI:beam:ellipticity:Maps:217GHz}{1.06}
\setsymbol{HFI:beam:ellipticity:Maps:353GHz}{1.05}
\setsymbol{HFI:beam:ellipticity:Maps:545GHz}{1.14}
\setsymbol{HFI:beam:ellipticity:Maps:857GHz}{1.19}


\setsymbol{HFI:FWHM:Mars:100GHz:units}{9\parcm37}
\setsymbol{HFI:FWHM:Mars:143GHz:units}{7\parcm04}
\setsymbol{HFI:FWHM:Mars:217GHz:units}{4\parcm68}
\setsymbol{HFI:FWHM:Mars:353GHz:units}{4\parcm43}
\setsymbol{HFI:FWHM:Mars:545GHz:units}{3\parcm80}
\setsymbol{HFI:FWHM:Mars:857GHz:units}{3\parcm67}

\setsymbol{HFI:FWHM:Mars:100GHz}{9.37}
\setsymbol{HFI:FWHM:Mars:143GHz}{7.04}
\setsymbol{HFI:FWHM:Mars:217GHz}{4.68}
\setsymbol{HFI:FWHM:Mars:353GHz}{4.43}
\setsymbol{HFI:FWHM:Mars:545GHz}{3.80}
\setsymbol{HFI:FWHM:Mars:857GHz}{3.67}


\setsymbol{HFI:beam:ellipticity:Mars:100GHz}{1.18}
\setsymbol{HFI:beam:ellipticity:Mars:143GHz}{1.03}
\setsymbol{HFI:beam:ellipticity:Mars:217GHz}{1.14}
\setsymbol{HFI:beam:ellipticity:Mars:353GHz}{1.09}
\setsymbol{HFI:beam:ellipticity:Mars:545GHz}{1.25}
\setsymbol{HFI:beam:ellipticity:Mars:857GHz}{1.03}


\setsymbol{HFI:CMB:relative:calibration:100GHz}{$\lsim 1\%$}
\setsymbol{HFI:CMB:relative:calibration:143GHz}{$\lsim 1\%$}
\setsymbol{HFI:CMB:relative:calibration:217GHz}{$\lsim 1\%$}
\setsymbol{HFI:CMB:relative:calibration:353GHz}{$\lsim 1\%$}
\setsymbol{HFI:CMB:relative:calibration:545GHz}{}
\setsymbol{HFI:CMB:relative:calibration:857GHz}{}


\setsymbol{HFI:CMB:absolute:calibration:100GHz}{$\lsim 2\%$}
\setsymbol{HFI:CMB:absolute:calibration:143GHz}{$\lsim 2\%$}
\setsymbol{HFI:CMB:absolute:calibration:217GHz}{$\lsim 2\%$}
\setsymbol{HFI:CMB:absolute:calibration:353GHz}{$\lsim 2\%$}
\setsymbol{HFI:CMB:absolute:calibration:545GHz}{}
\setsymbol{HFI:CMB:absolute:calibration:857GHz}{}


\setsymbol{HFI:FIRAS:gain:calibration:accuracy:statistical:100GHz}{}
\setsymbol{HFI:FIRAS:gain:calibration:accuracy:statistical:143GHz}{}
\setsymbol{HFI:FIRAS:gain:calibration:accuracy:statistical:217GHz}{}
\setsymbol{HFI:FIRAS:gain:calibration:accuracy:statistical:353GHz}{2.5\%}
\setsymbol{HFI:FIRAS:gain:calibration:accuracy:statistical:545GHz}{1\%}
\setsymbol{HFI:FIRAS:gain:calibration:accuracy:statistical:857GHz}{0.5\%}


\setsymbol{HFI:FIRAS:gain:calibration:accuracy:systematic:100GHz}{}
\setsymbol{HFI:FIRAS:gain:calibration:accuracy:systematic:143GHz}{}
\setsymbol{HFI:FIRAS:gain:calibration:accuracy:systematic:217GHz}{}
\setsymbol{HFI:FIRAS:gain:calibration:accuracy:systematic:353GHz}{}
\setsymbol{HFI:FIRAS:gain:calibration:accuracy:systematic:545GHz}{7\%}
\setsymbol{HFI:FIRAS:gain:calibration:accuracy:systematic:857GHz}{7\%}


\setsymbol{HFI:FIRAS:zero:point:accuracy:100GHz:units}{0.8\MJysr}
\setsymbol{HFI:FIRAS:zero:point:accuracy:143GHz:units}{}
\setsymbol{HFI:FIRAS:zero:point:accuracy:217GHz:units}{}
\setsymbol{HFI:FIRAS:zero:point:accuracy:353GHz:units}{1.4\MJysr}
\setsymbol{HFI:FIRAS:zero:point:accuracy:545GHz:units}{2.2\MJysr}
\setsymbol{HFI:FIRAS:zero:point:accuracy:857GHz:units}{1.7\MJysr}

\setsymbol{HFI:FIRAS:zero:point:accuracy:100GHz}{0.8}
\setsymbol{HFI:FIRAS:zero:point:accuracy:143GHz}{}
\setsymbol{HFI:FIRAS:zero:point:accuracy:217GHz}{}
\setsymbol{HFI:FIRAS:zero:point:accuracy:353GHz}{1.4}
\setsymbol{HFI:FIRAS:zero:point:accuracy:545GHz}{2.2}
\setsymbol{HFI:FIRAS:zero:point:accuracy:857GHz}{1.7}


\setsymbol{HFI:unit:conversion:100GHz:units}{0.2415\MJysrmK}
\setsymbol{HFI:unit:conversion:143GHz:units}{0.3694\MJysrmK}
\setsymbol{HFI:unit:conversion:217GHz:units}{0.4811\MJysrmK}
\setsymbol{HFI:unit:conversion:353GHz:units}{0.2883\MJysrmK}
\setsymbol{HFI:unit:conversion:545GHz:units}{0.05826\MJysrmK}
\setsymbol{HFI:unit:conversion:857GHz:units}{0.002238\MJysrmK}

\setsymbol{HFI:unit:conversion:100GHz}{0.2415}
\setsymbol{HFI:unit:conversion:143GHz}{0.3694}
\setsymbol{HFI:unit:conversion:217GHz}{0.4811}
\setsymbol{HFI:unit:conversion:353GHz}{0.2883}
\setsymbol{HFI:unit:conversion:545GHz}{0.05826}
\setsymbol{HFI:unit:conversion:857GHz}{0.002238}


\setsymbol{HFI:colour:correction:alpha=-2:V1.01:100GHz}{0.9893}
\setsymbol{HFI:colour:correction:alpha=-2:V1.01:143GHz}{0.9759}
\setsymbol{HFI:colour:correction:alpha=-2:V1.01:217GHz}{1.0007}
\setsymbol{HFI:colour:correction:alpha=-2:V1.01:353GHz}{1.0028}
\setsymbol{HFI:colour:correction:alpha=-2:V1.01:545GHz}{1.0019}
\setsymbol{HFI:colour:correction:alpha=-2:V1.01:857GHz}{0.9889}


\setsymbol{HFI:colour:correction:alpha=0:V1.01:100GHz}{1.0008}
\setsymbol{HFI:colour:correction:alpha=0:V1.01:143GHz}{1.0148}
\setsymbol{HFI:colour:correction:alpha=0:V1.01:217GHz}{0.9909}
\setsymbol{HFI:colour:correction:alpha=0:V1.01:353GHz}{0.9888}
\setsymbol{HFI:colour:correction:alpha=0:V1.01:545GHz}{0.9878}
\setsymbol{HFI:colour:correction:alpha=0:V1.01:857GHz}{1.0014}